\documentclass[a4paper, 11pt]{article}
\usepackage{amssymb}
\usepackage{epsfig}
\usepackage{graphics}
\usepackage{afterpage}
\usepackage{enumerate}
\psfigdriver{dvips}

\begin{document}
%
%
%


\let\jnlstyle=\rm
\def\refjnl#1{{\jnlstyle#1}}

\def\aj{\refjnl{AJ}}                   
\def\araa{\refjnl{ARA\&A}}             
\def\apj{\refjnl{ApJ}}                 
\def\apjl{\refjnl{ApJ}}                
\def\apjs{\refjnl{ApJS}}               
\def\ao{\refjnl{Appl.~Opt.}}           
\def\apss{\refjnl{Ap\&SS}}             
\def\aap{\refjnl{A\&A}}                
\def\aapr{\refjnl{A\&A~Rev.}}          
\def\aaps{\refjnl{A\&AS}}              
\def\azh{\refjnl{AZh}}                 
\def\baas{\refjnl{BAAS}}               
\def\jrasc{\refjnl{JRASC}}             
\def\memras{\refjnl{MmRAS}}            
\def\mnras{\refjnl{MNRAS}}             
\def\pra{\refjnl{Phys.~Rev.~A}}        
\def\prb{\refjnl{Phys.~Rev.~B}}        
\def\prc{\refjnl{Phys.~Rev.~C}}        
\def\prd{\refjnl{Phys.~Rev.~D}}        
\def\pre{\refjnl{Phys.~Rev.~E}}        
\def\prl{\refjnl{Phys.~Rev.~Lett.}}    
\def\pasp{\refjnl{PASP}}               
\def\pasj{\refjnl{PASJ}}               
\def\qjras{\refjnl{QJRAS}}             
\def\skytel{\refjnl{S\&T}}             
\def\solphys{\refjnl{Sol.~Phys.}}      
\def\sovast{\refjnl{Soviet~Ast.}}      
\def\ssr{\refjnl{Space~Sci.~Rev.}}     
\def\zap{\refjnl{ZAp}}                 
\def\nat{\refjnl{Nature}}              
\def\iaucirc{\refjnl{IAU~Circ.}}       
\def\aplett{\refjnl{Astrophys.~Lett.}} 
\def\apspr{\refjnl{Astrophys.~Space~Phys.~Res.}}
\def\bain{\refjnl{Bull.~Astron.~Inst.~Netherlands}} 
\def\fcp{\refjnl{Fund.~Cosmic~Phys.}}  
\def\gca{\refjnl{Geochim.~Cosmochim.~Acta}}   
\def\grl{\refjnl{Geophys.~Res.~Lett.}} 
\def\jcp{\refjnl{J.~Chem.~Phys.}}      
\def\jgr{\refjnl{J.~Geophys.~Res.}}    
\def\jqsrt{\refjnl{J.~Quant.~Spec.~Radiat.~Transf.}}
\def\memsai{\refjnl{Mem.~Soc.~Astron.~Italiana}}
\def\nphysa{\refjnl{Nucl.~Phys.~A}}   
\def\physrep{\refjnl{Phys.~Rep.}}   
\def\physscr{\refjnl{Phys.~Scr}}   
\def\planss{\refjnl{Planet.~Space~Sci.}}   
\def\procspie{\refjnl{Proc.~SPIE}}   

\let\astap=\aap
\let\apjlett=\apjl
\let\apjsupp=\apjs
\let\applopt=\ao

\title{Deep optical imaging of the field of PC1643+4631A\&B, I:
Spatial distributions and the counts of faint galaxies.}

\author{Toby Haynes$^{\! 1}$, Garret Cotter$^{\! 1}$,\\ 
Joanne C. Baker$^{\! 2}$, Steve Eales$^{\! 3}$, Michael E. Jones$^{\! 1}$,\\ 
Steve Rawlings$^{\! 4}$, Richard Saunders$^{\! 1}$\\ \\
$^1$ Astrophysics, Cavendish Laboratory,\\ 
Madingley Road, Cambridge, CB3 0HE, UK\\
$^2$ Astronomy Department, University of California, \\
Berkeley CA 94720, USA\\
$^3$ Department of Physics and Astronomy,\\ 
University of Cardiff, Cardiff CF2 3YB, UK\\
$^4$ Astrophysics, Department of Physics, Keeble Road,\\
Oxford, OX1 3RH, UK\\
}

\maketitle

\begin{abstract}
We present deep optical images of the PC1643+4631 field obtained at
the WHT. This field contains two quasars at redshifts z=3.79 \& 3.83
and a cosmic microwave background (CMB) decrement detected with the
Ryle Telescope. The images are in $U,G,V,R$ and $I$ filters, and are
complete to 25th magnitude in $R$ and $G$ and to 25.5 in $U$. The
isophotal galaxy counts are consistent with the results of
\cite{MSC96}, \cite{Hogg97}, and others. We find an excess of robust
high--redshift Ly--break galaxy candidates with $25.0<R<25.5$ compared
with the mean number found in the fields studied by Steidel et al.\ --
we expect 7 but find 16 -- but we do not find that the galaxies are
concentrated in the direction of the CMB decrement. However, we are
still not sure of the distance to the system causing the CMB
decrement. We have also used our images to compare the commonly used
object--finding algorithms of FOCAS and SExtractor: we find FOCAS the
more efficient at detecting faint objects and the better at dealing
with composite objects, whereas SExtractor's morphological
classification is more reliable, especially for faint objects near the
resolution limit. More generally, we have also compared the flux lost
using isophotal apertures on a real image with that on a noise--only
image: recovery of artificial galaxies from the noise--only image
significantly overestimates the flux lost from the galaxies, and we
find that the corrections made using this technique suffer a
systematic error of some 0.4 magnitudes.
\end{abstract}

\section{Introduction}

In April 1997 we reported (\cite{Jones97}) the detection, using the
Ryle Telescope, of a decrement in the cosmic microwave background
(CMB) towards the $3.1'$ separation quasar pair PC1643+4631A\&B
(\cite{ptgs94}) which have redshifts $z=3.79$ \& $3.83$
respectively. We suggested that the decrement is most likely to be
caused by the Sunyaev--Zel'dovich {(S--Z)} effect of the ionising gas
in a $10^{15}M_{\odot}$ system. The decrement fortuitously lay in the
field of a ROSAT pointed observation, and from the X--ray upper limit
we concluded that such an S--Z producing system must lie at $z>1$ if
it were similar to known massive clusters, or, if it lay closer, it
must be rarefied (with a targetted ROSAT observation, Kneissl et
al. report an increased redshift limit). We simultaneously reported
(\cite{Saun97}) $R$, $J$ and $K$ imaging of the field together with
new optical spectroscopy of the quasar pair, which further supported
the notion that the system responsible for the decrement must be
either very distant or dark, and we also suggested how, despite the
apparent redshift difference, a massive system at $z\sim2$ could
gravitationally lens a single quasar and create the images A and
B. Thus there appear to be three possibilities: (a) there is a massive
system at $z = 3.8$ traced by quasars A and B; (b) there is a massive
system at $ z \approx 2 $ with quasars A and B separate objects
gravitationally lensed by the massive system; and (c) there is a
massive system at $z \approx 2$ causing multiple imaging of a single
background quasar.  If there is a significant population of such
systems, this may challenge theories of structure formation in the
universe. We therefore embarked on a further series of follow--up
observations to try to find the system responsible for the CMB
decrement. Here we present deep, multicolour optical images. Unless
otherwise stated, we take {$H_0=50$ km s$^{-1}$ Mpc$^{-1}$},
$\Omega_0=1.0$ and $\Lambda=0$.  All magnitudes are given in AB mags
(see \cite{OG83}).

We note that Richards et al. (1997) find a CMB decrement in the field
of a $z = 2.561$ quasar pair. Recently Campos et al. (1998) have found
spectroscopic evidence for a galaxy system at the same redshift as the
quasars in this field.

\section{Observing Strategy}

Given the lack of any evidence for a cluster in our previous optical
and infrared observations to $R\approx23$, $J\approx22$ \&
$K\approx20$, it was clear that very deep observations would be
essential to identify this system.

We aimed to find Lyman--break galaxies (see, for example
\cite{SGP96}), whose colour signatures can be found using broadband
filters. For example, $U$ dropouts will occur as the Lyman limit moves
into the $U$ filter at $z\approx3$. At intermediate redshifts
($1<z<2.5$) there will not be such clear colour signatures, but
many--colour work does provide some constraints given colour modelling
(see eg.\ \cite{HuR94} \&\ \cite{StII}).

There are other indicators of possible structures in the field we
which also aimed to follow up. The spectrum of Quasar A shows a damped
Ly-$\alpha$ absorption system at $z=3.14$, which has been the focus
of several previous investigations (e.g.\ \cite{FBB94}). Narrow--band
imaging at the wavelength of this Ly--$\alpha$ absorption might
provide evidence for a concentration of galaxies at this
redshift. Similarly, narrow--band imaging at the wavelength of
Ly-$\alpha$ emission in the quasar spectra might identify any
concentration of galaxies associated with the quasars. However, the
failure of many previous searches for high--redshift galaxies by
Ly-$\alpha$ imaging, see eg \cite{TDT95}, and additionally the small
equivalent widths of \mbox{Ly-$\alpha$} in those galaxies discovered
by Steidel et al, indicate the difficulties of finding a
high--redshift cluster this way. Nevertheless, we had two custom
narrow--band interference filters constructed. The first, L5840, is a
2\%\ fractional bandwidth filter centred on 5840\AA\ to detect
Lyman-$\alpha$ emission in the interval $3.79<z<3.81$. The second,
S5040, is a 1\%\ filter centred on 5040\AA\ and corresponds to
Ly-$\alpha$ at $z=3.14$.

Our previous observations on UKIRT and WHT had relied on the mosaicing
of seven small $1'\times1'$ images. Data reduction and analysis are
significantly simplified by imaging the whole field at once, and for
this we used the large format Tek 2 CCD positioned at prime focus at
the WHT. This gives a $1024\times1024$ pixel image, covering a field
of view of $7'\times7'$, with a pixel size of $0.422''$.

\section{Data Acquisition}

Imaging was done in parts of four consecutive nights at the WHT, from
15th April 1996.

Five broadband filters were used: $U,V,R$ \&\ $I$ filters were taken
from the Harris set and were supplemented with the $G$ filter (lent to
us by Richard McMahon), which has a characteristic wavelength of
4900\AA\ and a width of 1000\AA. This was used instead of the Harris B
filter for its superior transparency and as a more sensitive probe of
Lyman limit imaging as undertaken by Steidel and collaborators. These
broadband filters were complemented by our two custom--made narrow-band
filters.

Images in each filter were taken at dithering positions separated by
$15''$ in a square grid arrangement. This allows an accurate mapping
of the background level over most of the image, although it does not
perfectly produce a complete background map. Since only the
overlapping areas of the images can be used for cataloguing and
analysis, this method reduced the effective field of view from
$7'\times7'$ to $5'\times5'$. Individual exposure times ranged from
300s in $I$ to 900s in $U$ and were chosen to be long enough to be
background limited, while ensuring that a minimum of pixels were
saturated. Sky flats were taken for each filter used on each night to
aid correction of the background levels in each image.

Conditions generally were photometric -- no observation made during
periods of poor transparency and/or high airmass was used in the
production of the final images. In total, over 33 ks of exposures were
used (see Table~\ref{tab:obslog}). Only the best transparency $U$
images were used in making the final images and subsequent catalogues.

Landolt calibrators were observed in each filter on each night to
allow calibration of the $U,V,R$ \& $I$ filters, while
spectrophotometric calibrators were observed in order to calibrate the
non-standard $G$ filter and the two narrow--band filters $L5840$ and
$S5040$.

We decided to work on the AB magnitude scale which allows direct
comparison between magnitude and spectral energy distribution.
Conversion of the calibrators from Johnson to AB magnitudes was done
by comparing, in each colour, the calibrator flux with that of a
bright Johnson calibrator for which a spectrum was also available.

Where we had observations in the same filter on different nights with
independent calibrations, we used these to check the photometric
consistency of the observations. In the case of the $U$ images, data
from one of the three nights showed a discrepancy in the calibrated
magnitude of objects in the field against the rest of the $U$
calibrations. This was corrected for by calibrating that night's
observations using direct comparison with the other self--consistent
composite $U$ images.

\section{Data Reduction}

\subsection{Processing of images} 
\label{sec:reduct}            

The raw images were reduced using the standard IRAF package
(\cite{IRAF93}). Bias exposures taken on each night were stacked and
checked for structure and consistency with the overscan regions of the
CCD -- no structure was found and the bias level was determined for
each image using the bias strip and subtracted out. All images were
initially flattened with sky flats and the fluxes, point spread
function and background levels of bright but unsaturated stars in each
image were measured. This allowed the identification of images which
were adversely affected by atmospheric conditions, such as high cirrus
cloud, by comparing the trends in these measurements as a function of
airmass. Images which showed unexpected behaviour, such as a
significant drop in flux, were discarded. Sky background levels and
rms pixel noise were also measured and used to eliminate
non-photometric frames. However, acceptably levelled backgrounds (i.e.
in which the standard deviation of the background level was consistent
with the pixel noise) could not, initially, be obtained by flattening
the image frames with the sky flats. Comparison of the sky flats
between nights also revealed that there were changes in the
illumination pattern between observations, requiring that all frames
be processed on a night--by--night basis.

The non--level background was caused by four bright stars, with $V$
magnitudes between $12-16$, which produce a noticeable distortion of
the background level out to $40''$ radius surrounding each star. Using
the approach of \cite{StII}, of combining the dithered
frames to produce a median frame devoid of objects and containing only 
the background level, results in an overestimate of the sky background
level around these stars and a corresponding bowling of the flattened
images after this illumination correction is made. In these regions,
the spacing of the dithered positions is too small to provide a good
estimate of the background level. This problem would have been
alleviated by much larger dithering steps, but this would have meant
that the central overlap region for the dithered position was reduced
to an unacceptably small size.

We overcame this problem as follows. The individual frames in each
filter on each night were averaged together, discarding the highest
and lowest counts for each pixel stack, to create a data image
flat. This has a few sources left from the field, but has a lower
noise than the individual frames. We then used the cataloguing program
SExtractor (Source Extractor, \cite{Sext}, see
section~\ref{sec:catcreat}) to produce a heavily smoothed background
image. This still contains an overestimated background level in the
regions around the three brightest stars; these regions were inspected
individually and replaced with an estimate of the background
level. This was a surface fitted with a second--order polynomial to an
annulus immediately adjoining the replaced region. The resulting frame
was then used as an illumination correction to all the images in that
filter on that night. This gave a background across the whole image,
including the affected areas, that was acceptably flat -- i.e. the
standard deviation of the sky level was consistent with that expected
from the pixel noise.

The $I$ images were additionally contaminated with sky fringes. After
processing the images as described, the fringes were examined and seen
to be similarly distributed in each image. They were mapped using a
median image and subtracted from each $I$--image. Linear fitting of
the fringes to each image prior to subtraction was investigated but no
significant improvement in the signal:noise ratio was gained.

Once flattened, the images were aligned and stacked. Observations of
photometric and spectrophotometric standard stars were finally used to
obtain flux calibration for all images in the AB magnitude system.
Limiting magnitudes for these images were calculated using the rms
pixel noise (see Table\ref{tab:calibrat}).

\subsection{Catalogue creation -- comparison of FOCAS and SExtractor}
\label{sec:catcreat}

\sloppy{ We used two cataloguing programs -- FOCAS (\cite{FOCAS81}) and
SExtractor -- in order to provide a comparison between the available techniques
for detecting, measuring and classifying the objects in the fields. In
both cases isophotal apertures were used with the isophotal levels
being defined at three times the rms noise in the image. The
parameters for subdivision of multiple objects in both pieces of
software were optimised to correctly separate real objects from close
companions. This was done with particular reference to separating
Quasar A and its close companion and also the fainter objects nearer
to the brightest stars in the image. Identical convolution filters
(i.e. the built-in FOCAS detection filter) were used in both cases for
detection and allocation of isophotal apertures.}

Isophotal apertures were made using FOCAS in each of the five
broadband filters $U$,$G$,$V$,$R$ and $I$, and applied to each of the
seven images to create 35 individual catalogues. Sources with
isophotal apertures of smaller size than the effective seeing disc
were rejected from the isophotal apertures. From these catalogues, six
matched catalogues were then created: five catalogues based on each
set of broadband isophotal apertures, and one catalogue in which each
broad--band image used its own set of isophotal apertures, with the
two narrow--band filters using the isophotes defined by the deep R
image. (No significant bias is expected by imposing the isophotes from
the $R$ image on the much less deep narrow--band images -- only
objects with $G-R>4$ would be lost from the $S5040$ filter -- an
extreme colour not observed in the catalogues.)

A second set of catalogues was also made using SExtractor, also using
isophotal apertures. From each image, SExtractor was used to create a
background map and a segmentation map. The segmentation map shows how
the software has detected and, in the case of composite objects, split
the objects in the field. This can be used as isophotal apertures in a
similar way to FOCAS. Catalogues were then constructed by measuring
the counts inside each aperture from each background--subtracted
image, for each of the seven images using the five isophotal aperture
sets as before, to create 35 catalogues. Other parameters, such as
intensity weighted position and photometry errors were calculated from
the data, using the same algorithms as SExtractor. Because this method
relies on the accurate mapping of the background level over a large
area rather than immediately around each object as is the case with
FOCAS, photometric measuring errors are likely to be worse with
SExtractor than FOCAS for crowded areas of the plate.

\subsection[Different isophotal apertures]{Comparison of magnitudes measured using different isophotal apertures}

\begin{figure}
\begin{center}
\epsfig{figure=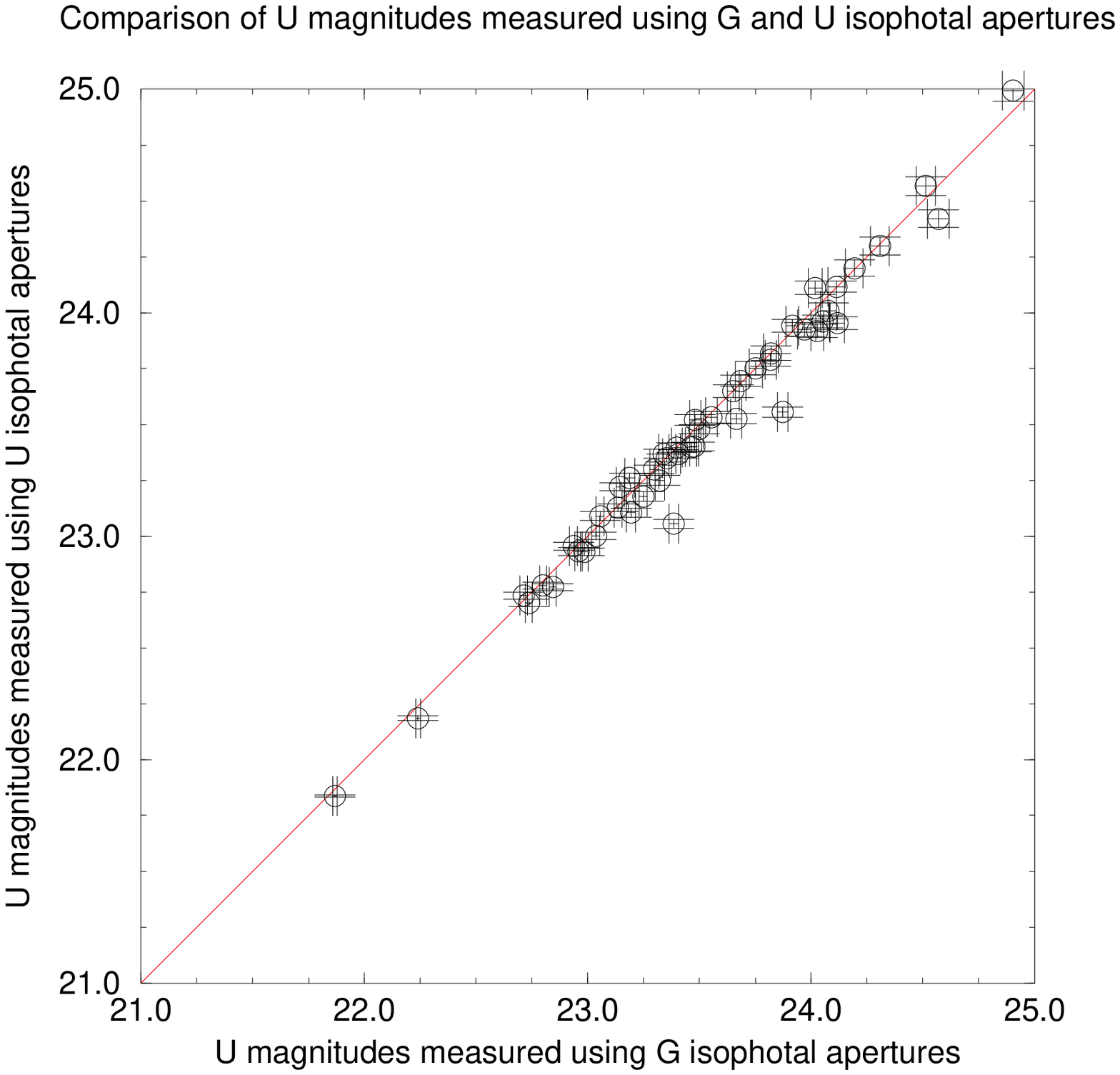,angle=270,width=0.8\linewidth}
\caption{Comparison of $U$ magnitudes measured using isophotal
apertures made from the $U$ and $G$ images. The objects shown here all
have $21<R<22$ and $U<25$.
\label{fig:gucomp}
}
\end{center}
\end{figure}

\begin{figure}
\begin{center}
\epsfig{figure=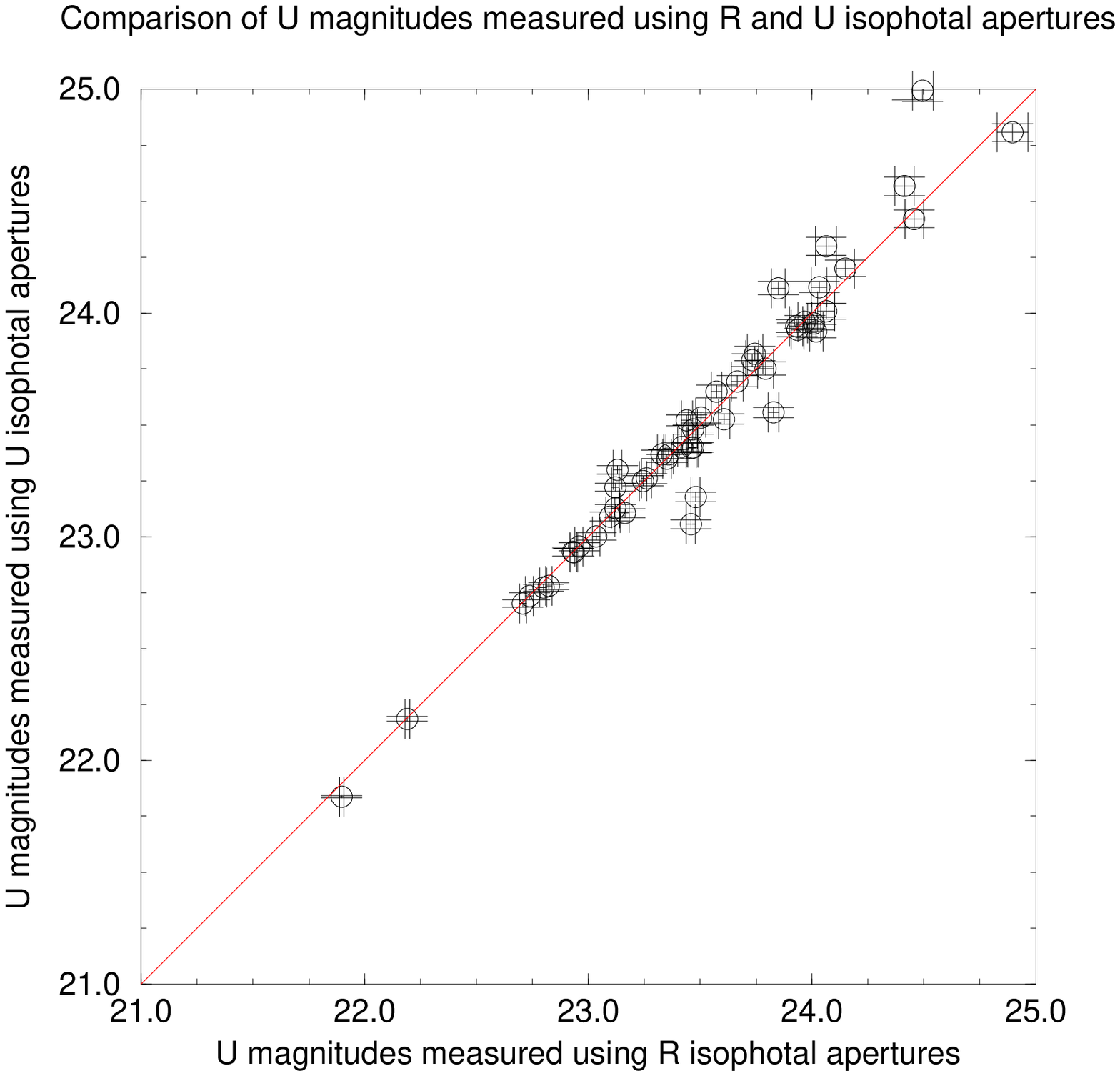,angle=270,width=0.8\linewidth}
\caption{Comparison of $U$ magnitudes measured using isophotal
apertures made from the $U$ and $R$ images. The objects shown here all
have $21<R<22$ and $U<25$.
\label{fig:rucomp}
}
\end{center}
\end{figure}

To check that the choice of isophotal apertures did not produce a
systematic offset in the measured flux recorded from each image, we
compared the magnitudes measured from isophotes based on each
broadband image. Examples of these comparisons are shown in
Figures~\ref{fig:gucomp} \& \ref{fig:rucomp}. There is no evidence for
a systematic offset in the magnitudes recorded between any of the
isophotal apertures for the majority of the sample. However, it seems
that occasionally, in five percent of cases at the most, objects have
magnitudes which are sensitive to the choice of isophotal
aperture. This is most marked in Figure~\ref{fig:rucomp}, where some
six objects show a greater difference in $U$ magnitude than can be
accounted for by photometric errors. Since these colour--dependencies
will only arise in resolved objects, this is only likely to affect
those galaxies at low redshift. Additionally, we stress that the
comparison between $R$ and $U$ is the most extreme case in this
catalogue.

\subsection{Determination of differential galaxy counts}

\begin{figure}
\begin{center}
\epsfig{figure=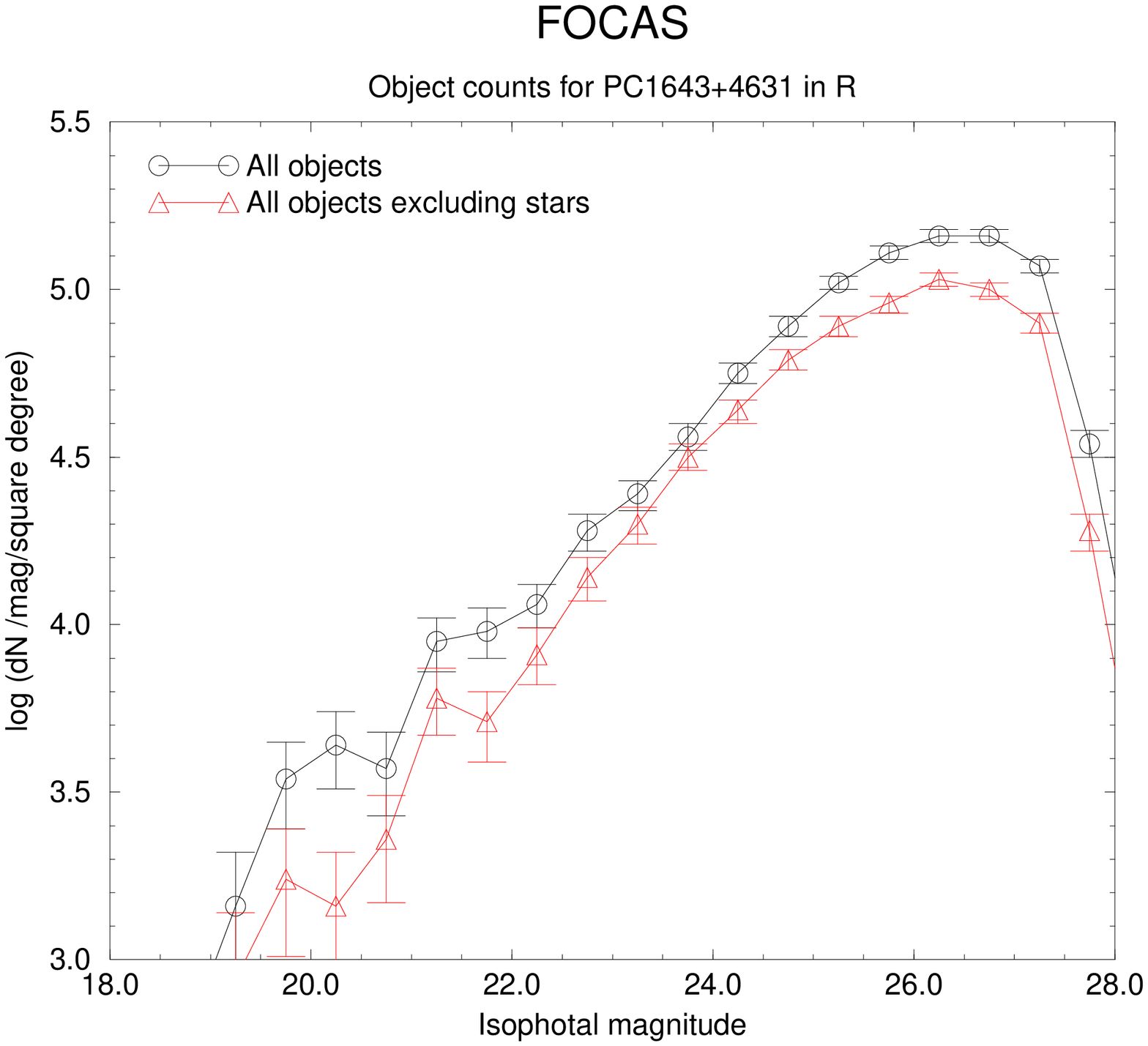,width=0.65\linewidth,angle=270}
\epsfig{figure=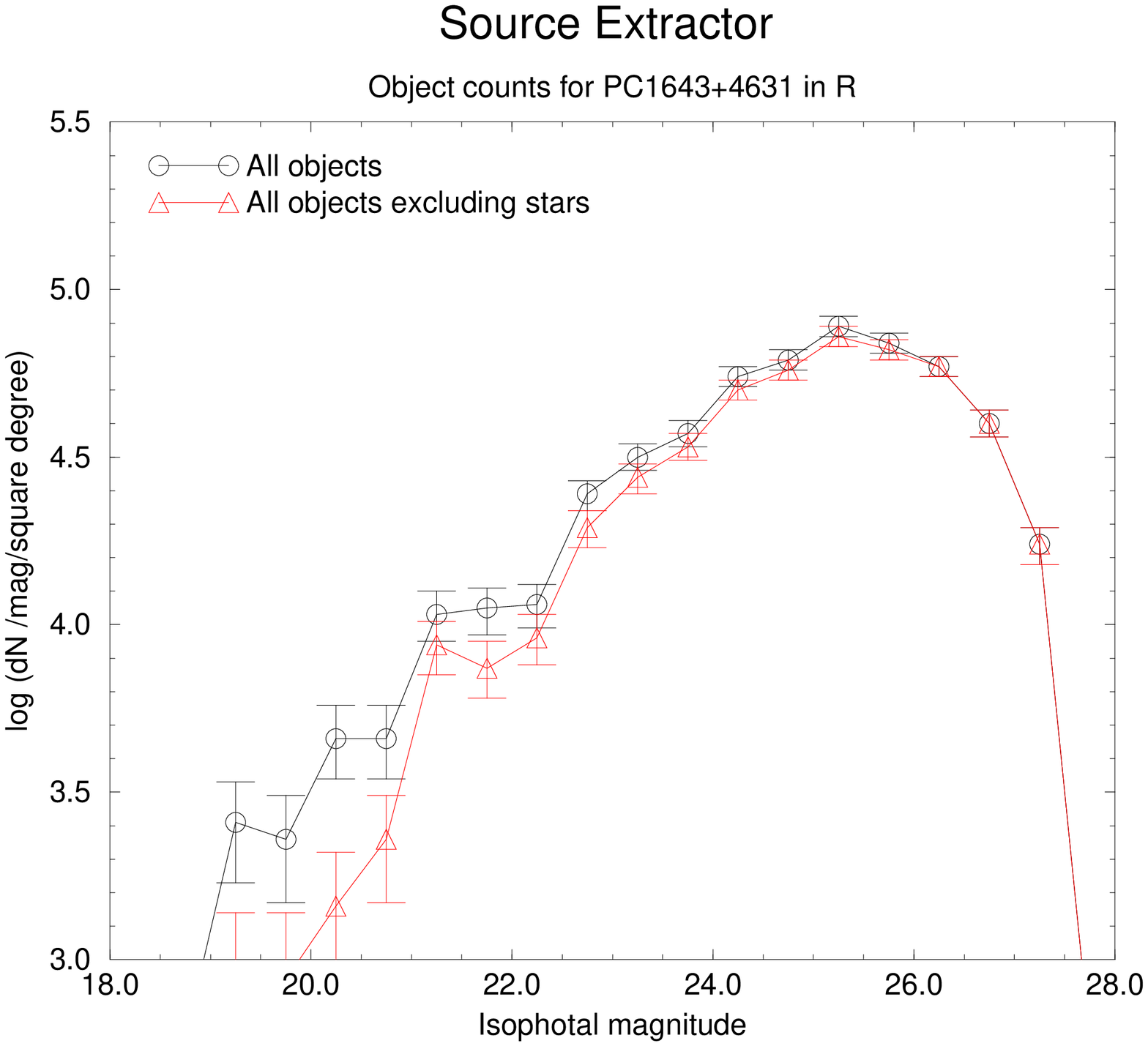,width=0.65\linewidth,angle=270}
\caption{Comparison of the object counts before and after star
rejection for FOCAS and SExtractor
\label{fig:rcounts}
}
\end{center} 
The error bars on the counts are derived from the number of objects in
each magnitude bin assuming a Poisson distribution.
\end{figure}

To compare our imaging of the PC1643 field with other deep
observations, we first investigated the differential galaxy
counts. This requires that all stellar objects in the field are
removed from the catalogues (see, for example, \cite{Tyson88} ), and
the two programs use different approaches to stellar
classification. The FOCAS software uses the point spread function to
differentiate between stellar and non-stellar features and allocates a
type to the object depending on the geometry of the
object. SExtractor's approach relies on a previously--trained neural
network to assess the light distribution of the object and assigns a
`stellar index' to each object; this is a confidence estimate on the
stellar-like nature of the object, ranging between 0 (galaxy) and 1
(star). We have taken all objects with a stellar index of greater than
0.8 to be stars and these objects have been excluded from the galaxy
counts. The results of this procedure are illustrated in
Figure~\ref{fig:rcounts}.
                                                       
Both these approaches to classification have difficulty at faint
magnitudes -- the SExtractor algorithm is unable to give high
confidence levels for objects within about 2 magnitudes of the
catalogue limit, as one might expect given the reduced signal to
noise. FOCAS appears on the other hand to continue to classify very
faint objects as stars almost down to the noise level -- since many of
the fainter galaxies are effectively unresolved, they appear as
point--like objects and are misclassified as stars. FOCAS is therefore
almost certainly overzealous in its allocation of stellar
classifications, since for $R>24$ it classifies over half the objects
as stars, whereas here one would expect the galaxies to dominate: the
number of stars per unit area of sky is roughly equal to the number of
galaxies per unit area of sky at 20th magnitude at high galactic
latitude (\cite{Sext}). The very brightest objects, with $R<21$, are
predominantly stars, although the increase in the total and
galaxy--only counts at $R\sim20$ is misleading since these objects are
increasingly saturated in the CCD image for $R<20$, leading to an
artificial excess of galaxy counts at this magnitude.

FOCAS appears to be capable of detecting more faint
objects than SExtractor, as demonstrated by the magnitude at which the
differential counts begin to turn over
(Figure~\ref{fig:rcounts}). This may be due to the differing
approaches to splitting multiple objects employed by the catalogue
programs -- the magnitudes at which this effect is most noticeable are
close to the limit of the images. Closer examination of the faintest
detected objects in both catalogues suggests further
possibilities. The validity of the faintest objects in the FOCAS
catalogue is questionable, as there is a significant increase in the
density of objects around the brightest stars in the image, where the
background noise is higher. This is almost certainly due to the global
threshold value above the background which FOCAS uses to determine its
intensity threshold -- in areas of higher background noise, such as
around the brightest objects, the local rms noise level is higher,
leading to a higher probability of the software identifying noise
peaks as real objects.

SExtractor suffers difficulties in separating faint objects from
brighter companions, even with the highest allowed level of contrast,
and may therefore wrongly aggregate such objects together. This is
most noticeable around the brightest stars in the image, where it
fails to separate objects out of the wings of the stars, even on the
most extreme contrast settings. Similarly, considerable care in
setting the detection parameters is necessary to ensure that faint
objects near to bright galaxies are correctly found.

\afterpage{\clearpage}

\subsection{Determination of photometric measurement errors}
\label{sec:sim}

To quantify the ability of the software to recover galaxies from the
field, we carried out simulations using artificial galaxies created
with the IRAF package ARTDATA. This package allows the user to create
and place galaxies with either de Vaucoleurs or exponential profiles
in random, clustered or user-defined distributions with various
magnitude distributions.

We followed two approaches. First, 100 sample galaxies with a 40:60
mix of ellipticals and spirals, of fixed magnitude and a comparable
range of sizes to those in the actual image but otherwise random
orientation and aspect, were added to a noise--only image. This
noise--only image mimicked the background level in the actual field,
including the raised background level around the stars. Second,
model galaxies generated in the same way were also added directly to
the real image. FOCAS was then used to perform the photometry on both
the new images and the recovered fluxes were compared with the
starting values.

\begin{figure}
\begin{minipage}[hp]{0.5\linewidth}
\epsfig{figure=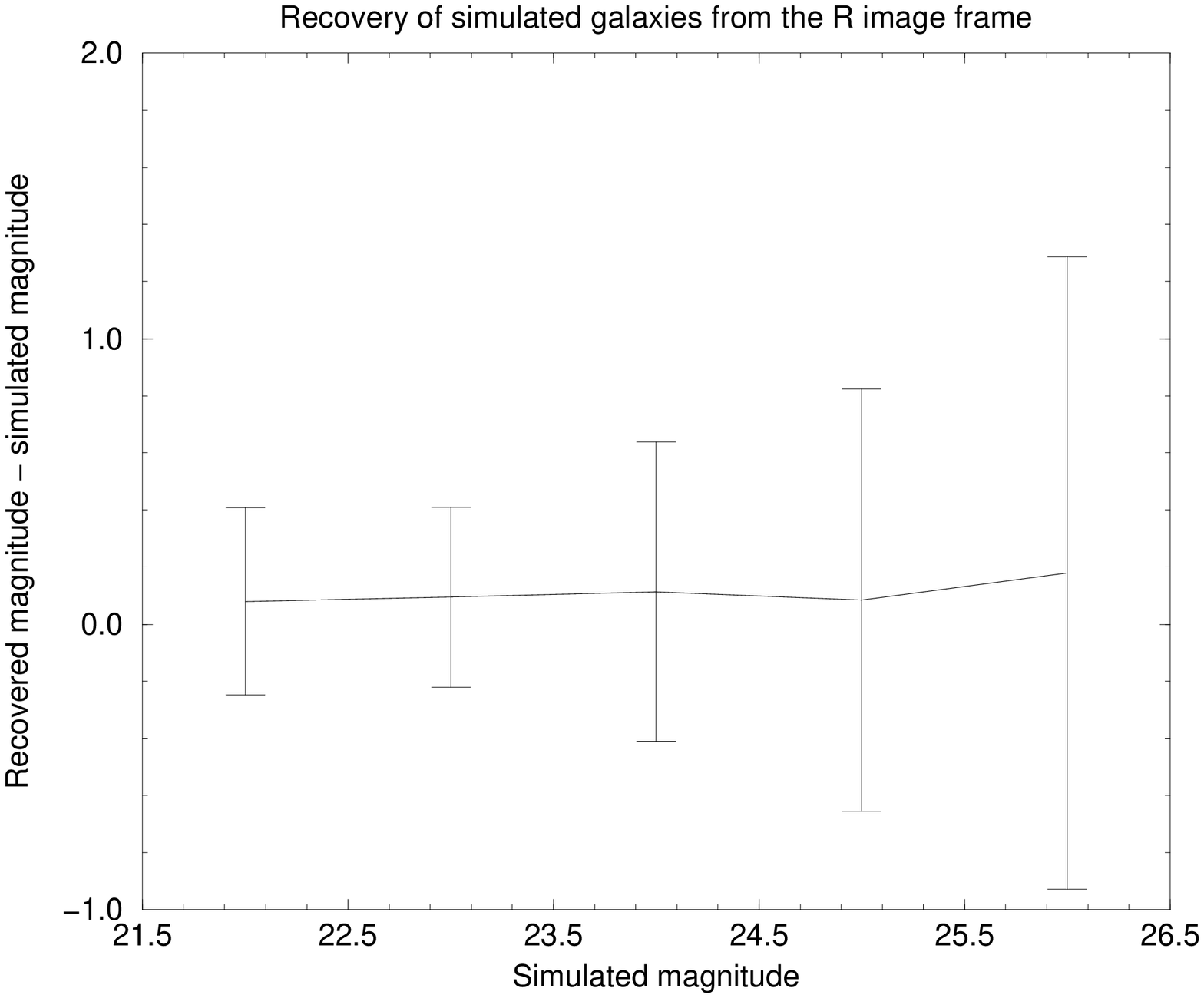,width=0.8\linewidth,angle=270}
\end{minipage}\hfill
\begin{minipage}[hp]{0.5\linewidth}
\epsfig{figure=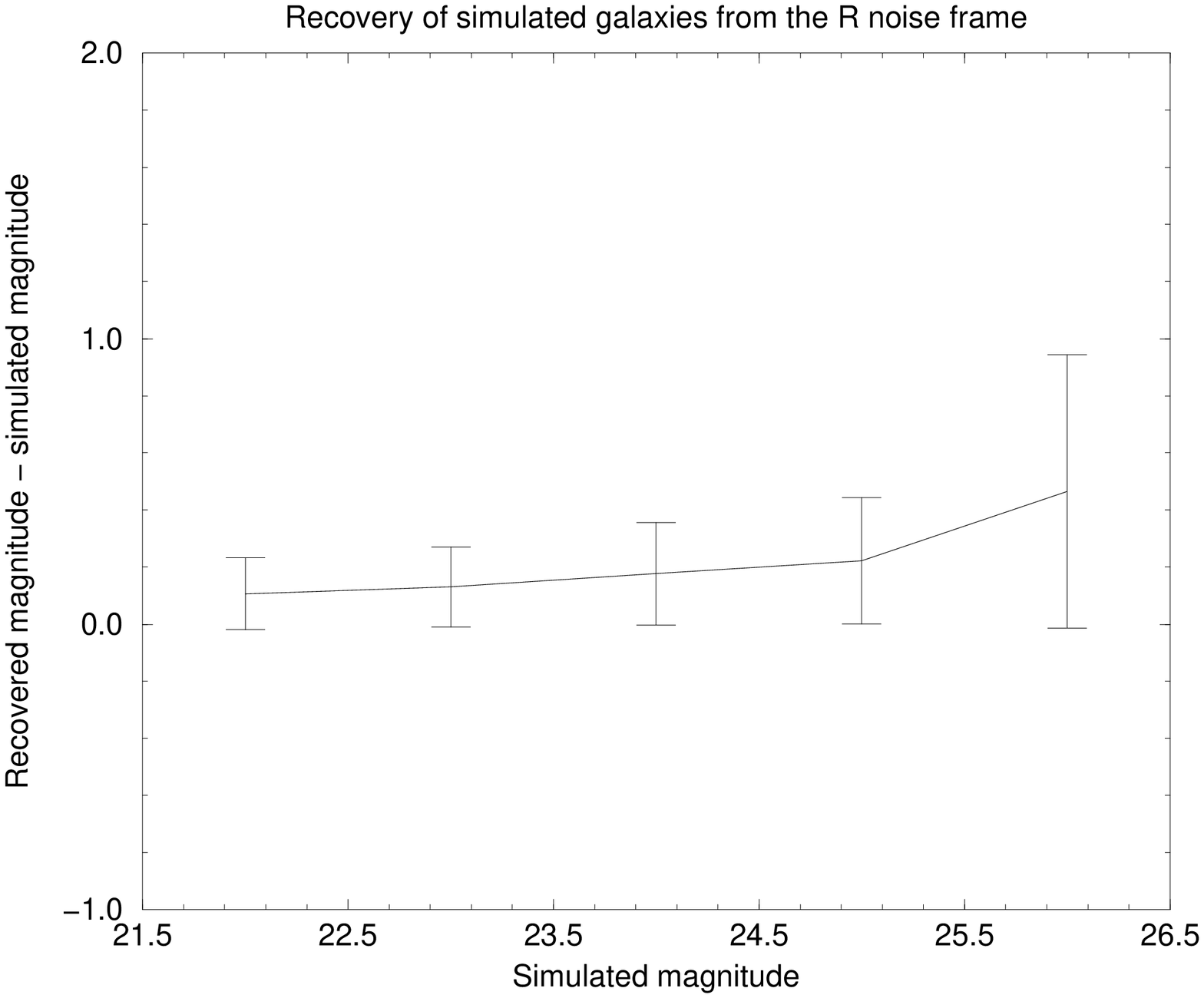,width=0.8\linewidth,angle=270}
\end{minipage}
\caption{Comparison of the recovery of
simulated galaxies from real (left) and noise-only (right) images in
R. The error bars shown here are the standard deviation of the
magnitudes of the recovered galaxies from the original magnitudes, and
do not take into account the skewness of the distributions. The lines
merely join the means of the distributions.
\label{fig:comprecov}
}
\end{figure}

\begin{figure}
\begin{minipage}[hp]{0.5\linewidth}
\epsfig{figure=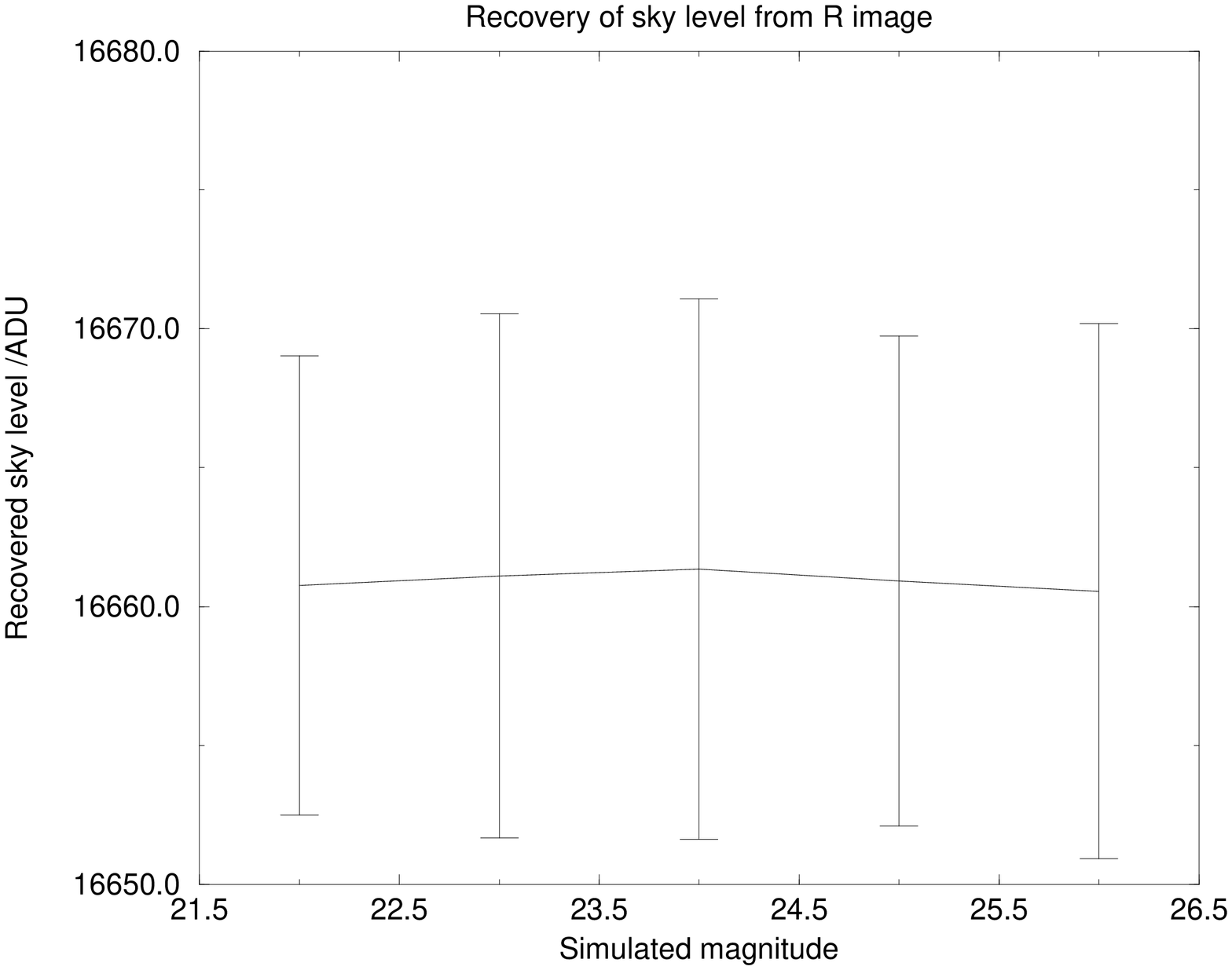,width=0.8\linewidth,angle=270}
\end{minipage}\hfill
\begin{minipage}[hp]{0.5\linewidth}
\epsfig{figure=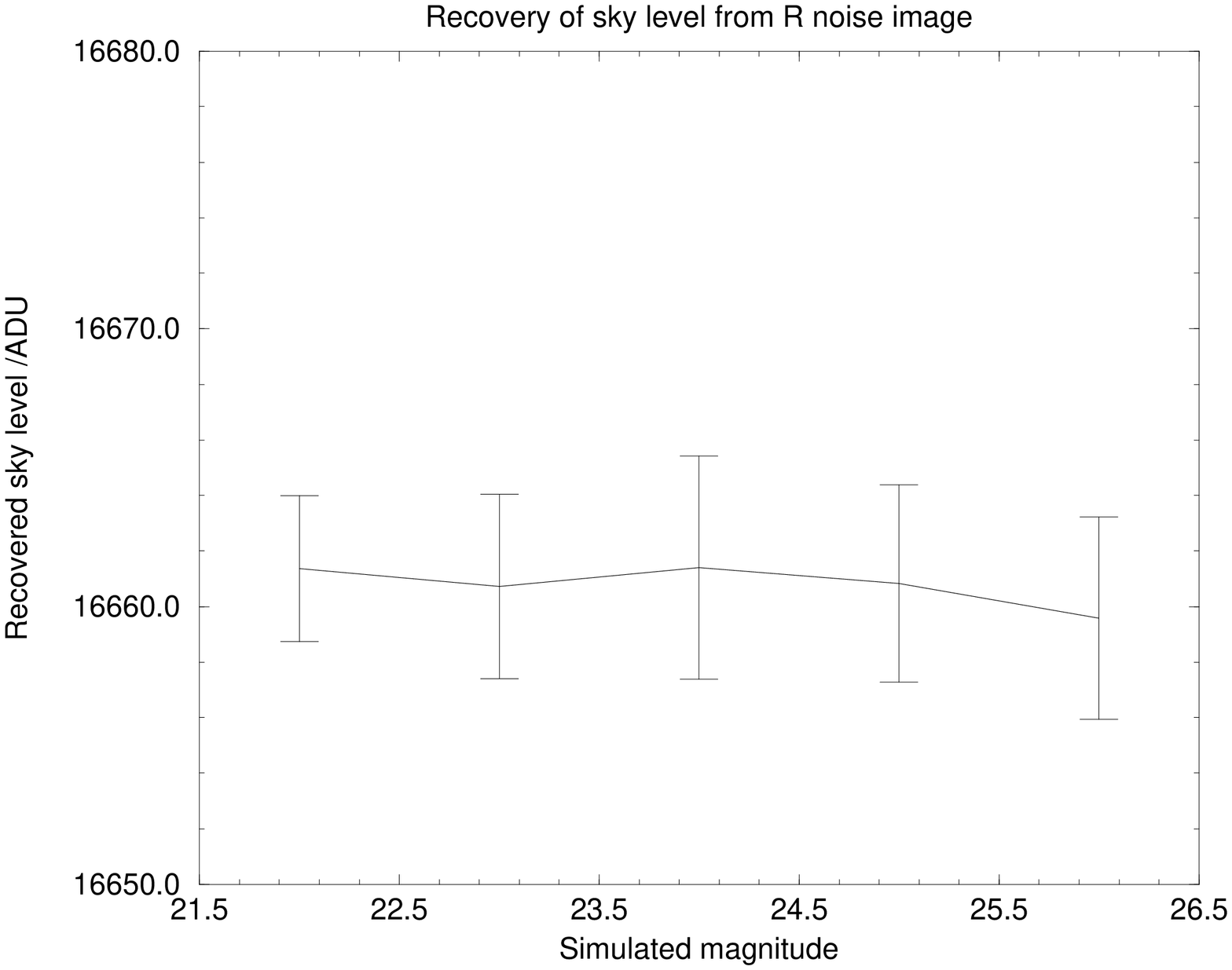,width=0.8\linewidth,angle=270}
\end{minipage}
\caption{Comparison of the recovery of sky level around simulated
galaxies from real (left--hand figure) and noise-only images in $R$
\label{fig:skyrec}
}
\end{figure}

As can be seen from Figure~\ref{fig:comprecov}, there are significant
differences between these two approaches. Recovery of objects from a
noise-only image suffers from a serious discrepancy of as much as half
a magnitude from the real magnitude (as indicated by the error bars in
Figure~\ref{fig:comprecov}), whereas the results of recovery of the
simulated galaxies from the real field are, on average, much closer to
the expected value. This is not a failure of the software to determine
accurately the sky level surrounding the objects in question as
demonstrated in Figure~\ref{fig:skyrec}, since there is no evidence
for the measured sky background level around the objects being a
function of simulated magnitude. As expected, recovery from a real
field shows a much wider spread of magnitudes due to contamination by
neighbouring galaxies and this inevitably leads to an increase in the
measured fluxes of the artificial objects. This confusion between
between objects on the sky is almost certainly responsible for
reversing the loss of flux beyond the isophotal apertures. Since this
effect will occur for real as well as the simulated galaxies, using
simulations based on recovery from noise-only images or mosaics
significantly overestimates the loss of flux from the apertures.

\begin{figure} 
\begin{minipage}[hp]{0.5\linewidth}
\epsfig{figure=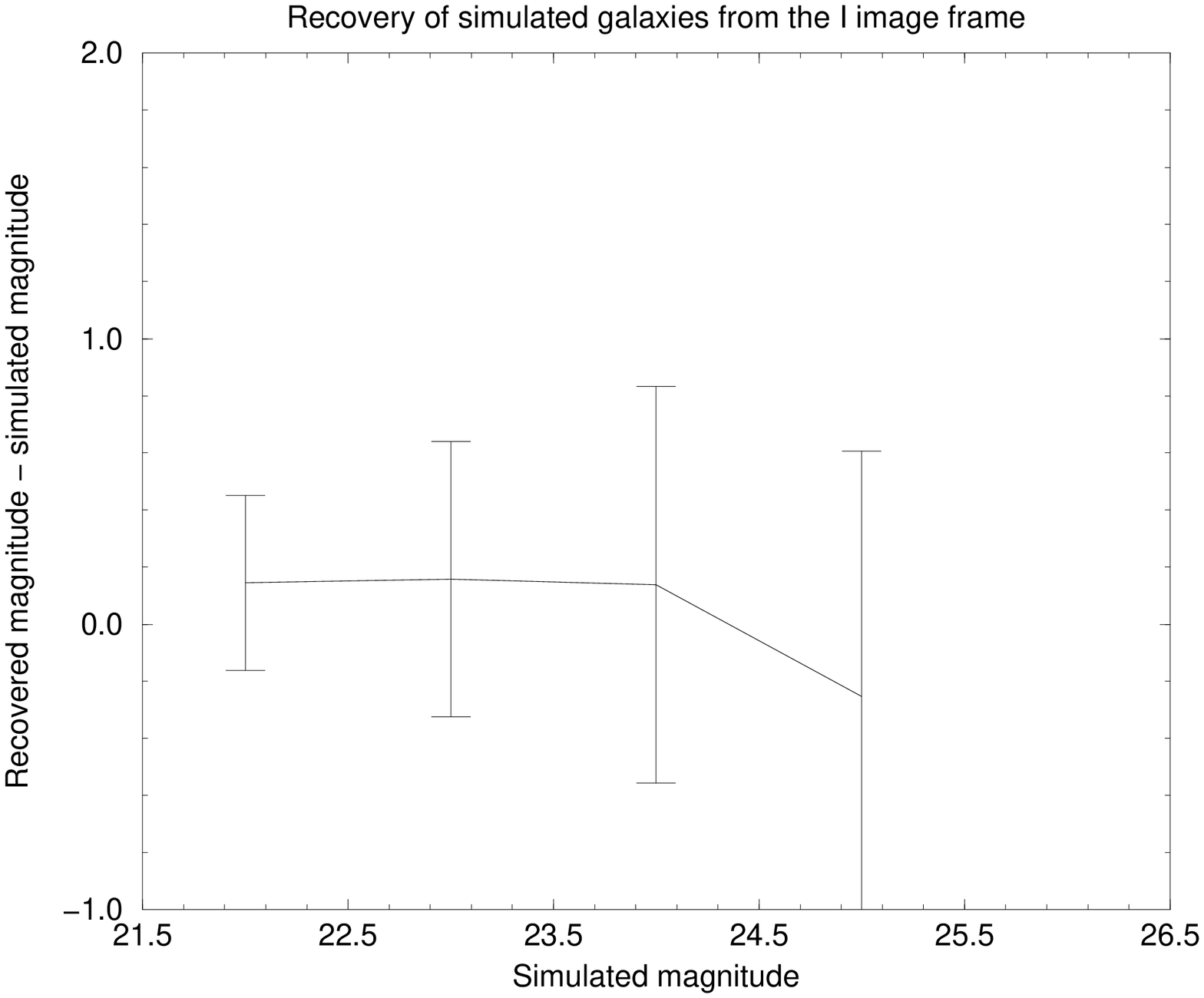,width=0.8\linewidth,angle=270}
\end{minipage}\hfill
\begin{minipage}[hp]{0.5\linewidth}
\epsfig{figure=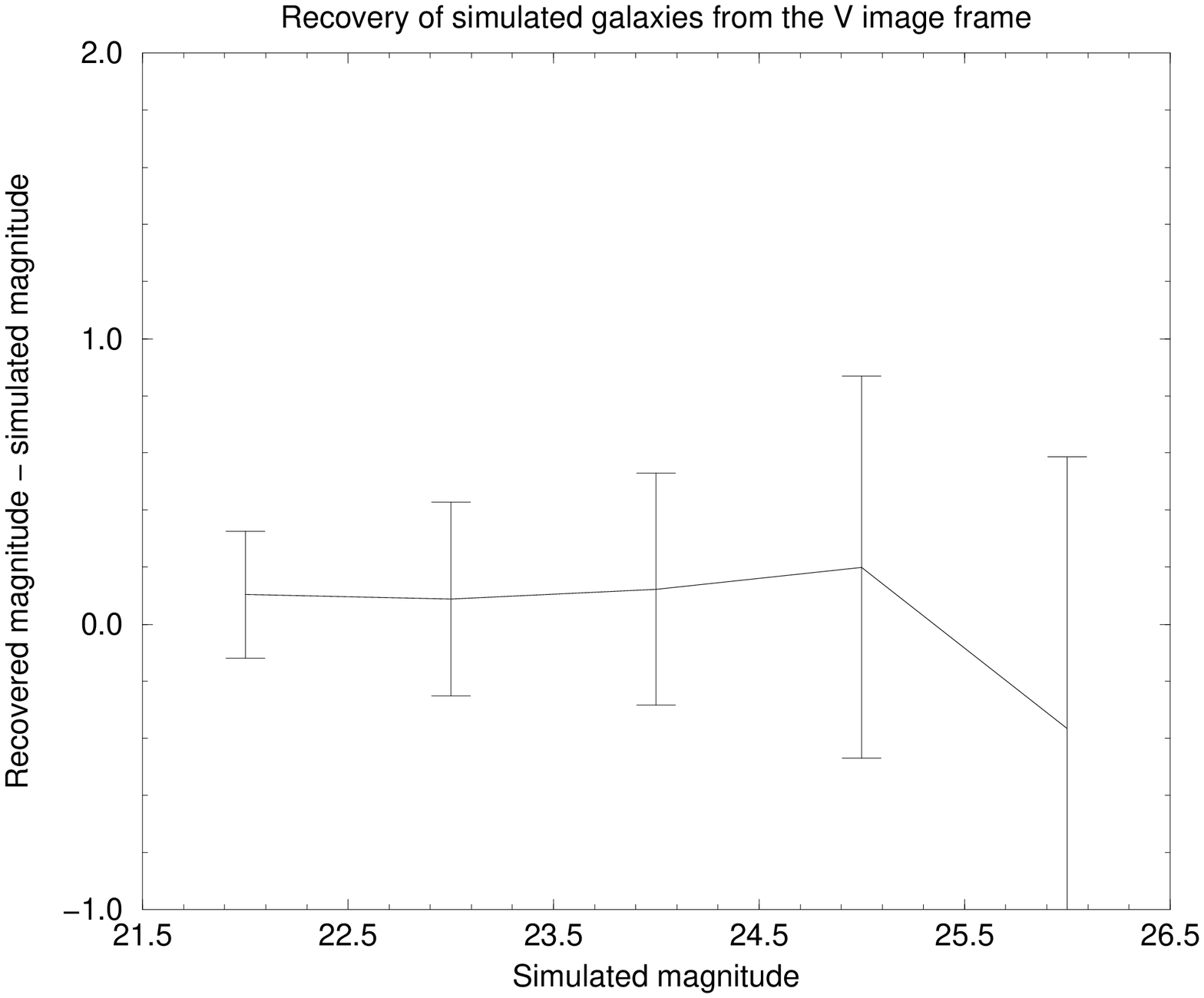,width=0.8\linewidth,angle=270}
\end{minipage}
\begin{minipage}[hp]{0.5\linewidth}
\epsfig{figure=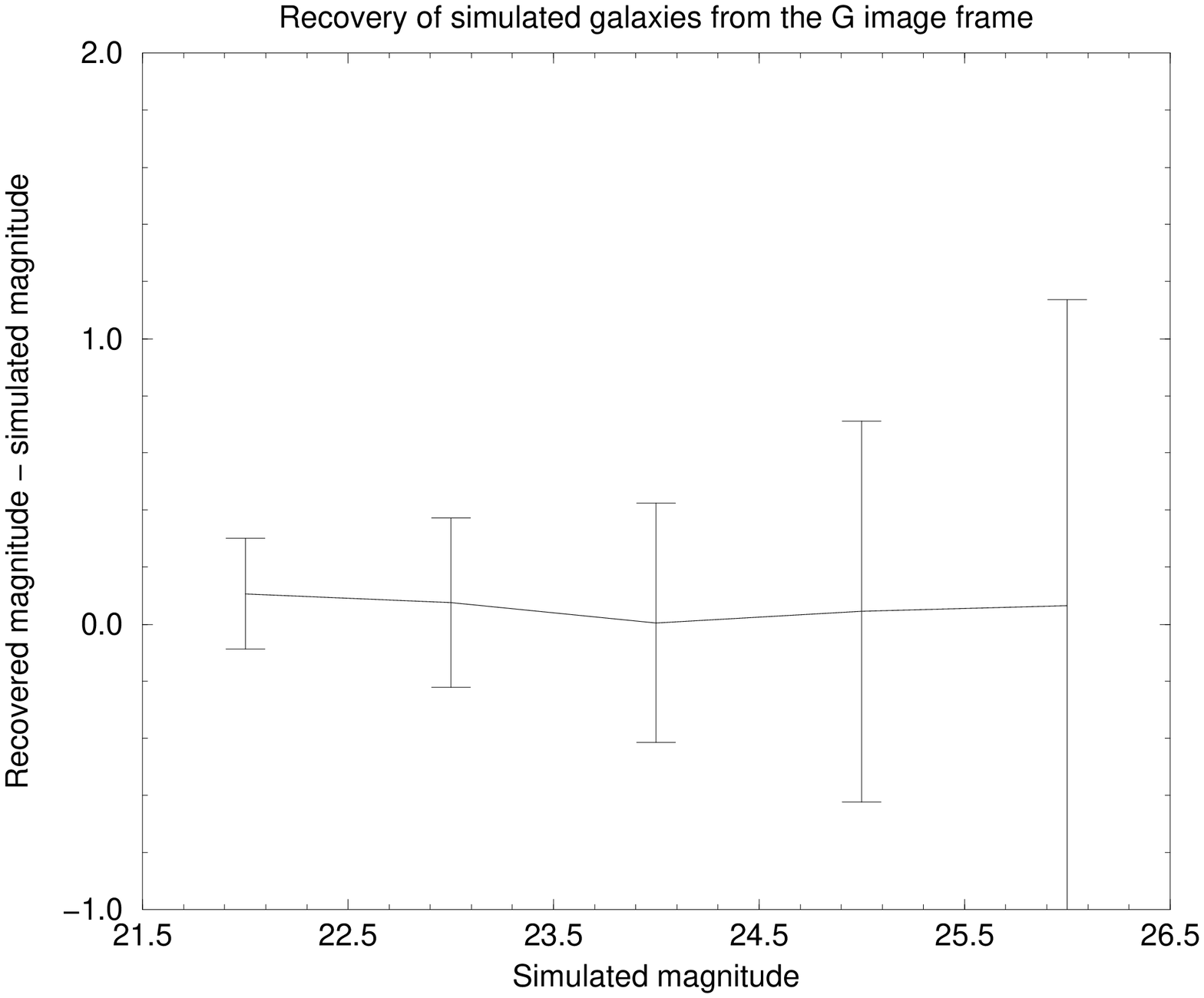,width=0.8\linewidth,angle=270}
\end{minipage}\hfill
\begin{minipage}[hp]{0.5\linewidth}
\epsfig{figure=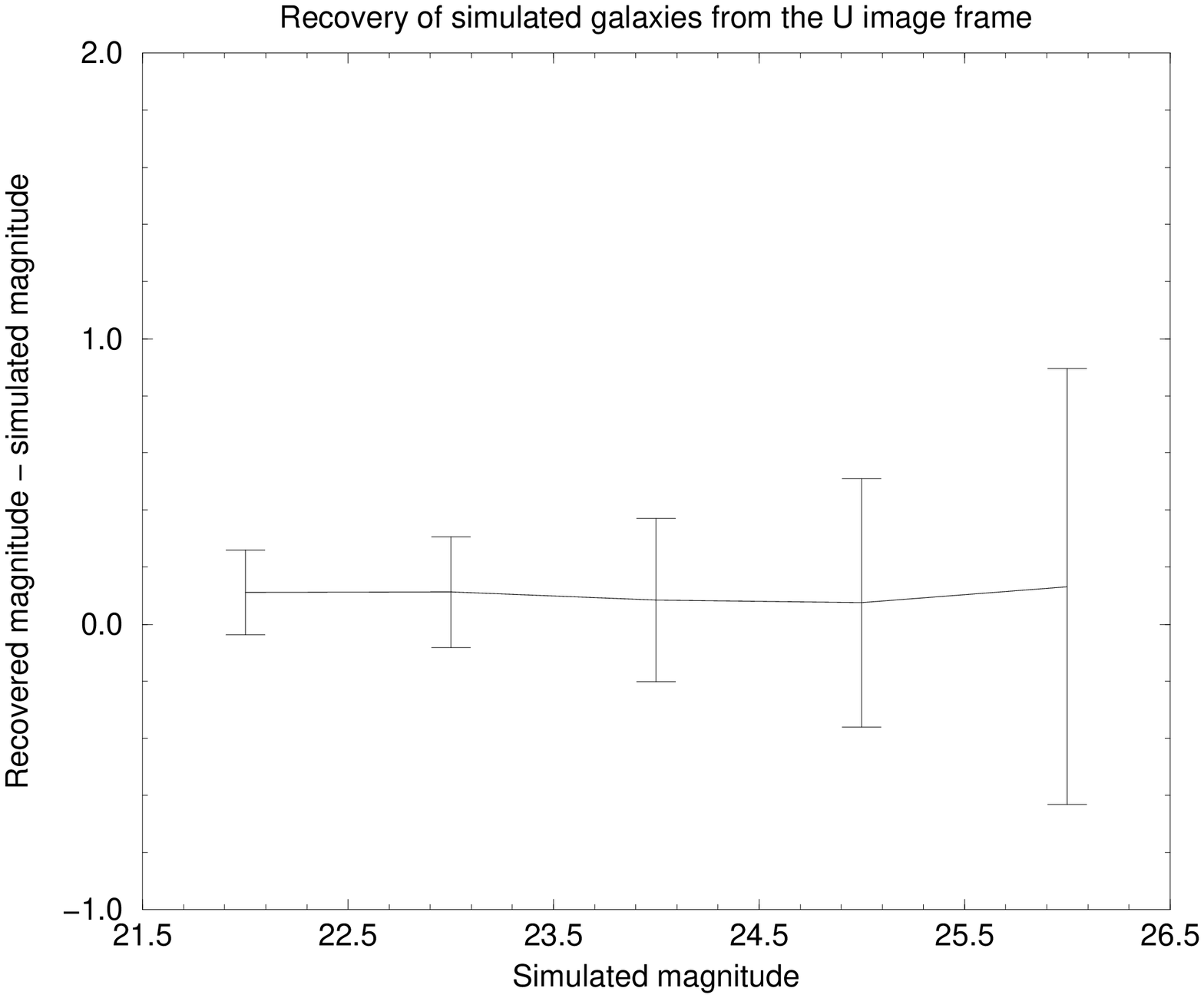,width=0.8\linewidth,angle=270}
\end{minipage}
\caption{Recovery of simulated galaxies from actual $U$,$G$,$V$ and $I$ images
\label{fig:recovugvi}
}
\end{figure}

The effects of confusion with faint sources in the field are most
pronounced at the faintest magnitudes measured, resulting in a
significant overestimate of the brightness of the source -- the
faintest sources show the greatest increase in isophotal magnitude
when they overlap with real field objects. This is demonstrated in the
$V$ and $I$ image simulations in Figure~\ref{fig:recovugvi}, where
$I=25$ and $V=26$ galaxies are comparatively brightened on
average. The $R$, $G$ and $U$ images are almost 1.5 magnitudes deeper
than the $I$ observations and such effects are not seen in the range
22nd--26th magnitude.

In summary, it appears that using magnitude corrections based on
detecting and performing photometry on simulated galaxies placed in a
noise--only field suffers a significant systematic discrepancy when
compared to recovery of similar galaxies from the actual image. This
difference can be as high as 0.4 magnitudes for objects with low
signal--to--noise. For these fainter objects, the presence of many
brighter neighbouring objects counteracts the loss of flux from the
edges of an object measured using isophotal apertures. It is also
worth noting that in recovering simulated galaxies from the real
image, the measured isophotal magnitudes show a greater spread of
values about the simulated magnitude than those recovered from the
noise image, further reinforcing that there is no evidence for the
need to make any corrections to the isophotal magnitudes.

\afterpage{\clearpage}

\subsection{Completeness of galaxy counts}

To estimate the completeness of the catalogue, we examined the ability
of the software algorithms to recover sets of simulated galaxies added
to the field. Although maximising the number of galaxies added in each
iteration reduces computing time, it is important that the number of
galaxies added is not so great that a significant number of galaxies
overlap. We chose to add 100 galaxies per iteration: the probability
of any two of these galaxies being coincident is approximately one
percent and is therefore insignificant.

Accordingly, 100 simulated galaxies were added to the real image, with
the magnitudes distributed to imitate the real distribution, and with
the 40:60 elliptical/spiral mix as before. The indices of the
power--laws were taken directly from the real raw galaxy counts for
this field, and were based on linear regression of the linear section
of the galaxy counts fainter than 20th magnitude. This procedure was
repeated 20 times for each broadband image; this was enough to show
clear trends.

\begin{figure*}
\begin{center}
\begin{minipage}[h]{0.45\linewidth}
\epsfig{figure=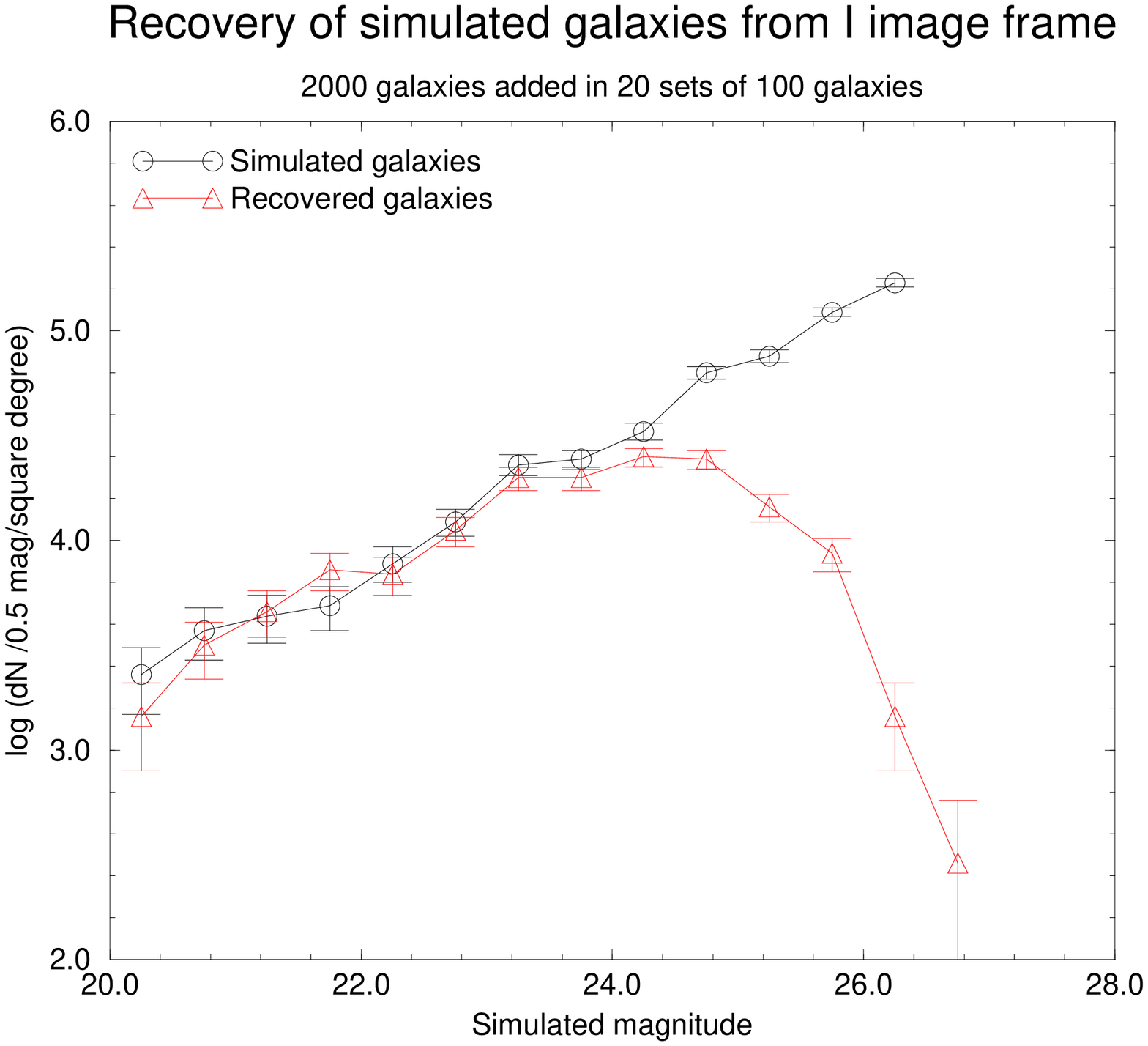,width=0.8\linewidth,angle=270}
\end{minipage}
\begin{minipage}[h]{0.45\linewidth}
\epsfig{figure=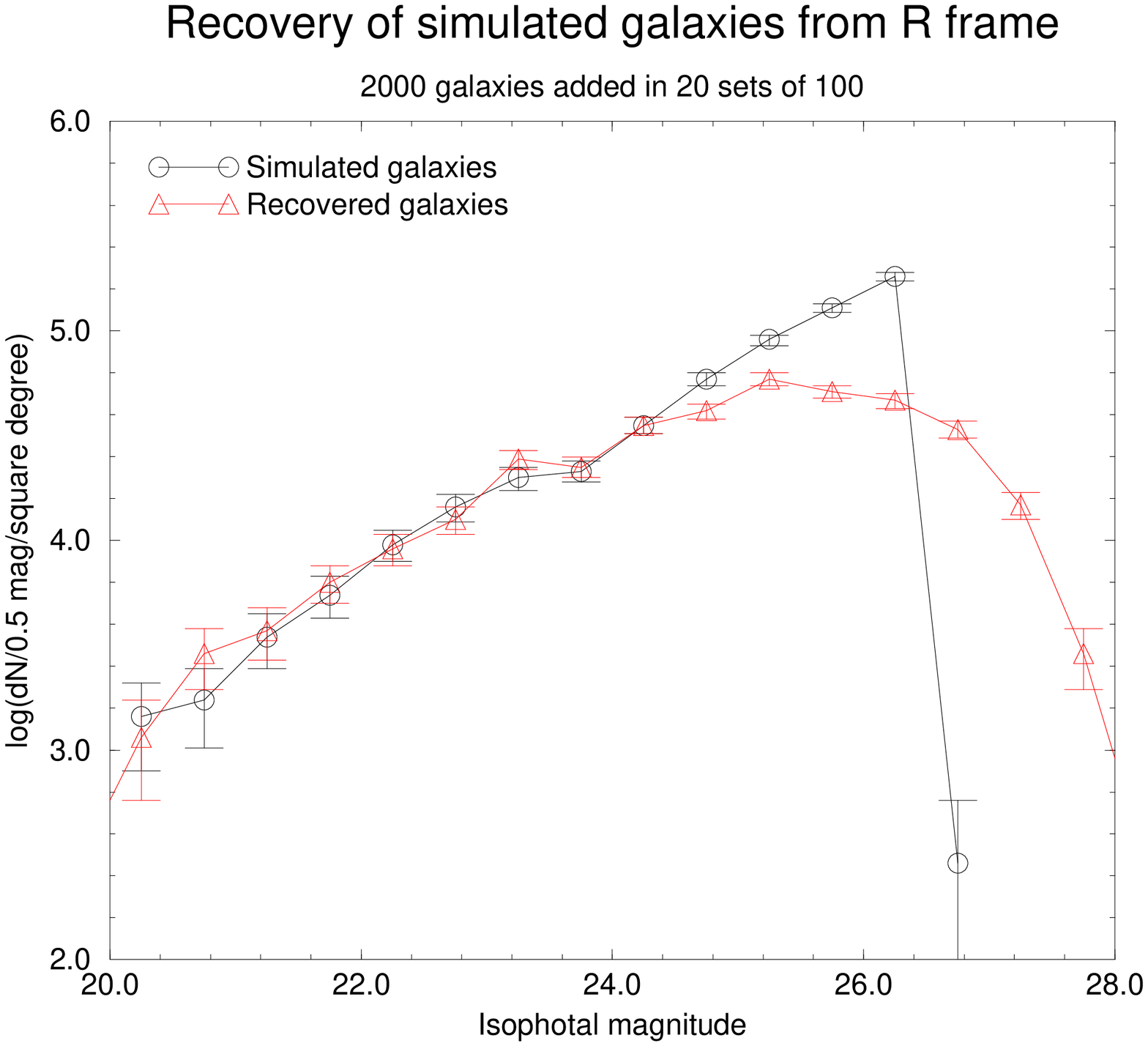,width=0.8\linewidth,angle=270}
\end{minipage}\hfill
\begin{minipage}[h]{0.45\linewidth}
\epsfig{figure=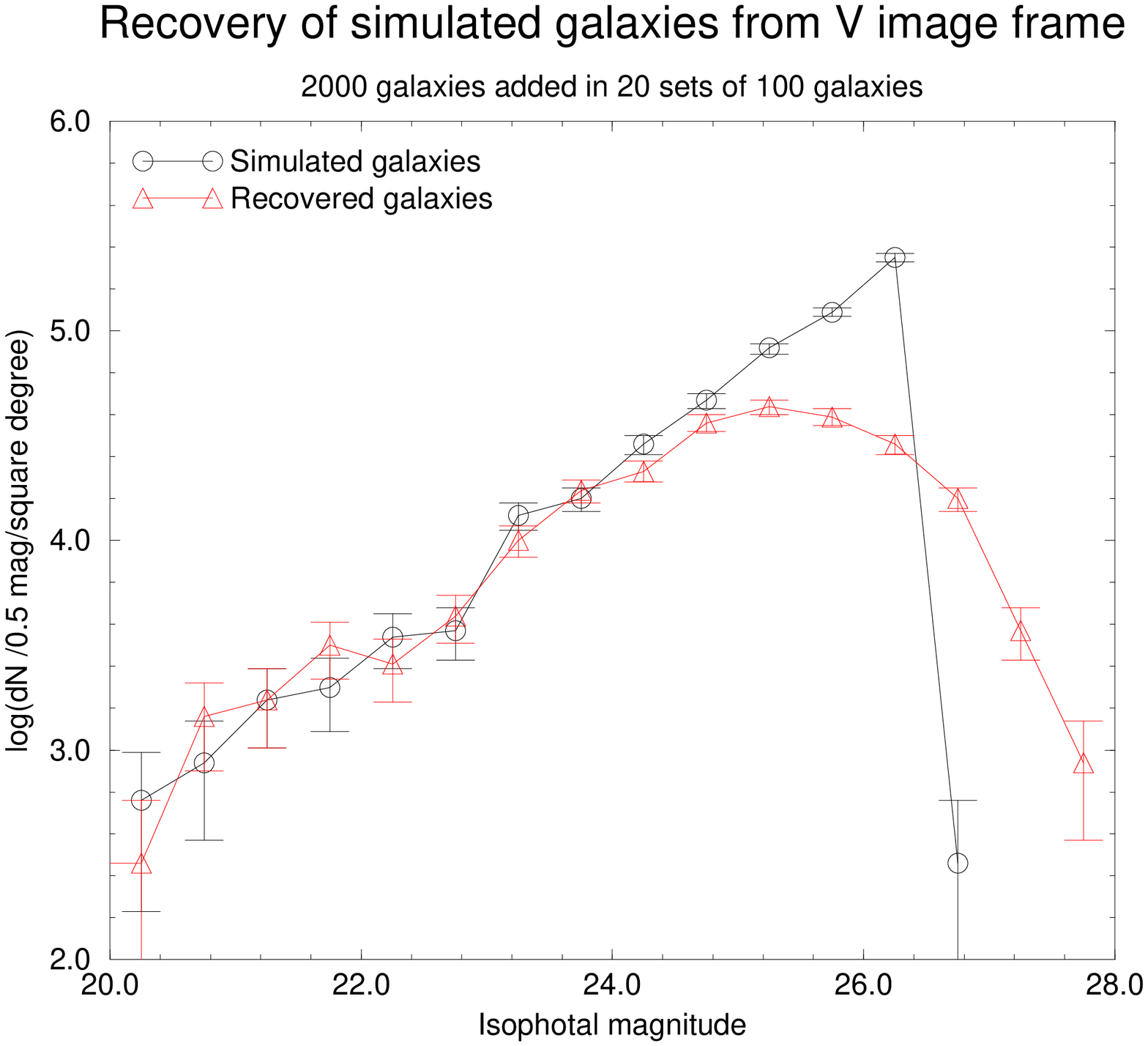,width=0.8\linewidth,angle=270}
\end{minipage}
\begin{minipage}[h]{0.45\linewidth}
\epsfig{figure=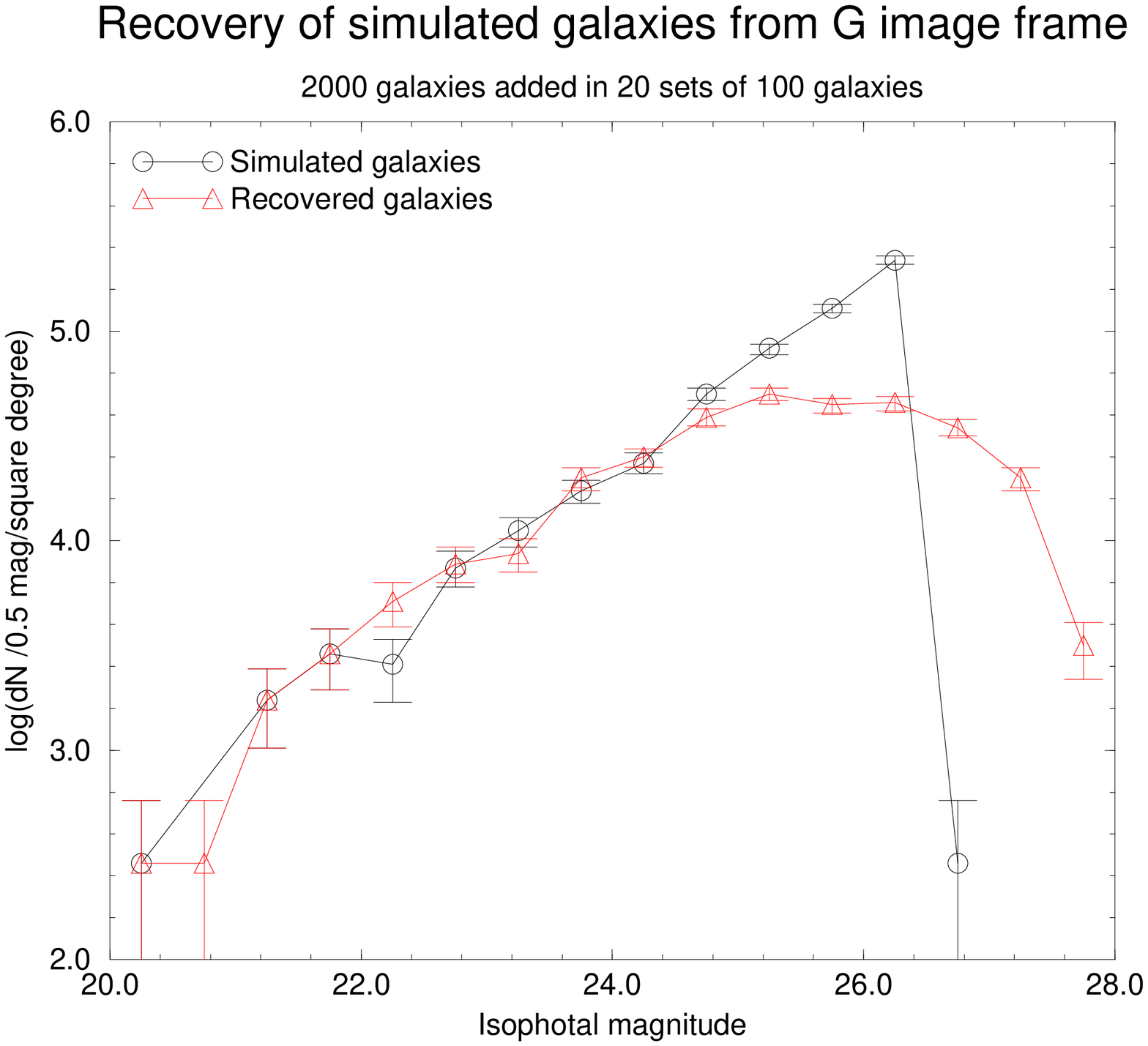,width=0.8\linewidth,angle=270}
\end{minipage}\hfill
\begin{minipage}[h]{0.45\linewidth}
\epsfig{figure=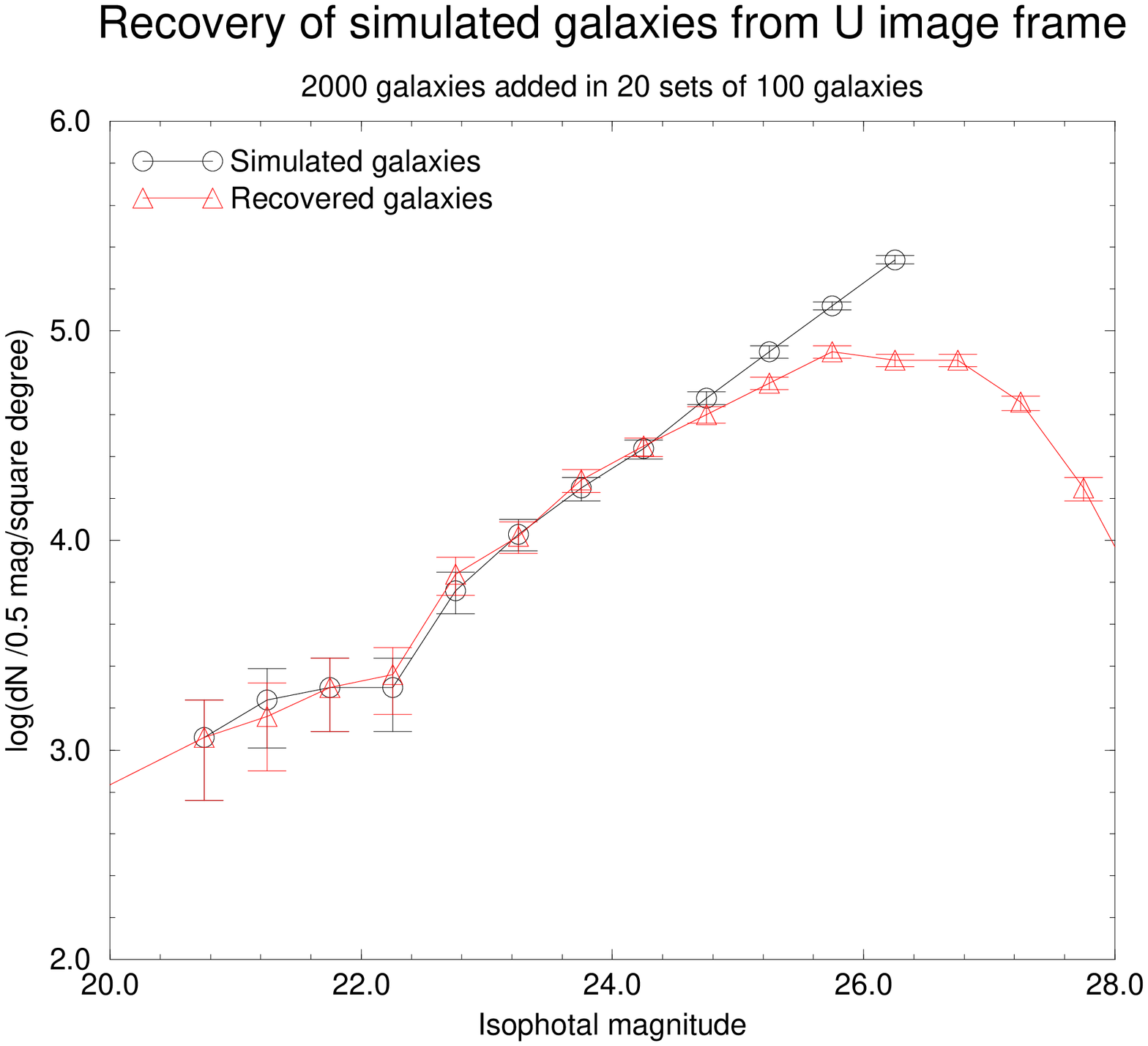,width=0.8\linewidth,angle=270}
\end{minipage}
\caption{Simulations on the completeness of the five broadband catalogues
\label{fig:powersim}
}
\end{center}
\end{figure*}

The results of these simulations (Figure~\ref{fig:powersim}) give
direct information on the `loss' of galaxies from their real magnitude
bin. This loss occurs in two forms: failure to detect a faint galaxy,
usually due to it falling below the surface brightness limits of the
images; and failure to determine the magnitude of the galaxy to an
accuracy of less than the bin width in the histogram, as explored in
section~\ref{sec:sim}.

It is worth noting that for the brighter end of the simulation, there
is little or no deviation between the simulated galaxy magnitude
histogram and that of the recovered galaxy histogram, entirely
consistent with the accurate recovery of individual galaxies as seen
in section~\ref{sec:sim}. The point at which the catalogues become
significantly incomplete (which we take to be as losing more than half
the real number of galaxies) is tightly correlated with the accuracy
with which the galaxy magnitude can be measured, and a completeness
limit of roughly 50\% is reached when the photometry errors reach a
magnitude.

We also point out that these simulations are not suitable for
accurately estimating the completeness of the catalogue at all
magnitudes as they do not go faint enough, despite going close to the
measurable limit of the catalogues. This is clearly demonstrated in
the $U$ simulations, where the simulated $U$ counts turn over before
the real $U$ counts do. If these simulations are used to attempt to
correct the raw differential counts to the actual galaxy counts, the
resulting counts are over--estimates because the simulations
themselves do not cover all the real spread of
magnitudes. Additionally, because these simulations rely on prior
knowledge to provide a distribution from which to simulate the galaxy
counts, there may be a tendency for the results of these simulations
to merely confirm the starting hypothesis when the catalogues become
markedly incomplete. In summary, unless the raw counts themselves are
effectively complete, the corrections often made to the raw counts to
account for the incompleteness may prove to be erroneous if the
starting hypothesis is incorrect.

\section{Images}

A full--colour image comprising all the five broad band filters is
shown in Figure~\ref{fig:fullcolour}. There is no
obvious cluster in this field, which might be visible if the cluster
were similar to a rich Abell cluster at a redshift of $z<1$ (cf
\cite{LK97}) -- such a cluster would have $I\approx20$ for the
brightest cluster member, and have several members brighter than
$I\approx23$. The band of bright objects across the centre of the
field evidently consists of objects with several different colours and
is therefore not a system at one specific redshift. The quasars A \& B
appear yellow in this image, which is due to the absence of any
observed flux in $U$, and no unequivocal third image candidate is seen
in this colour image, although there are some faint yellow objects in
the centre of the field. It is also notable that the faintest objects
are predominantly blue.

\section{Differential Galaxy counts}
\label{sec:counts}

\begin{figure}
\begin{center}
\epsfig{figure=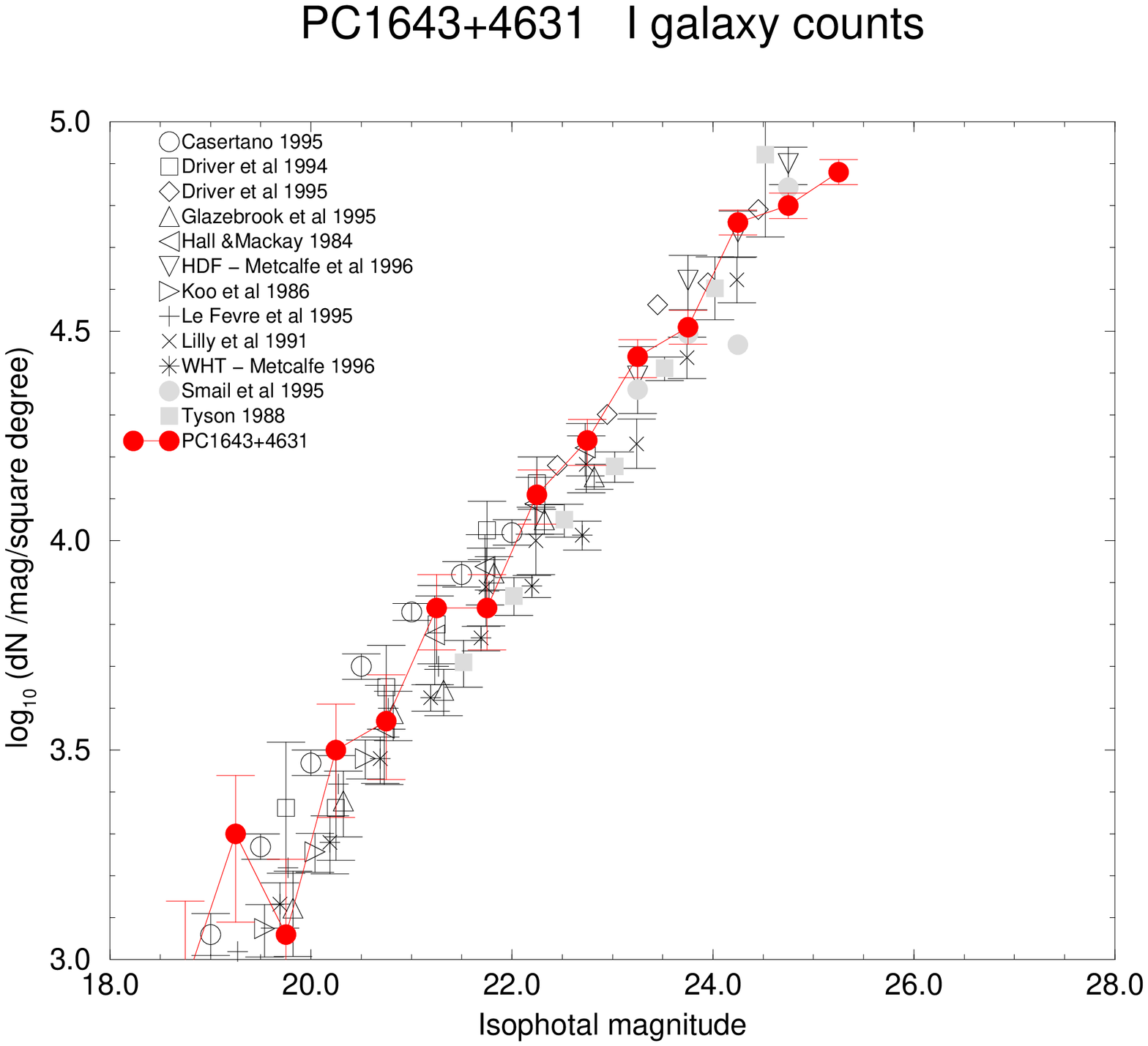,width=0.65\linewidth,angle=270}
\end{center}
\caption{Galaxy counts for $I$ filters, compared with previously published results
\label{fig:icount}
}
\end{figure}

\begin{figure}
\begin{center}
\epsfig{figure=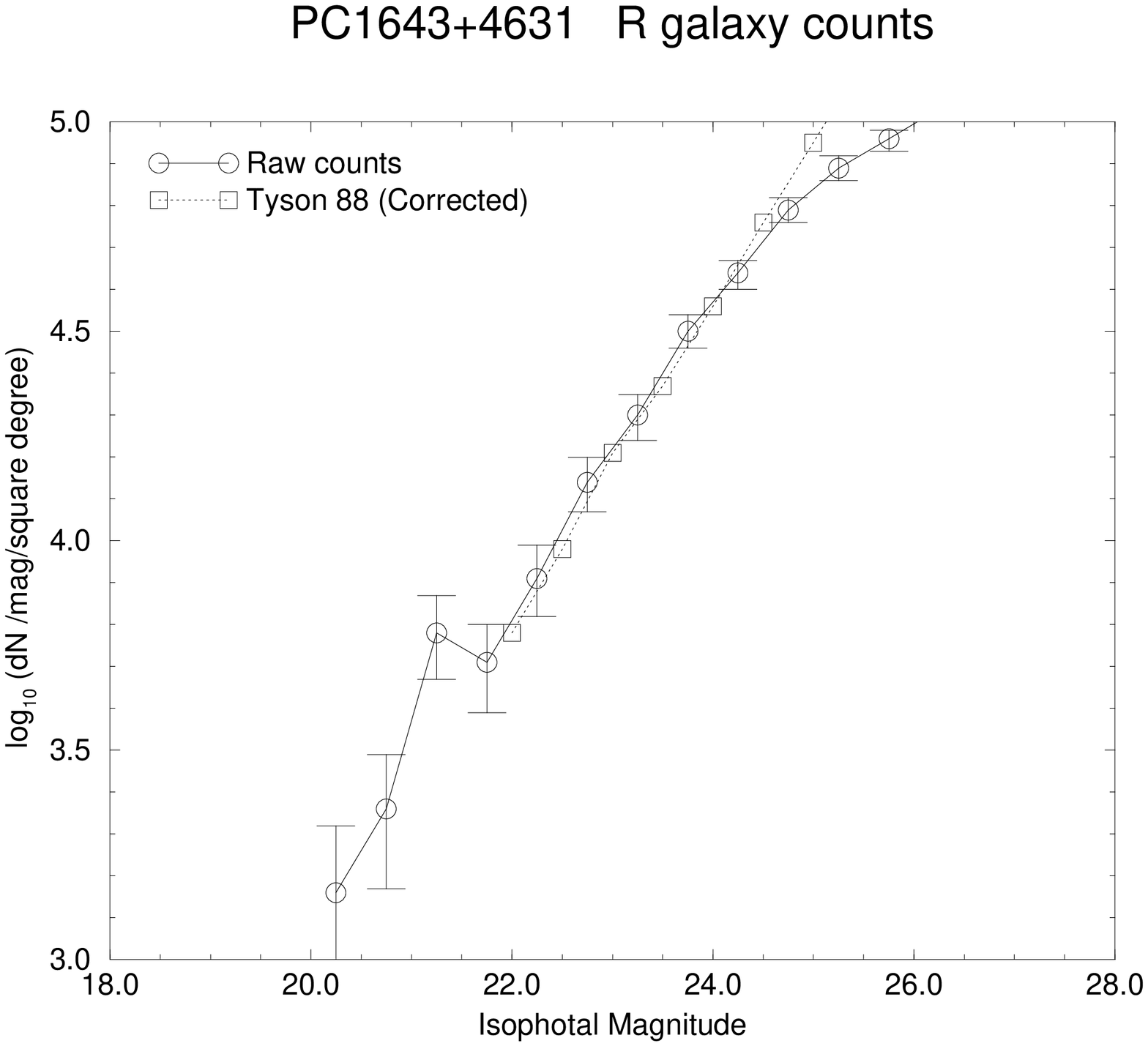,width=0.65\linewidth,angle=270}
\end{center}
\caption{Galaxy counts for $R$ filters, compared with previously published results
\label{fig:rcount}
}
\end{figure}

\begin{figure}
\begin{center}
\epsfig{figure=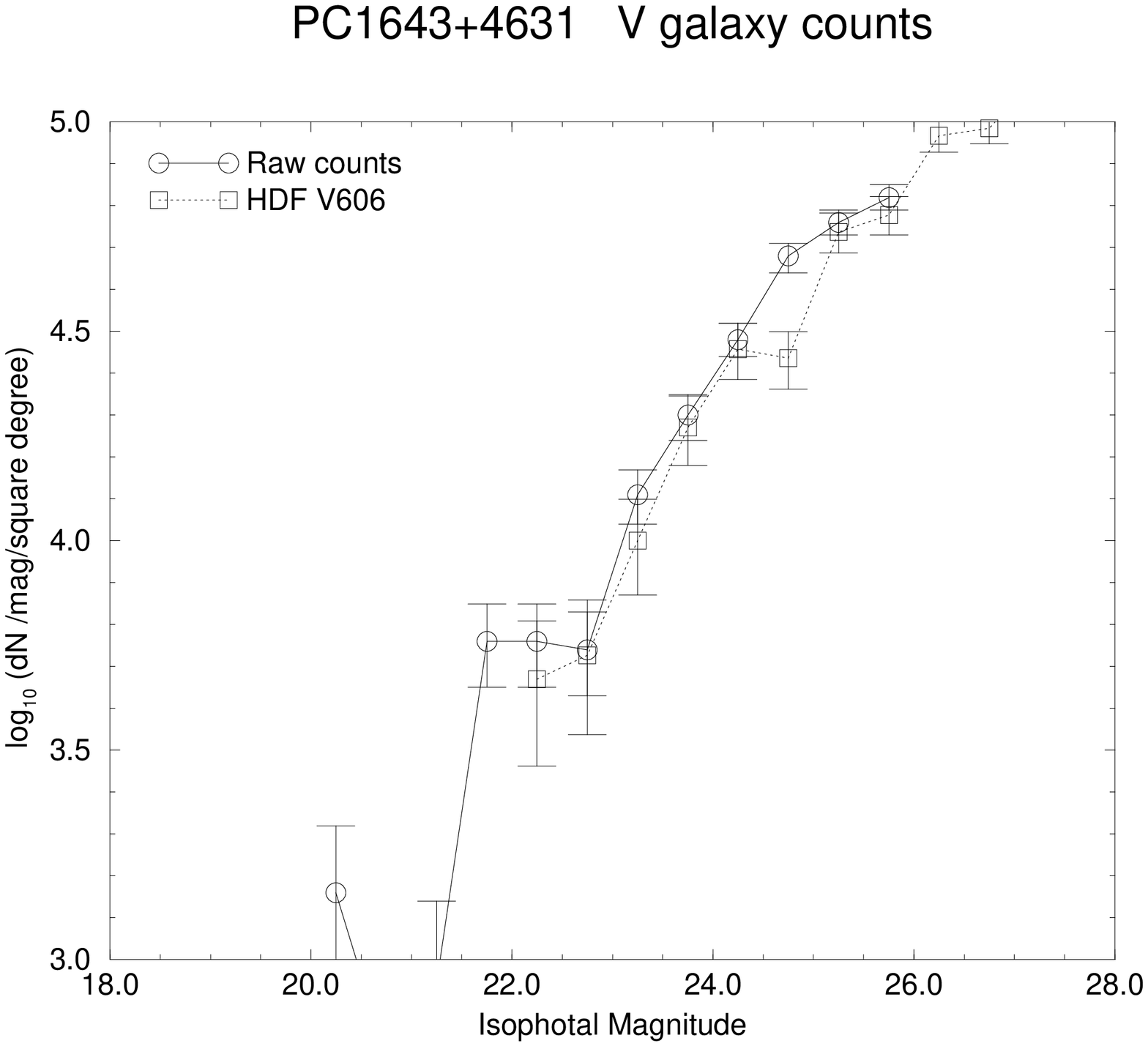,width=0.65\linewidth,angle=270}
\end{center}
\caption{Galaxy counts for $V$ filters, compared with previously published results
\label{fig:vcount}
}
\end{figure}

\begin{figure}
\begin{center}
\epsfig{figure=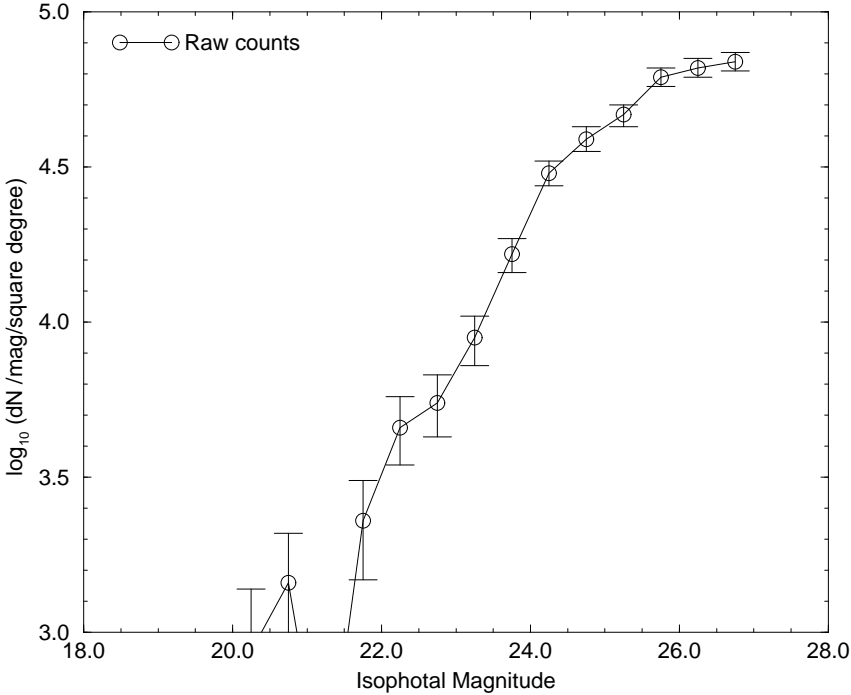,width=0.65\linewidth,angle=270}
\end{center}
\caption{Galaxy counts for $G$ filters
\label{fig:gcount}
}
\end{figure}

\begin{figure}
\begin{center}
\epsfig{figure=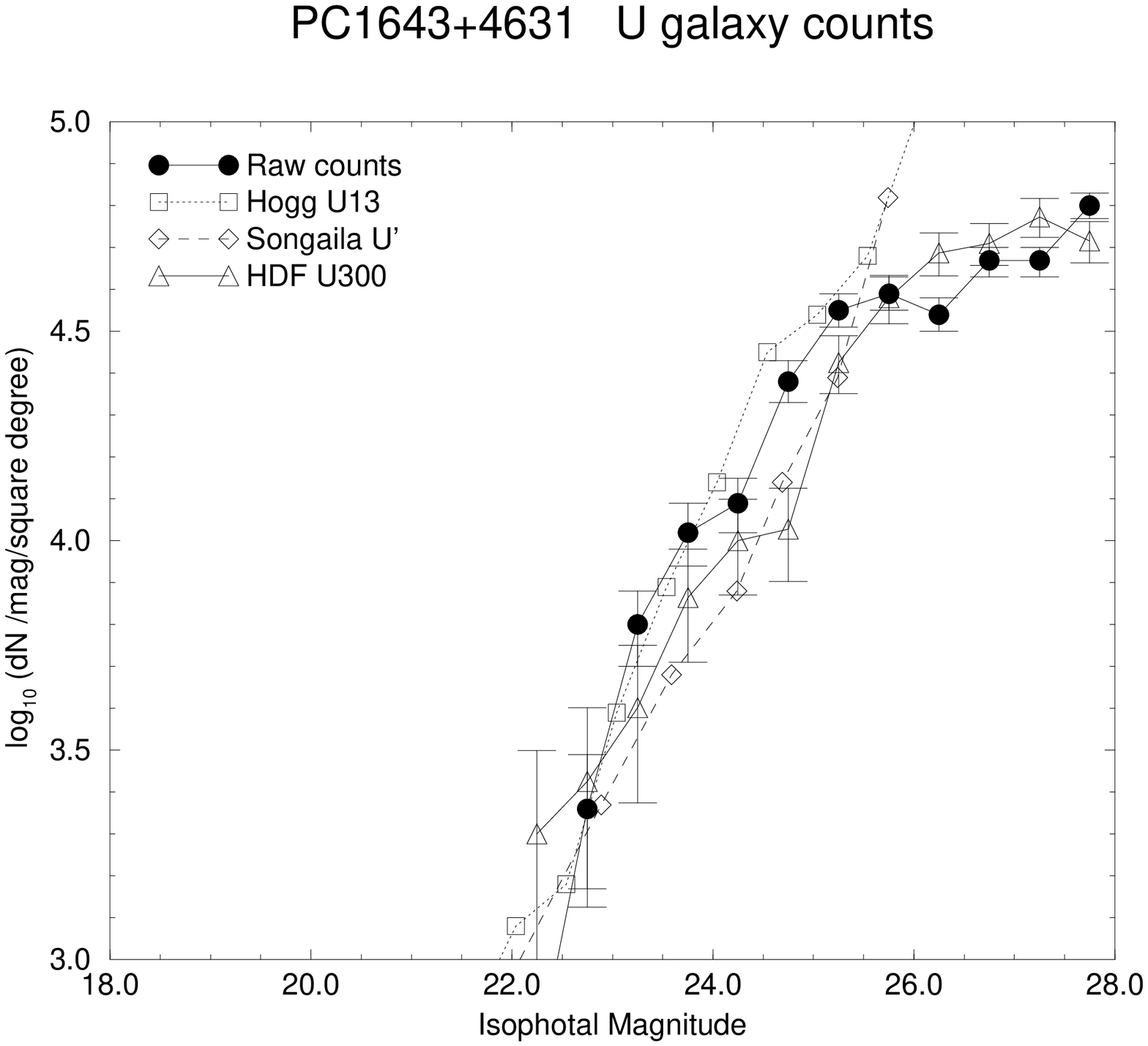,width=0.65\linewidth,angle=270}
\end{center}
\caption{Galaxy counts for $U$ filters, compared with previously published results
\label{fig:ucount}
}
\end{figure}

I--band galaxy counts already exist for several
fields. Figure~\ref{fig:icount} shows some of these together with the
countes for PC1643; there is very good agreement. In
Figures~\ref{fig:rcount}--\ref{fig:ucount} we present the PC1643
counts in $R$, $G$, $V$ \& $U$, with comparison counts where
possible. Four points are of note:\\
\nocite{Cas95}\nocite{Dri95}\nocite{Dri94}
\nocite{Gla95}\nocite{HM84}\nocite{MSC96}
\nocite{Koo86}\nocite{LeF95}\nocite{SHY95}

\begin{figure}
\begin{center}
\epsfig{figure=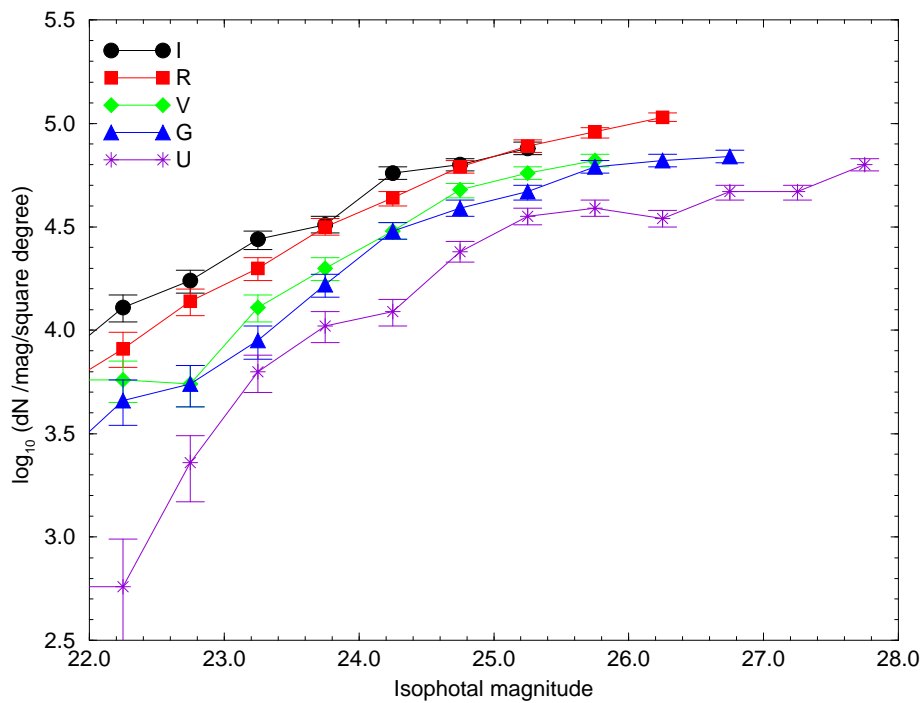,width=0.65\linewidth,angle=270}
\end{center}
\caption{Galaxy counts for all broadband filters used here
\label{fig:allcounts}
}
\end{figure}

\begin{enumerate}[(1)]

\item{The field of PC1643 is similar in its counts to all the other
ground--based deep fields used in the literature for measuring
differential galaxy counts. This is true in each broadband filter.}

\item{Down to $\textrm{magnitudes}\sim24$ the slope of the counts
increases with decreasing wavelength, caused by the presence of the
faint blue population which becomes a significant proportion of the
sample at the fainter magnitudes. This is clearly seen in
Figure~\ref{fig:allcounts}. Beyond a critical magnitude, the counts
fall off due to the difficulties in detecting the fainter objects
above the pixel noise. No attempt has been made in these galaxy counts
to `correct' for these discrepancies.}

\item{Also clear is the flattening of the $U$--band counts at fainter
magnitudes -- this may be partly due to the Lyman limit of
high--redshift galaxies at $z\sim3$ moving into the $U$ filter,
resulting in a drop in perceived counts.}

\item{It is noticeable that the results published by \cite{Hogg97} and
\cite{SCL90} continue to rise steeply where our counts level off at
$U\approx26$. Since our results are raw counts, this may be evidence
of incompleteness in our sample. However, it is interesting to note
that the $U_{300}$ counts published in \cite{HDF96} from the Hubble
Deep Field do not continue to rise steeply past $U\approx26$. The HST
$U_{300}$ filter has a wavelength range of approximately from 2000\AA\
to 4000\AA\, compared with ground--based $U$ filters which are
effectively limited by atmospheric absorption to a shortest wavelength
of approximately 3000\AA . The $U$ bandpass (including the effects of
atmospheric transmission, CCD response and telescope lightpaths) of
these observations gives us a range from 3000\AA\ to 4000\AA , i.e. a
subset of the HST bandpass. The greater range of wavelength of the HST
$U_{300}$ observations means that the $U_{300}$ counts will start
levelling off at slightly different magnitudes to the $U$ used here.}
\end{enumerate}

The method used here to test completeness, i.e.\ of comparing the
expected differential counts with those actually recovered from the
images, relies on the actual counts following the trend set out in the
model -- in this case, a power--law behaviour. Other completeness
models (eg \cite{Hogg97}) also rely on there being no sharp changes in
the slopes of the real differential counts. With the results of the
HDF counts, we can be confident that this is indeed the case for the
counts presented here, with the exception of the $U$ counts. It seems
entirely feasable that this sharp change results in an overestimate of
the counts compared with two ``traditional'' completeness estimators.

\afterpage{\clearpage}

\section{Searches for line emission using the narrow--band filters}

We searched for candidates showing strong line--emission around
$z=3.14$ \& $z=3.81$ by comparison of the magnitudes of the objects in
the narrow--band images against the magnitudes of the same objects in
the broad--band images.  Similarly we also searched for strong
absorption at the two redshifts, such as is seen in the spectrum of
Quasar A. Our search criteria was based on identifying objects which
have $(\textrm{broadband}-\textrm{narrowband})$ colours which lie at
least $4\sigma$ away from the expected continuum value. Additionally,
these objects should be extremely faint in $U$, since the column
density of neutral hydrogen will absorb any radiation below 912\AA\ in
the rest frame of the galaxy, and at a redshift $z>3$ this falls
inside our $U$ filter. With the exception of the quasars there are no
strong candidates with highly significant
$(\textrm{broadband}-\textrm{narrowband})>6\sigma$ values. More
marginal candidates exist with in both cases, with
$(\textrm{broadband} - \textrm{narrowband}) \approx 4\sigma$,
$R\gtrsim24$ and faint $U$ magnitudes.

\subsection{Searches for galaxies at $z\sim3.8$ using the L5840\AA\ filter}
\label{sec:L5840}

\begin{figure*}
\begin{center}
\epsfig{figure=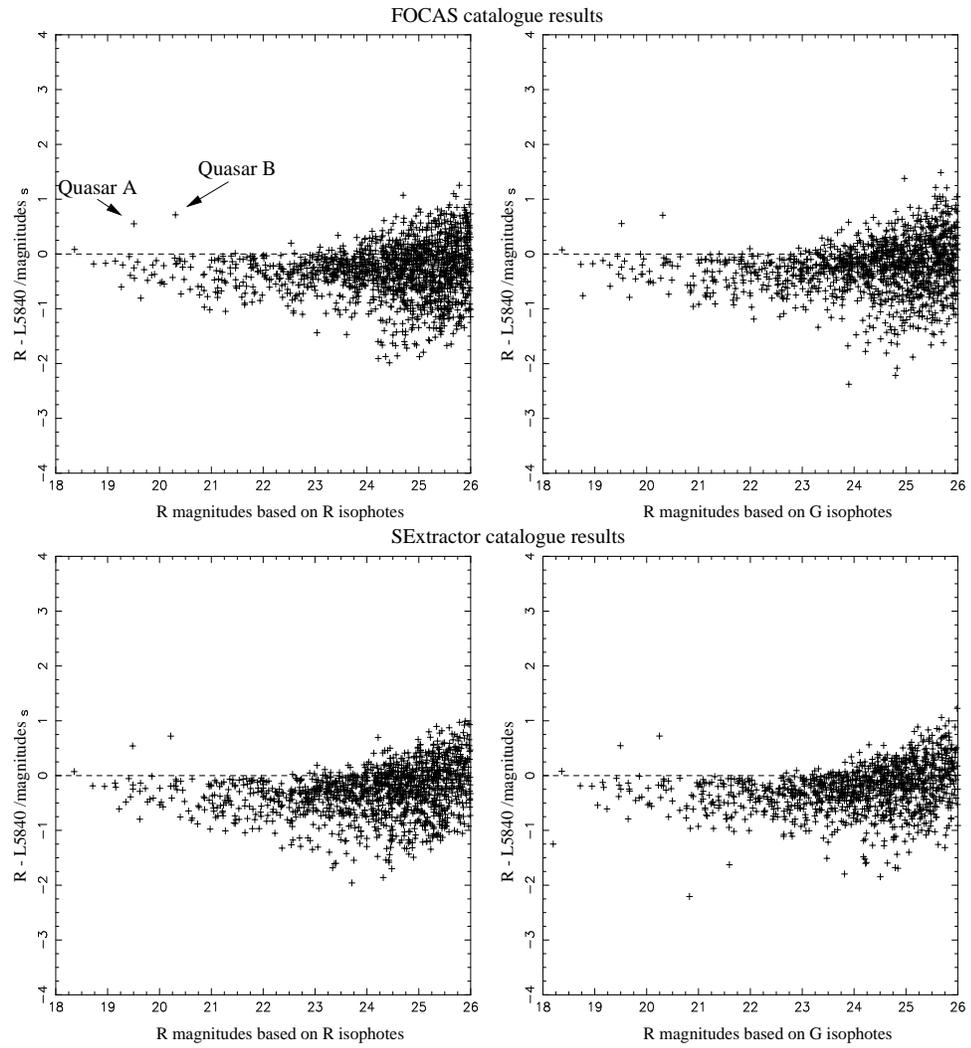,width=0.8\linewidth,angle=0}
\caption{Broadband -- narrowband magnitudes for the search for
Ly-$\alpha$ emission at $z\approx3.81$. The two graphs on the left
have magnitudes determined from isophotal apertures defined in the $R$
image, while the two on the right rely on isophotal apertures defined
in the $G$ image. The upper two are from the FOCAS catalogue, with the
lower two using the SExtractor results.
\label{fig:L5840}
}
\end{center}

\end{figure*}

Figure~\ref{fig:L5840} shows the $R-L5840$ colours for catalogues
based on $R$ and $G$ isophotal apertures constructed using FOCAS and
SExtractor for comparison. The quasars are marked and are the only
objects standing clear of the rest of the catalogue. The increase in
photometry errors can account for the spread of the candidates at
faint magnitudes, and the range of $G-R$ colours contribute to the
spread of $R-L5840$ colours at all magnitudes. Choice of isophotal
apertures or photometry software appears to make little difference,
although there are more faint--magnitude objects in the FOCAS
catalogues. Based on the FOCAS catalogue using $R$ isophotal
apertures, we find 25 objects with $R-L5840>4.5\sigma$ in the range
$19<R<26$. Of these candidates, eight objects (including the quasars)
have $U-G>2$ suggesting that they may be at $z>3$.

The apparently increased scatter in $R-L5840$ suggesting absorption at
5840\AA\ (ie increasingly negative $R-L5840$ values) is not
significant, since the fainter $L5840$ magnitudes have larger
photometric uncertainties. This also explains the apparent rise in the
median $R-L5840$ values at $R>24$, since the fainter $L5840$
candidates are increasingly undetectable for the larger negative
$R-L5840$ values. 

\subsection{Searches for galaxies at $z\sim3.14$ using the S5040\AA\ filter}

\begin{figure*}
\begin{center}
\epsfig{figure=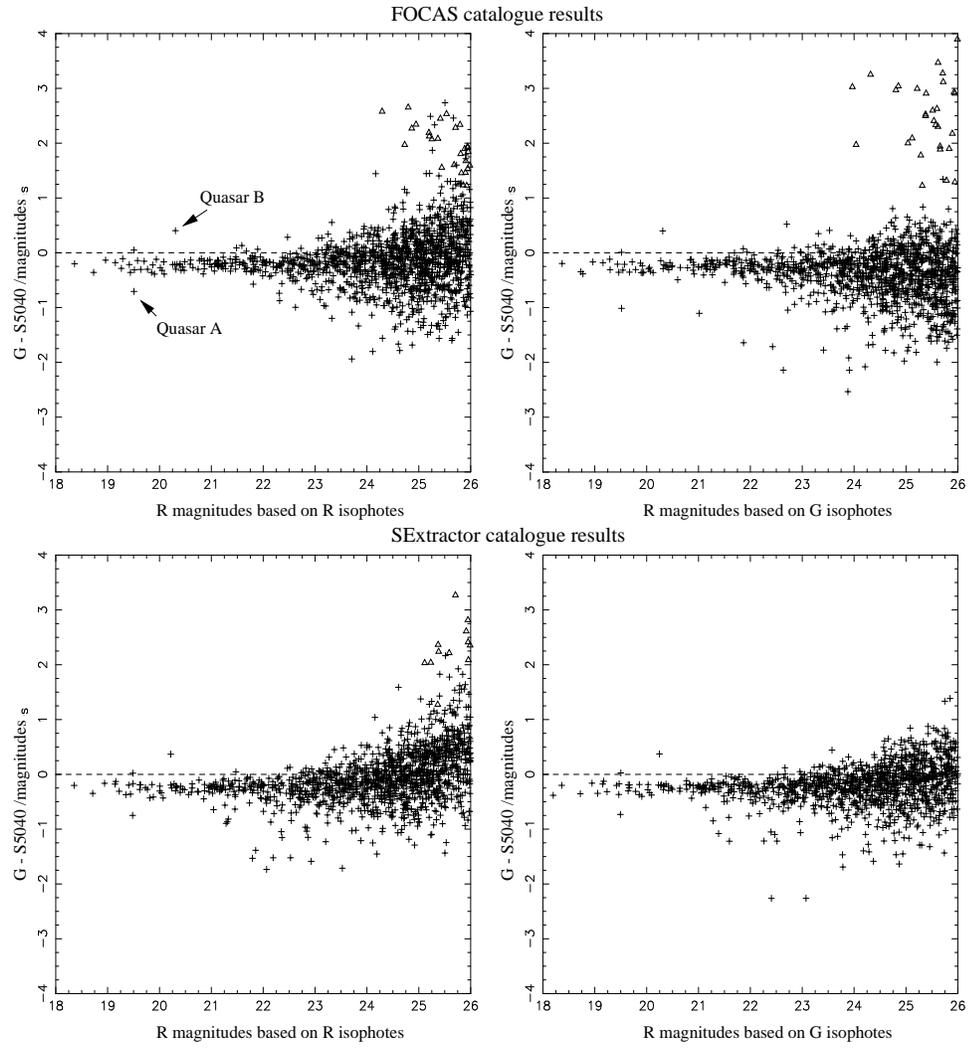,width=0.8\linewidth,angle=0}
\caption{Broadband -- narrowband magnitudes for the search for
Ly-$\alpha$ emission at $z\approx3.14$
\label{fig:S5040}
}
\end{center}
\end{figure*}

In Figure~\ref{fig:S5040}, we plot the $G-S5040$ colours for
catalogues based on $R$ and $G$ isophotal apertures constructed using
FOCAS and SExtractor for comparison. Quasar A stands clearly out from
the rest of the catalogue showing the Ly-$\alpha$ absorption
feature. However, Quasar B shows a distinct excess over the G band.
Examination of the spectrum of Quasar B (\cite{Saun97}) reveals this
is consistent -- there is some emission between 4950\AA\ and 5100\AA\
over the contiuum level but this is almost certainly due to Ly-$\beta$
(perhaps with some OVI) emission from the quasar itself rather than
Ly-$\alpha$ emission from an object at $z=3.14$.

In the range $19<R<26$ we find 70 objects with $G-S5040>3\sigma$ and 6
objects with $G-S5040>4.5\sigma$. Of the 70 objects, 3 have
$U-G>2$. The differences between the FOCAS and SExtractor--derived
catalogues at the faintest magnitudes ($R\approx25$) shows a number of
more extreme $G-S5040$ candidates for both sets of isophotal
catalogues, but these objects have large photometric uncertainties.

In comparison with other narrowband searches for Ly-$\alpha$ emitters
at $z>3$, we would have expected about 6 Ly$-\alpha$ emitters in a
field this size (\cite{Hu98}). Our broadband images are sufficiently
deep to have detected continuum from all but the faintest galaxies
assuming a similar distribution to Cowie \& Hu's sample
(\cite{CHu98I}). However, our narrowband images are not as deep and it
is likely that there may be other faint galaxies with strong emission
lines which are buried by the photometric errors. Given that
$R-L5840\lesssim1$ even for a strong Ly-$\alpha$ emitter, using
objects detected in the much deeper broad--band filters rather than
using catalogues based on isophotal apertures in the narrow--band
images should not have missed a significant number of galaxies. The
range of broadband filters we have here also allows us to discriminate
strongly against other possible emission lines, such as [OII], by
ensuring that the continuum measurements are consistent with
Lyman--break high--redshift galaxies.

\section{Colours and spatial distributions}

\begin{figure}
\begin{center}
\epsfig{figure=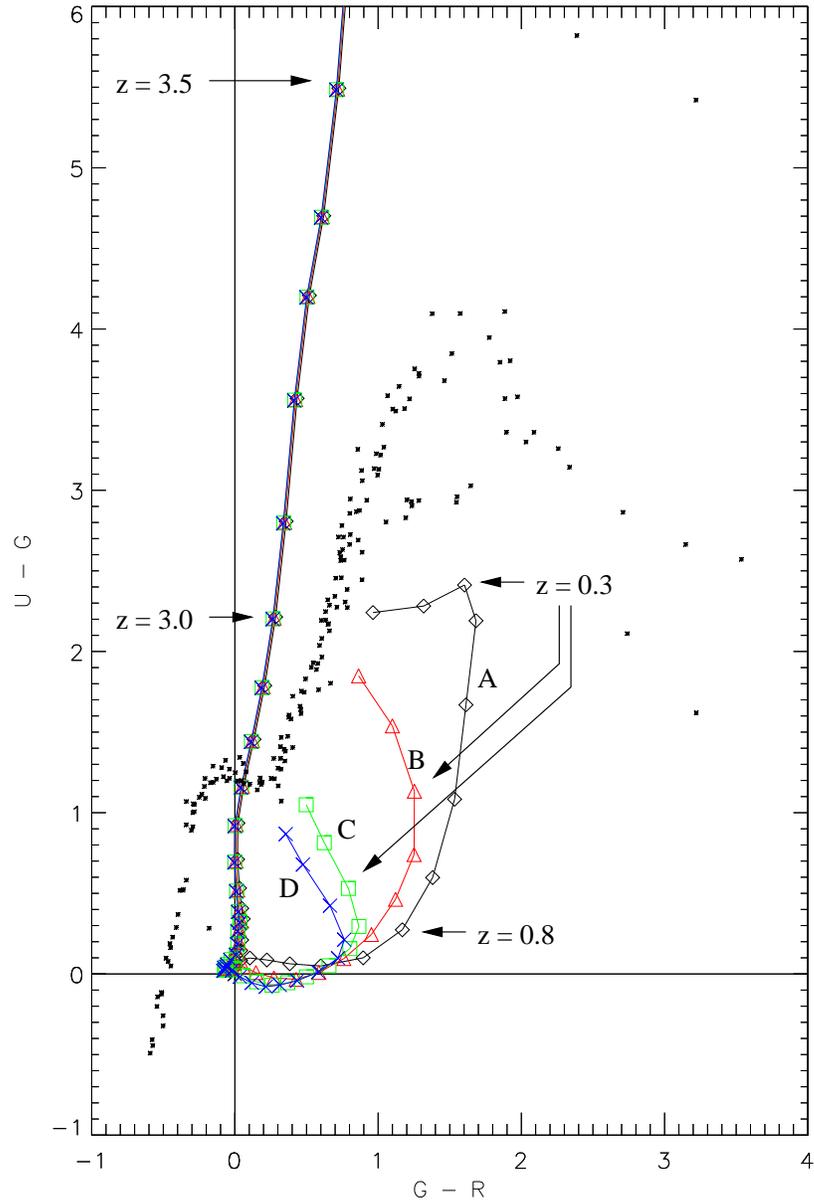,width=0.75\linewidth,angle=0}
\end{center}
\caption{\label{fig:bc} Simulated galaxy colours as a function of
redshift. The crosses represent stars taken from the stellar database
(\cite{GS83}) ranging from about O5 to M3. These models are derived
using a formation redshift $z_f = 5.0$, using
bc96\_0p0200\_sp\_ssp\_kl96.ised in the B\&C 96 distribution. The
various tracks A-D are based on the following star formation
histories:}

\begin{center}
\begin{minipage}[hp]{0.85\linewidth}
A, B \&\ C have exponentially decreasing star formation rates $e^{-\tau}$\\
\hspace{1in}A : $\tau =$ 1 Gyr\\
\hspace{1in}B : $\tau =$ 2 Gyr\\ 
\hspace{1in}C : $\tau =$ 7 Gyr\\ 
D has a constant star formation rate
\end{minipage}
\end{center}

\end{figure}

We used the GISSEL code developed by Bruzual \& Charlot \cite{BC93} to
model the colours of distant galaxies observed through our filters, in
a similar manner to that adopted by \cite{StII}. The standard
evolution curves for three simulated galaxies in the $U-G$ and $G-R$
colours are shown in Figure~\ref{fig:bc}. Note the close coincidence
of all of these evolution loci beyond a redshift of 2.6, and the
extreme red $U-G$ colours for higher redshift galaxies.

In general, trying to select objects at some redshift by optical
colour is only efficient for specific redshift ranges. At low
redshift, the 4000\AA\ break provides a clear colour signature that
can be used to identify galaxies between $0<z\lesssim1.2$. Once the
4000\AA\ break moves out of the $I$ band around $z\approx1.2$, the
continuum spectrum covered in the 3000\AA\ -- 9000\AA\ range is fairly
flat, making identification more difficult. This trend continues until
the Ly--$\alpha$ forest and Lyman--break features move into the
shorter wavelength filters at $z\approx2.7$, resulting in a rapid
faintening of the observed $U$ magnitude and providing a clear $U-G$
colour signature by $z\approx3$. At still higher redshifts, the $U-G$
vs $G-R$ colour--colour diagrams would seem to suggest that
identification of such galaxies would be equally efficient, and in the
case of a extremely deep field, such as the Hubble Deep Field, this is
observed. However, our data are limited to $R\approx27$. As a result,
any population of $3.5<z<4.1$ galaxies, with $R\approx26$ and
$G-R\approx1$, are recorded with lower limits of $U-G\approx2$ and are
confused with the `body' of the colour--colour diagram. For redshifts
$z\gtrsim4.1$, the Lyman--break has moved into the $G$ filter, and
hence a similar search to that carried out for $z\approx3$ galaxies
can be done using $G-V$ vs $V-I$ or similar.

\subsection{Intermediate redshifts}

\begin{figure}
\begin{center}
\epsfig{figure=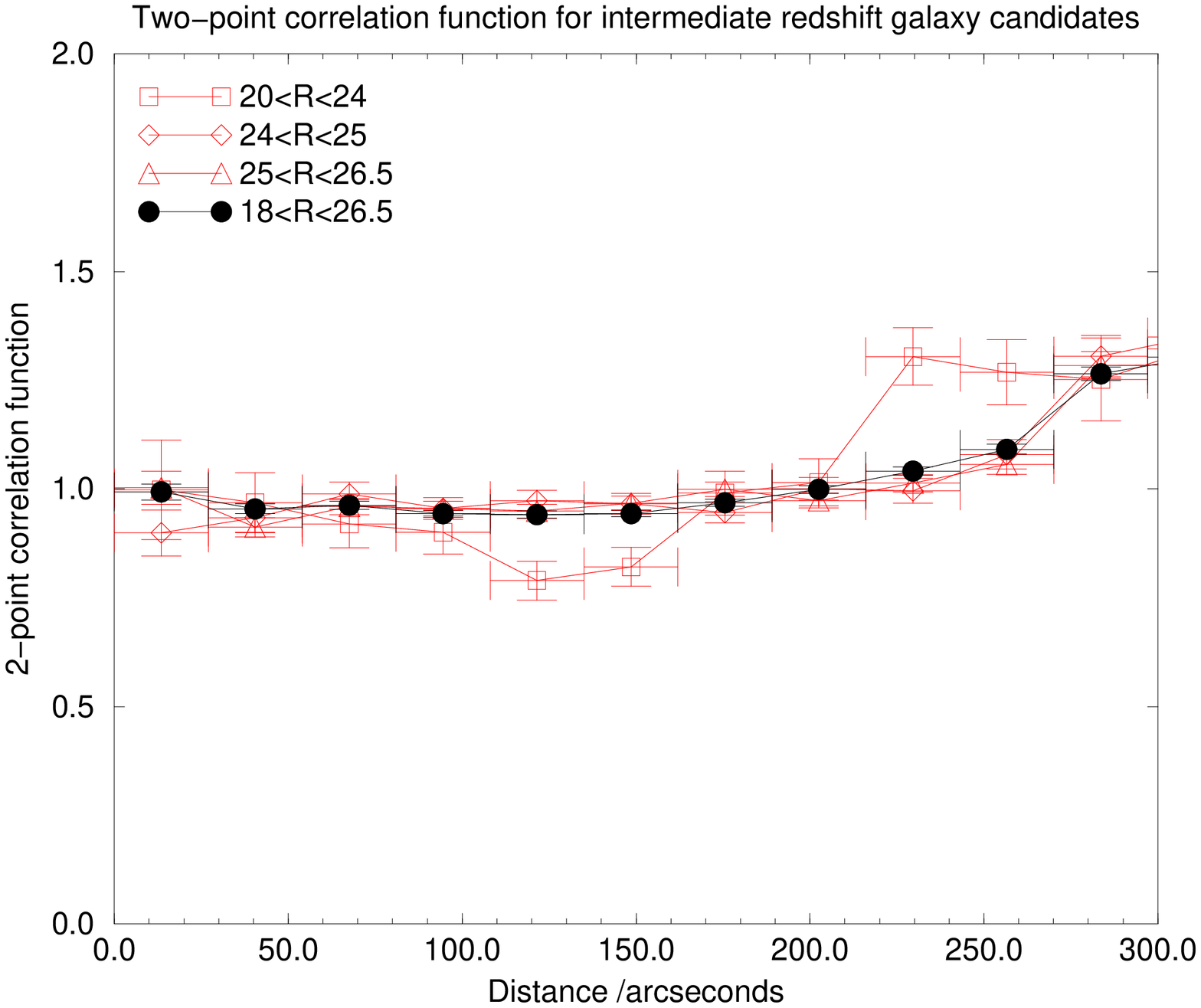,width=0.8\linewidth,angle=270}
\end{center}
\caption{Two--point correlation function for the intermediate redshift
candidates\label{fig:2pc-int}
}
\end{figure}

\begin{figure}
\begin{center}
\epsfig{figure=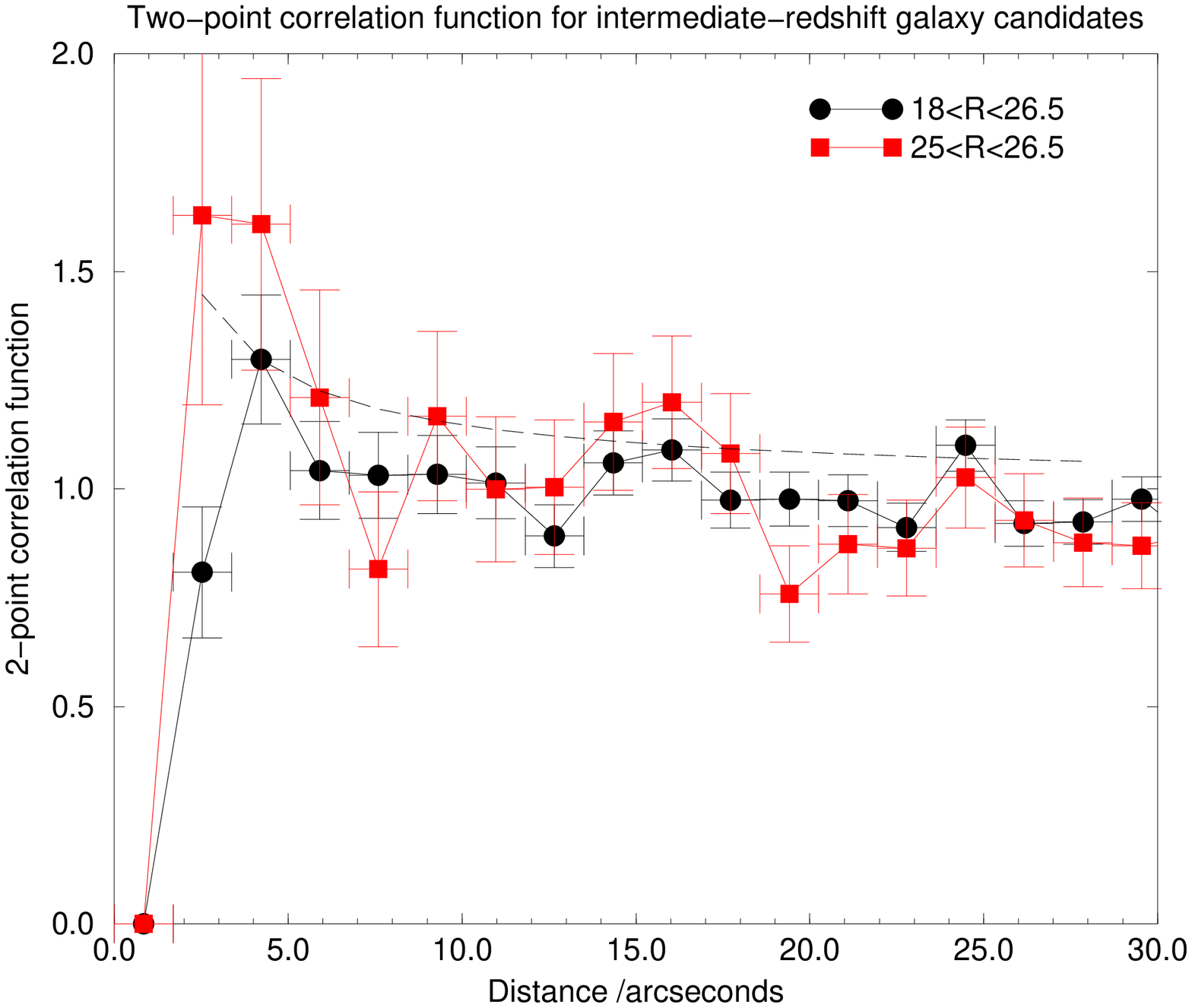,width=0.8\linewidth,angle=270}
\end{center}
\caption{Two--point correlation function for the faint intermediate--redshift candidates\label{fig:2pc-int-256} }
\end{figure}

Objects in the field at intermediate redshifts, $1<z<2.5$, are likely
to be very blue in $G-R$ (see Figure~\ref{fig:bc}), although reddening
of these galaxies may affect this colour difference. This also assumes
that they are still vigorously starforming at these
redshifts. Adopting limits of $0<(U-G)<2$,\ \ $0<(G-R)<0.5$ and
$R<26.5$ to limit contamination from low and high redshift galaxies as
far as possible, we find a total of 494 objects with suitable colours
in the central $5'\times5'$ region, using the catalogue based on the
$R$ isophotal apertures. Examination of the distribution of these
objects on the sky reveals no visually discernable dense region. The
two-point correlation function (given in Figure~\ref{fig:2pc-int}) for
all of these objects shows that the field is indeed remarkably
uniform, although there is a suggestion that there are more objects at
the edge of the field than in the middle . Splitting the sample into
magnitude--selected groups shows that there is more clustered
structure at the brighter magnitudes, while there is evidence of
small--scale ($<20$ arcsec) grouping of the candidates at the faint
magnitudes (Figure~\ref{fig:2pc-int-256}). Fitting the standard
power--law, $w(\theta)=A_\omega\theta^{-\beta}$, to the whole sample
over angular scales $\theta$ from 2 -- 30 arcseconds using non--linear
curve--fitting gives close--to--zero values for $A_\omega$. Repeating this
procedure for the faint end of the sample with $25<R<26.5$ gives
best--fit parameters of $A_\omega\approx3.3$ and $\beta\approx1.6$,
suggesting that these galaxies are more strongly clustered than galaxy
samples in the local universe. Values of $\beta=0.8$ and
$A_\omega\approx0.9$ (\cite{BS98}) can be well fitted with our data
for angular scales less than 20 arcseconds, but diverge for larger
scales at a 95\% confidence level.

\begin{figure}
\begin{center}
\epsfig{figure=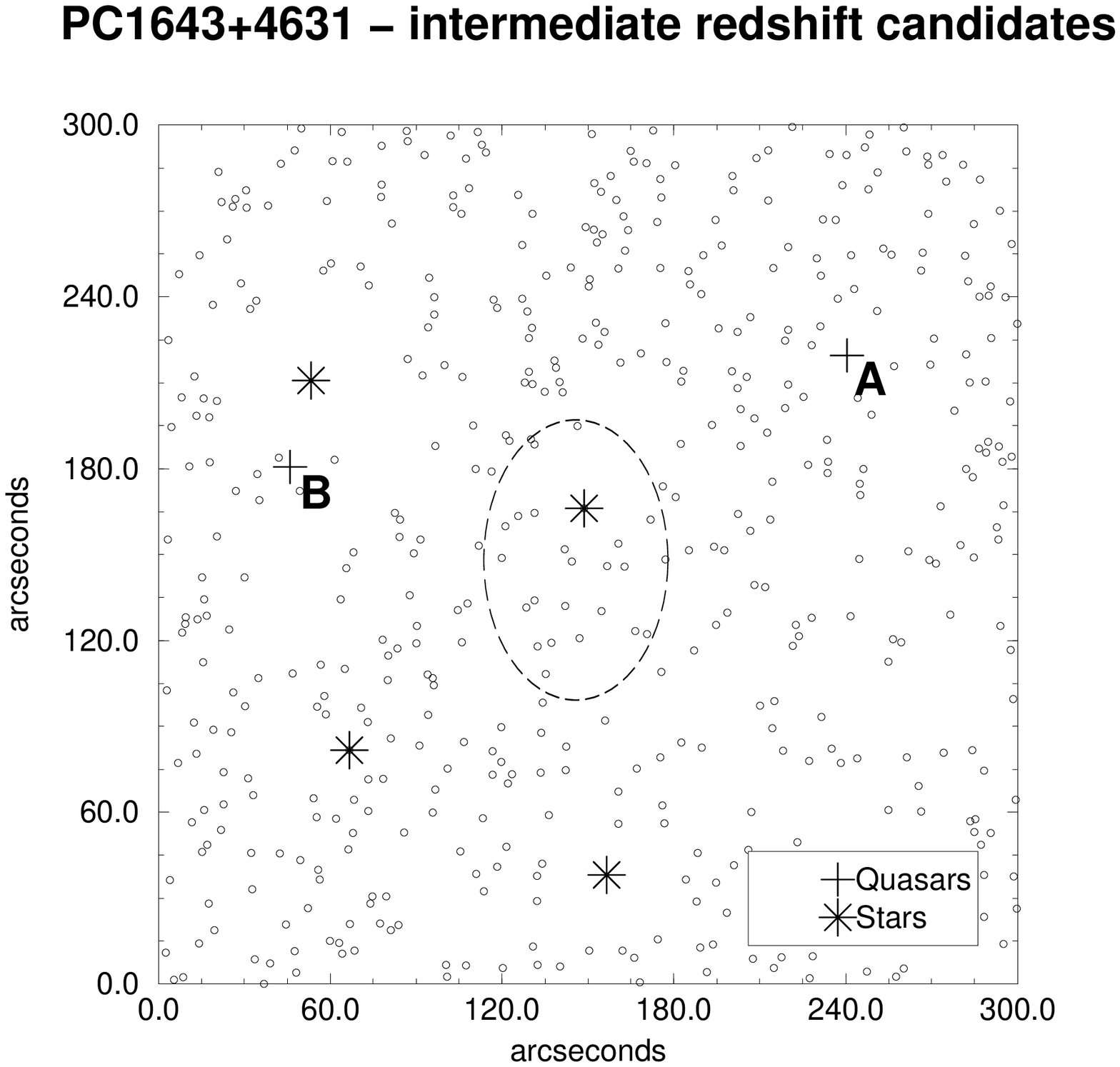,width=0.7\linewidth,angle=270}
\epsfig{figure=figure31.ps,width=0.7\linewidth,angle=270}
\end{center}
\caption{The spatial and magnitude distributions of the
intermediate--redshift candidates, based on $R$ isophotal
apertures. The dashed ellipse shows the $1\sigma$ limits of the
position of the centre of the S-Z decrement. The bright candidates
with $R<22$ are almost certainly blue stars contaminating the
sample\label{fig:intermediate} }
\end{figure}

As shown in Figure~\ref{fig:intermediate} the majority of these
intermediate redshift candidate objects are at faint magnitudes in the
range $24<R<26.5$ as expected; the 26 brightest objects with $R<22.5$
are almost certainly all stars. However the slope,
$\textrm{dN}/\textrm{dm}$, of these objects between $22.5<R<25$ is
higher than the slope of the whole sample (as shown in
section~\ref{sec:counts}) and is more consistent with the slope of the
$G$ band observations.

The exact number of intermediate redshift objects found depends on the
choice of isophotal apertures, as tabulated in
Table~\ref{tab:intersel}. Selection from the $G$ isophotal apertures
gives significantly (ie $3\sigma$ assuming Poisson statistics) more
candidates than from the $R$ isophotal apertures despite the
approximately equivalent magnitude limits of the images. Comparing the
magnitude distributions shows that this discrepancy occurs mainly at
magnitudes of $24<R<26.5$, suggesting that this effect is due to
photometric measurement errors. Selection of objects using the $G$
isophotal catalogue results in more faint $R$ objects being included
in the set. This may be due to the slightly larger isophotal apertures
obtained from the $G$ images increasing the photometric
errors. Comparison of these two matched catalogues based on isophotal
$R$ and $G$ apertures against the matched catalogue using isophotal
apertures made from each broad--band image for each filter suggests
that, in this case, the $R$ isophotal catalogue is giving a more
accurate set of results. While it would appear that using a catalogue
where each image uses its own isophotal set of apertures is optimal,
particularly since there is no guarantee that the isophotal apertures
need be colour independent, using $R$ isophotal apertures is more
effective here for several reasons:

\begin{enumerate}[(1)]

\item{the surface brightness limit of the $R$ image is fainter than
the other frames, with the exception of $U$;}

\item{by detecting objects in $R$, objects in filters bluer than $R$
are unlikely to be missed, which is the main reason not to use $U$ for
the isophotal apertures (e.g. Lyman--limit galaxies are not detected
in $U$);}

\item{where objects in less--deep images would not be recognised at
all using isophotes based on that image, using isophotal apertures
based on the deeper $R$ frame allows upper limits to be places on the
magnitude of that object.}

\end{enumerate}

\subsection{High redshifts}
\label{sec:steidel}

\begin{figure}
\begin{center}
\epsfig{figure=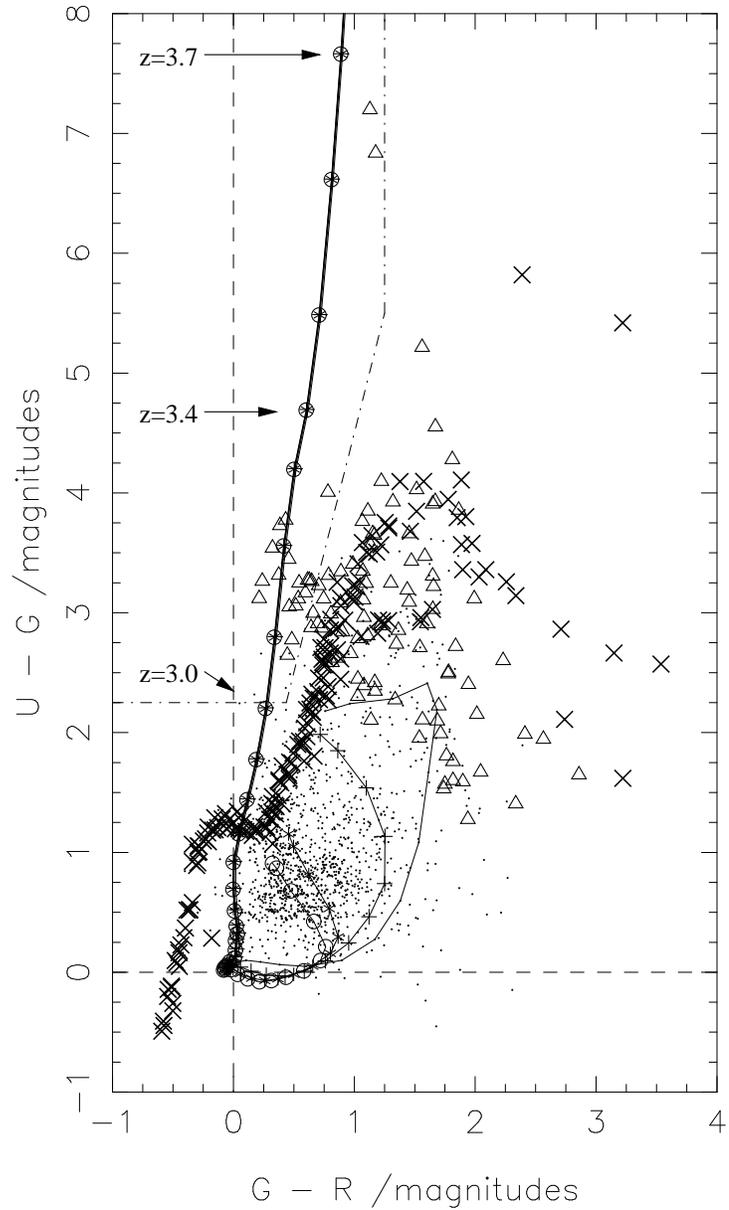,width=0.6\linewidth,angle=0}
\caption{ $U-G$ vs $G-R$ colour--colour graph for PC1643+4631 showing
the high--redshift selection criteria for all objects with $R<25.5$
and detections in $G$ of at least 2--$\sigma$. Dots denote detections
in $U$,\ $G$ \&\ $R$. Triangles denote 1--$\sigma$ lower limits in
$U-G$.  The crosses are stars taken from the Gunn \&\ Stryker
database, and the dot--dash lines is the bound for selecting the high
redshift candidates. The tracks on this graph are for model galaxies
derived using the Bruzual \&\ Charlot 1996 distribution, as in
Figure~\ref{fig:bc}.
\label{fig:steidel.uggr}
}
\end{center}
\end{figure}

We next investigated candidate Lyman--break galaxies at
$z\gtrsim3$. Using the same approach as Steidel and Hamilton (S\&H),
we have included objects which have $R<25.5$ (ie $R\gtrsim10\sigma$)
and are detected in $G$ to at least $2\sigma$. We then selected
`robust' candidates as having $U-G>2.25$, $G-R<1.25$ and
$(U-G)>4(G-R)+0.5$ as illustrated in Figure~\ref{fig:steidel.uggr}. A
second set of looser criteria to include more `marginal' detections
was also used, requiring $U-G>2.0$, $G-R<1.25$ and
$(U-G)>4(G-R)$. These are slightly different to the colour selection
criteria used by Steidel and Hamilton, and reflect the differences
between the filters we used and S\&H's -- most noticeably the track of
stars (taken from the Gunn \&\ Stryker atlas) runs at bluer $G-R$
colours than through the Steidel filters and our $U-G$ colours are
redder than those of S\&H's for the same galaxy at a similar
redshift. We stress that the colour criteria we use are at least as
conservative as that adopted by Steidel et al.: the lower limit on
$U-G>2.25$ cuts across the B\&C tracks at $z=3.0$, whereas the
`robust' Steidel criteria selects at $z\approx2.85$ (see eg
\cite{StII}). Similarly, by using $(U-G)>4(G-R)+0.5$, contamination
from any stars is minimised: the photometric errors for typical
Ly--limit candidates with $R\approx25$ do not extend to those colours
consistent with stars, whereas the criteria of Steidel et al. include
stars with $U_n-G\approx1.0$ and $G-\mathcal{R}\approx2.5$ (see
\cite{PSA97}).

\begin{figure}
\begin{center}
\epsfig{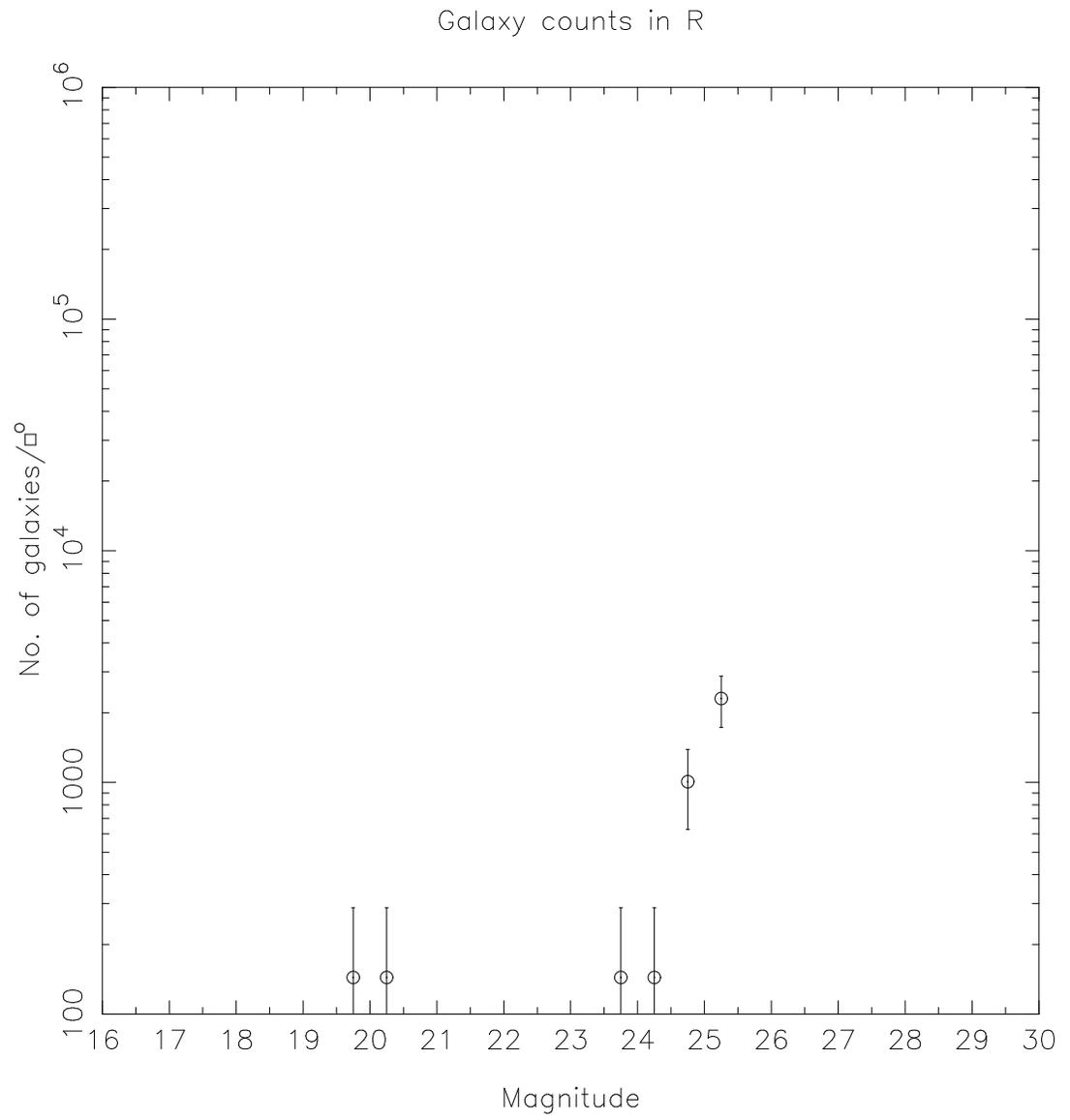}
\end{center}
\caption{$R$--magnitude distribution for the high--redshift galaxy
candidates
\label{fig:steidel.mag}
}
\end{figure}

We have identified 27 robust candidate high--redshift galaxies in
the central 5'x5' of our images, and a further 12 marginal
candidates. The two extreme $U-G\sim7$ objects in the field are
Quasars A \&\ B. Of the robust candidates, only 7 have detections in
$U$ of at least $1\sigma$, while 6 of the marginal candidates have $U$
detections of $1\sigma$, with at least one being almost certainly a
star. The magnitude distribution for the robust candidates is plotted
in Figure~\ref{fig:steidel.mag}, and shows that, with the exception of
the two quasars, all the objects have $R>23.5$, consistent with the
findings of Steidel et al.

We would expect to find 10 $U$--dropout galaxies with $R<25.0$
assuming a surface density of 0.4 per square arcmin
(\cite{SGP96})---and assuming Poisson-distributed galaxies---and
find 11. However, on lowering the limit to $R<25.5$, \cite{SAD98}
reports a mean surface density of $\approx0.7$ and so we expect only 7 with
$25.0<R<25.5$ but find 16 candidates -- a 3-$\sigma$ excess.

Comparison of the 27 candidates with $R<25.5$ in the central
$5'\times5'$ area with the surface density $\approx0.7$/square arcmin
for the SSA22 \cite{SAD98} field implies that we have a excess of
faint high--redshift objects in the field of PC1643.

Typical photometric measurement errors in $R$ are of the order of 0.05
magnitudes, with the uncertainty in the $G$ measurement being closer
to 0.1 magnitudes. This is not sufficient to drastically affect the
number of candidates included inside the criteria. Even allowing for
the additional errors due to photometric measurement, the robust
candidates are at least $1\sigma$ away from the stellar track in both
$U-G$ and $G-R$. This is consistent with the approach adopted by S\&H.

Using the different isophotal catalogues available has some effect on
the colours recorded for the candidates derived from the $R$
isophotes. Table~\ref{tab:steidel.diff} shows the fraction of the 27
candidates which are well matched in position to positions in the
different catalogues, along with the number which still fulfill the
original criteria for selection. Where cases have not fulfilled the
colour criteria, this is mainly because the isophotal apertures have
been taken from the less deep images, such as $V$ and $I$. The
deviation in the case of the $U$ isophotal apertures is due to the
strong colour dependance of this selection criterion.

\afterpage{\clearpage}

Most of the photometric uncertainty is in the determined value of
$U-G$, particularly the determination of the $U$ magnitude. Individual
scrutiny of the candidates indicates that some 10 candidates show
faint but discernible flux in $U$, and it would appear that these
detections can be supported by matching the positions of the
apparently $U$ devoid candidates against the catalogue based on the
$U$ isophotal apertures. Because galaxies beyond $z\approx3$ will show
significant extinction of photons below the Lyman--limit, such
candidates with $U$ flux are more likely to be at redshifts of less
than 3. However, it is worth noting that Quasar A has a close
companion within $2''$ with significant $U$ flux, underlining the fact
that the discarding of high-redshift candidates by this method may
actually remove real high-redshift objects through line--of--sight
confusion with foreground objects.

\begin{figure}
\begin{center}
\epsfig{figure=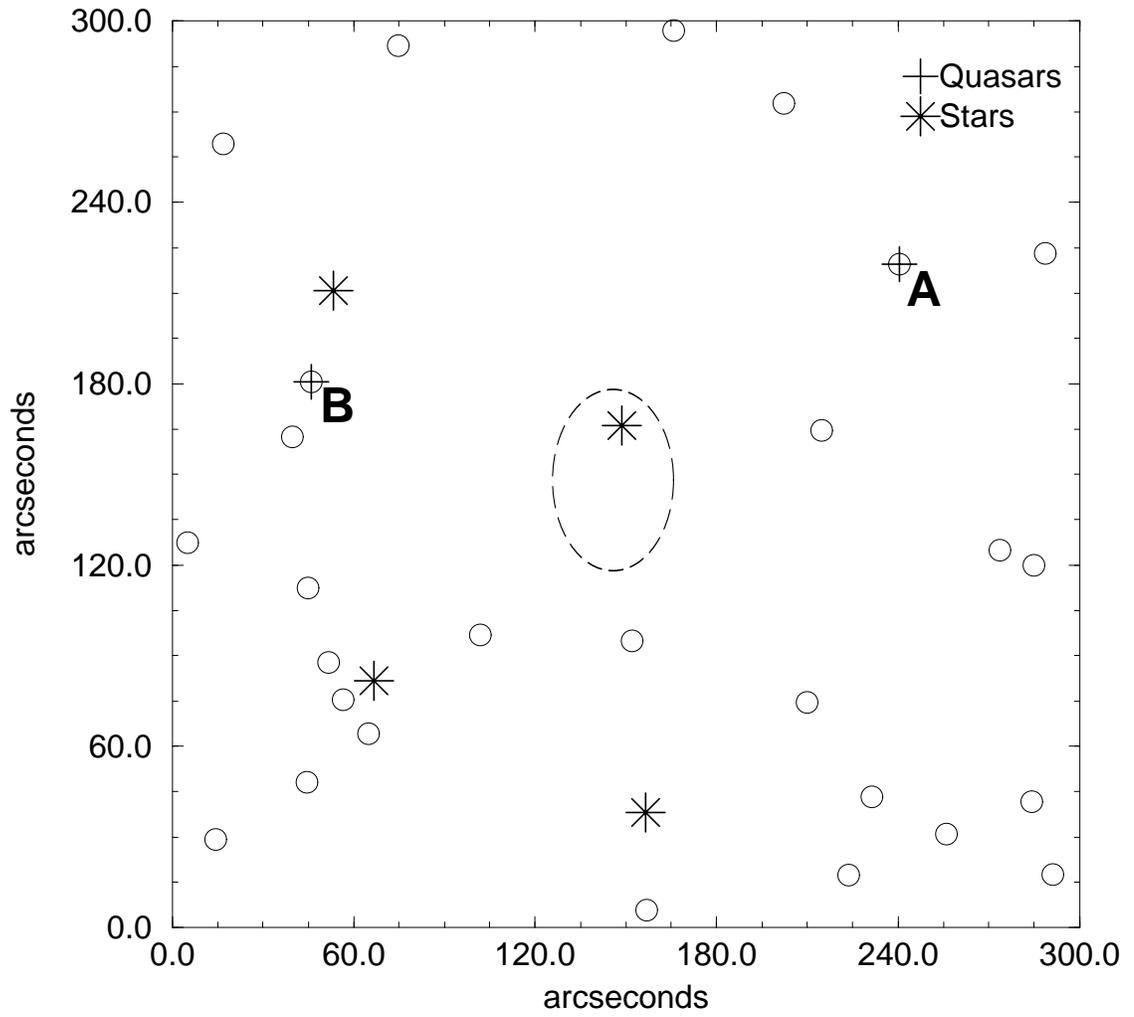,width=0.95\linewidth,angle=270}
\end{center}
\caption{Spatial distribution of the high--redshift candidates, as
determined using the colour criteria given in
section~\ref{sec:steidel}. The dashed ellipse shows the $1\sigma$
limits of the position of the centre of the S-Z decrement.
\label{fig:steidel.space}
}
\end{figure}

Examination of the positions of the 27 high--redshift candidates on
the sky shows a distinctly non-random distribution
(Figure~\ref{fig:steidel.space}). The region between the quasars
appears to be sparse in candidates -- taking a circle of radius 80''
about a position midway between the two quasars yields no objects,
while an annulus extending from 80'' to 160'' contains 20 objects. If
the objects in the field were uniformly distributed, one would expect
5 objects in the central circle and 15 objects in the outer
annulus. Applying a $\chi^2$--test to these data gives $\chi^2=6.6$
with 1 degree of freedom, which corresponds to a probability of
$>0.95$ that this distribution is not consistent with a uniform
distribution. The contrast with Figure~\ref{fig:intermediate}, the
intermediate--redshift candidates, is marked.

\begin{figure}
\begin{center}
\epsfig{figure=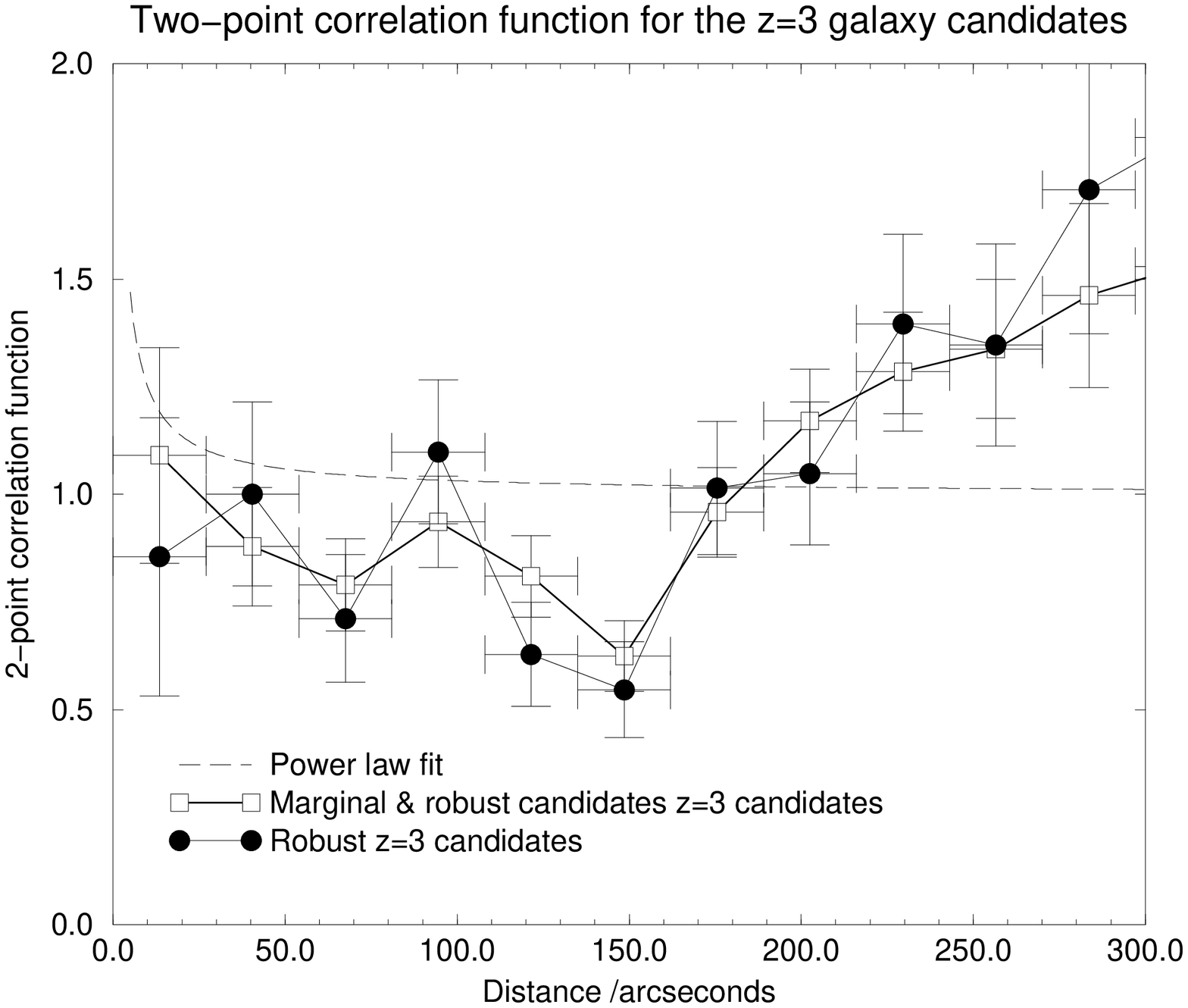,width=0.8\linewidth,angle=270}
\end{center}
\caption{Two--point correlation function for the high--redshift
candidates
\label{fig:2pc-steidel} 
} The dotted line plotted here is the power--law
$A_\omega\theta^{-\beta}$ with $A_\omega=2 \textrm{arcsec}^\beta$ and
$\beta=0.9$.
\end{figure}

We plotted the two--point correlation function, $\omega(\theta)$, for
the 27 high--redshift candidates (Figure~\ref{fig:2pc-steidel}) which
shows clustering on small scales and anti--correlation on scales of
$\approx 100''$, as expected given the lower surface density in the
centre of the field. The apparent clustering at $250''$ scales is due
to the characteristic length of separation of the sample across the
void. We plotted the best fit power--law as determined from the sample
of 871 Lyman--break galaxies (\cite{GSA98}), finding $A_\omega=2
\textrm{arcsec}^\beta$ and $\beta=0.9$, which is an extremely good fit
to our data here on scales shorter than 80 arcseconds. We tested
random distributions of points and determined that these features are
not edge effects and are most unlikely to have occurred by chance.

Overall, the field towards PC1643+4631A\&B appears to be at least as
densely populated as the most densely clustered $5'\times5'$ areas of
the SSA22 field, which covers $8.6'\times17.6'$. The presence of a
void in PC1643, with an area of 7 square arcminutes in the centre of
this field, appears unusual in comparison with SSA22.

It is difficult to determine whether this region does in fact have a
high--redshift structure based on this information alone -- the `wall'
at high-redshift found by Steidel et al following spectroscopy has an
angular size of at least 10 arcmin, about twice the field of view
on our images. However, the discovery that that the SSA `cluster' lies
at the same redshift as one of the quasars in that field is consistent
with the hypothesis that quasars act as markers of large--scale
structures at high redshifts, and that a similar correlation could be
present in the field of PC1643+4631.

The distribution of Lyman-break candidates, with an apparent deficit
between the quasars, could give rise to an apparent diminution of
radio surface brightness if all the Lyman-break galaxies are
sufficiently radioluminous, with 15 GHz flux $> 20 \mu$Jy, equivalent
to $> 40 \mu$Jy at 8 GHz. However, extremely deep 8-GHz VLA imaging of
the Hubble Deep Field (RIchards et al. 1998) detects no Lyman-brak
galaxies with a 5-$\sigma$ upper limit of 9$\mu$Jy. We conclude that
such a mechanism is not the cause of the microwave decrement. 

\afterpage{\clearpage}

\subsection{A third image of the quasar?}

Under the hypothesis that the two quasars PC1643+4631 A \& B are
indeed the same object gravitationally lensed by a $10^{15} M_{\odot}$
cluster of galaxies, we might expect to find a third image of the
quasars in the images taken. Such a third image would have the same
colours as the quasars (assuming no change of reddening), though if
the third image passes close to another system this might make the
colour signature unrecognisable. If A \& B are indeed magnified images
of a single quasar, then the third image should be demagnified and lie
closer to the centre of the S--Z detection (see, eg \cite{SGB98} and
\cite{KK98}).

To identify suitable candidates for the third image, we compared the
colours of Quasar B, which does not appear to have any nearby
companions on the sky which might pollute the isophotal aperture, to
all the catalogued objects detected in the $R$ isophotal
catalogue. The most obvious approach was to examine the various
colour--colour diagrams, particularly $U-G$ vs $G-R$ since this allows
the more extreme $U-G$ colours of high--redshift galaxies to stand out
from the rest of the objects in the field. As can be seen in
Figure~\ref{fig:steidel.uggr} there are no candidates with such
extremely red ($(U-G)\gtrsim5$) colours. This immediately gives us an
upper limit on the brightness of the third image, assuming that its
colours are not confused with a line--of--sight galaxy/star, of
$R\sim22$; i.e.\ three magnitudes fainter than the quasars themselves.

By comparing all the independent colours (i.e. $U-G$, $G-V$, $V-R$,
$R-I$, $S5040-G$ and $L5840-R$) available for all the objects against
Quasar B, we can obtain a collection of candidates for the third
image. We used the statistic

\begin{equation}
\chi^2= \sum_F\frac{(F_i-F_0)^2}{\sqrt{\sigma^2_i+\sigma^2_0}},
\label{eq:third}
\end{equation}
\begin{flushright}
\begin{tabular}{rr@{ = }l}
where   & $F_0$         & colour of quasar (e.g. $U-G$, $G-V$, etc), \\
        & $F_i$         & colour of object for comparison, \\
        & $\sigma^2_0$  & variance of colour of quasar, \\
and     & $\sigma^2_i$  & variance of colour of object, \\
\end{tabular}
\end{flushright}

\noindent to select the most similar objects, and included the known errors in
measuring the photometry (as discussed in section~\ref{sec:sim}) as
well as the actual statistical magnitude errors based on the counts
received inside the aperture. There are 7 candidates with
$22.0>R>26.0$ which have a $\chi^2<5.0$, equivalent to 95\% confidence
limit.

The errors in measuring faint objects near the limits of the catalogue
are significantly larger than the apparent statistical error; for
example, at $I=24$, the typical statistical errors ascribable to
Poisson errors in the background and aperture measurements are of the
order of 0.1 magnitudes, while the error due to measuring these faint
galaxies is of the order of 0.75 magnitudes. The search criteria
therefore tend to pick out the faintest objects in the field since
these produce significantly smaller values of $\chi^2$.

None of the objects selected is within the expected area of the sky
for a gravitationally lensed third image, given reasonable modelling
of the lensing potential constrained by the CMB decrement. However,
the lack of an obvious third image by no means rules out the
gravitational lensing hypothesis -- examples exist in the literature
of failed searches for a third image where the lensing hypothesis is
better constrained than here, such as the first FIRST gravitationally
lensed quasar (\cite{SGB98}) and the double quasar Q2138-431
(\cite{HCF97}). (The demagnification of the third image may be as
severe as to result in an image 5 or more magnitudes fainter than the
quasars \cite{SCR97}). The most probable reason for the absence of an
obvious third image is that it is confused with another object in the
line of sight.


\subsection{Extremely Red Objects}

Hu \& Ridgway have discovered two extremely red objects (EROs), HR10
\& HR14, within 1 arcmin of Quasar A (\cite{HuR94}). These were
interpreted as being dusty galaxies at an estimated redshift
$z\approx2.5$ according to analysis of the colours, with HR10 however
having a spectroscopic redshift of $z=1.44$ (\cite{GD96}). We have
compared our new optical observations with our previous infrared
observations (\cite{Saun97}) to investigate EROs in the other regions
of the field. Our J and K observations cover about 15 arcmin$^2$ of
the field, consisting of 7 pointed observations mosaiced together. The
limiting AB magnitudes are $J\approx22.1$ and $K\approx18.1$
(corresponds to $J\approx23$ and $K\approx20$ Johnson magnitudes).

\begin{figure}
\begin{center}
\epsfig{figure=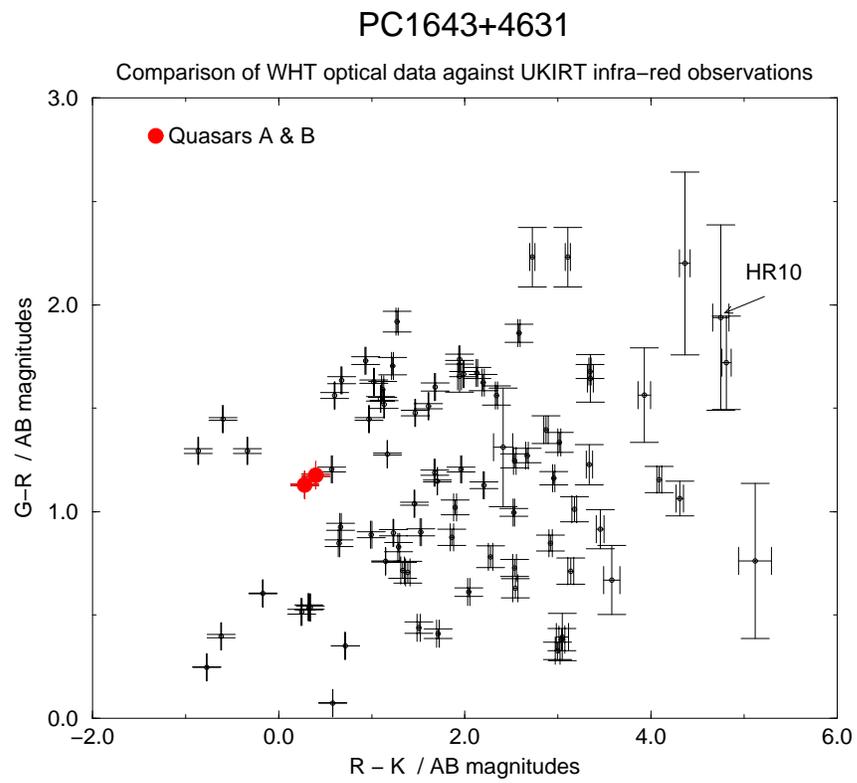,width=0.65\linewidth,angle=270}
\end{center}
\caption{Comparison of optical colour against infra--red colour.
\label{fig:rkgr}
}
\end{figure}

\begin{figure}
\begin{center}
\epsfig{figure=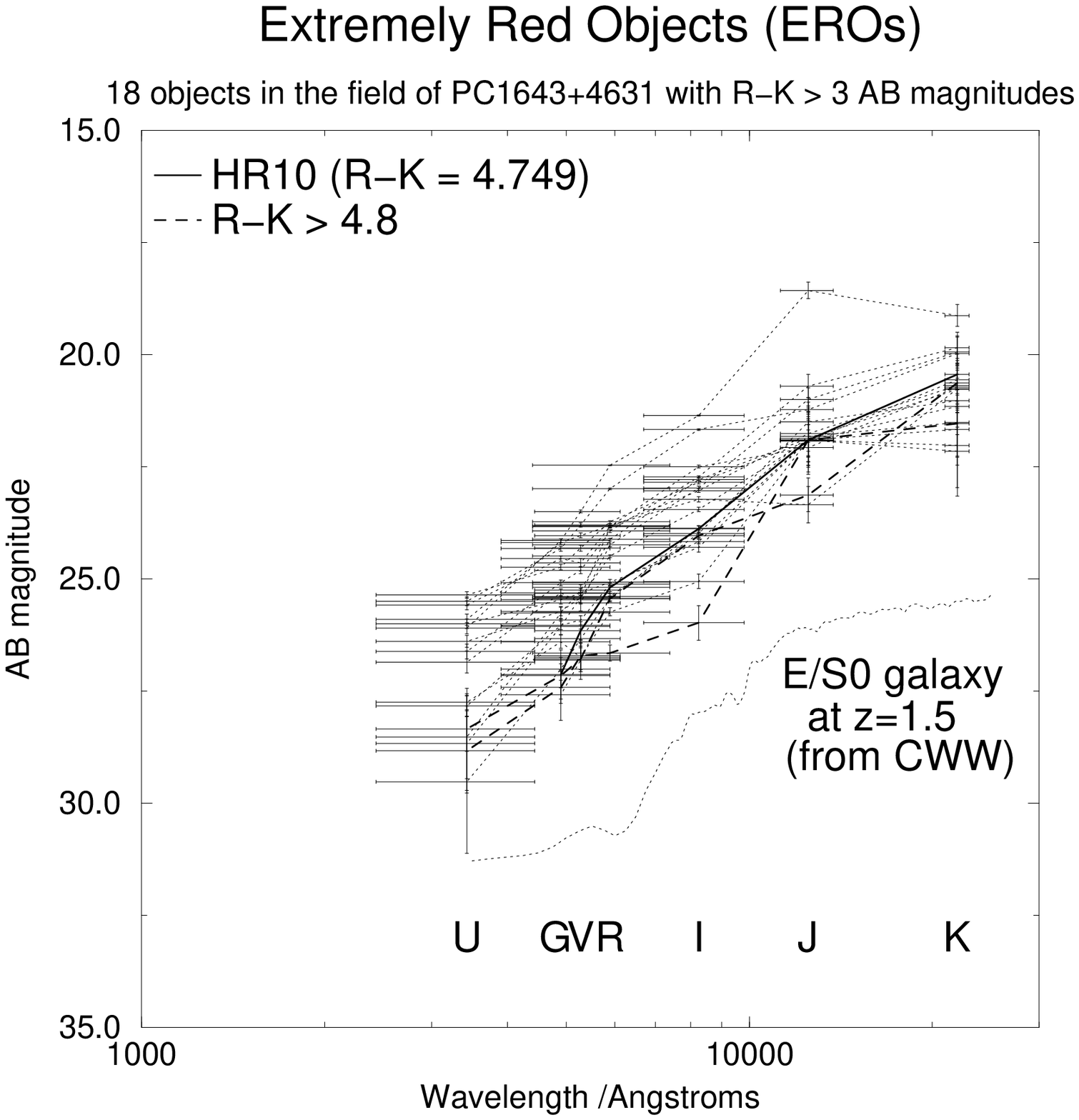,width=0.8\linewidth,
angle=270}
\caption{SEDs for the 18 objects with $R-K>3$
\label{fig:18eros}
}
\end{center}
\end{figure}

\begin{figure}
\begin{center}
\epsfig{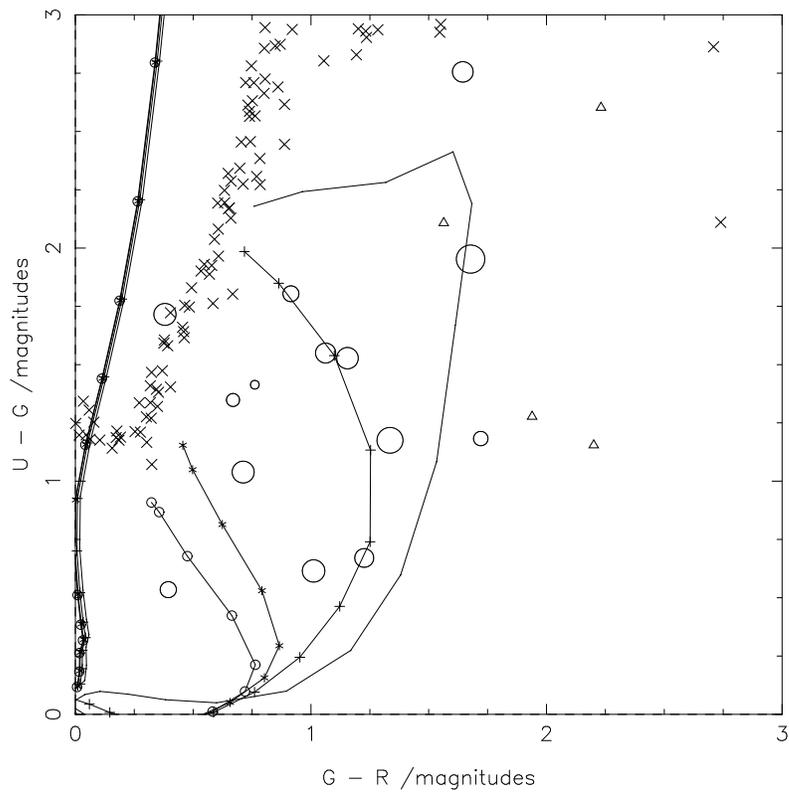}
\end{center}
\caption{Colour--colour plot of the $R-K>3$ candidates.
\label{fig:ero.uggr}
Circles denote detections in $U,G\ \&\ R$ and triangles are marked for
all objects with $U>1\sigma$. All objects are well detected in $R$
($R>>5\sigma$). The radii of the circles indicate the brightness --
smaller implies fainter. The tracks are the evolutionary tracks as in
Figure~\ref{fig:bc} and crosses are stars taken from the Gunn \&
Stryker 1983 database.}
\end{figure}

In Figure~\ref{fig:rkgr} we have plotted the optical colour against
the infrared colour. There are 18 objects with $R-K>3$, all of which
have $R>22$ (note that $R-K=3$ in the AB magnitude scheme corresponds
to $R-K=4.6$ in Johnson magnitudes); their SEDs are plotted in
Figure~\ref{fig:18eros}. Examining the $G-R$ and $U-G$ colours against
the simulations based on Bruzual \& Charlot algorithms suggests that
none of these fall into the high--redshift category defined in
section~\ref{sec:steidel}, being either too red in $G-R$ or being too
blue in $U-G$ (see Figure~\ref{fig:ero.uggr}). More critically, almost
all of them have measureable $U$ flux which would not be present if we
were imaging below the Lyman limit of these galaxies. All those
objects which have no measurable $U$ flux are at the faint limit of
the catalogue in $G$.

Note that while these objects have extreme $R-K$ colours, their
optical colours are unremarkable ($G-R\approx1$) and indistinguishable
from the rest of the catalogue. If we take an E/S0 galaxy SED, such as
that in Coleman, Wu \& Weedman \nocite{CWW} and examine the $R-K$
colours as we change the simulated source redshift, once the 4000\AA\
break moves between the $R$ and $K$ filters at $1.2<z<2.5$, we obtain
extremely red $R-K$ colours, peaking around $R-K\approx4.8$ at
$z\approx2$. At this redshift, the $G$ and $R$ filters are measuring
the small amount of UV flux in the E/S0 system, and have
$G-R\approx1$, consistent with the observed values here.

If the galaxy is undergoing star formation, the UV flux rises
dramatically, and to acheive the $R-K\approx5$ colours requires
significant reddening. If all galaxies at $z\approx2$ have a
significant amount of star formation, then the degree of reddening
required to produce the $R-K\approx5$ colours would result in further
reddening of the $G-R$ colours. Assuming a power--law reddening curve,
where\ $\textrm{extinction}\ \propto\ \lambda^{-1}$,
$\Delta(G-R)/\Delta(R-K)\approx0.4$. Given the $G-R$ colours and the
simulated galaxy colours derived from the Bruzual \& Charlot models,
this would suggest that extinction E($R-K$)$\approx2$. This is
assuming a uniform dusty extinction screen; it is more likely to be
clumpy (\cite{WTC92}) with the result that significant reddening of
the $R-K$ colours is possible while maintaining the $G-R\approx1$
colours.

To summarise, an ERO need not be a remarkable object. It can simply be
an E/S0 at $z\approx2$. Or if most galaxies at $z\approx2$ are
star--forming, reasonable amounts of reddening may in practice give
the observed spectra.

\subsection{HR10 and other galaxies with $R-K>4.5$}

We next compare our HR10 magnitudes with those already published
(\cite{HuR94}). HR10 is clearly detected in our $I,R$ \&\ $G$ filters,
and marginally in our $V$ filter given prior knowledge of its shape
and position from the other images. Because the magnitude of HR10 is
faint in all filters, the automated photometry routines are less
reliable than manual measurements, and it is these manual measurements
which are presented in Table~\ref{tab:hr10}. Where no detection is
made, $3\sigma$ upper limits are given.

\begin{figure}
\begin{center}
\epsfig{figure=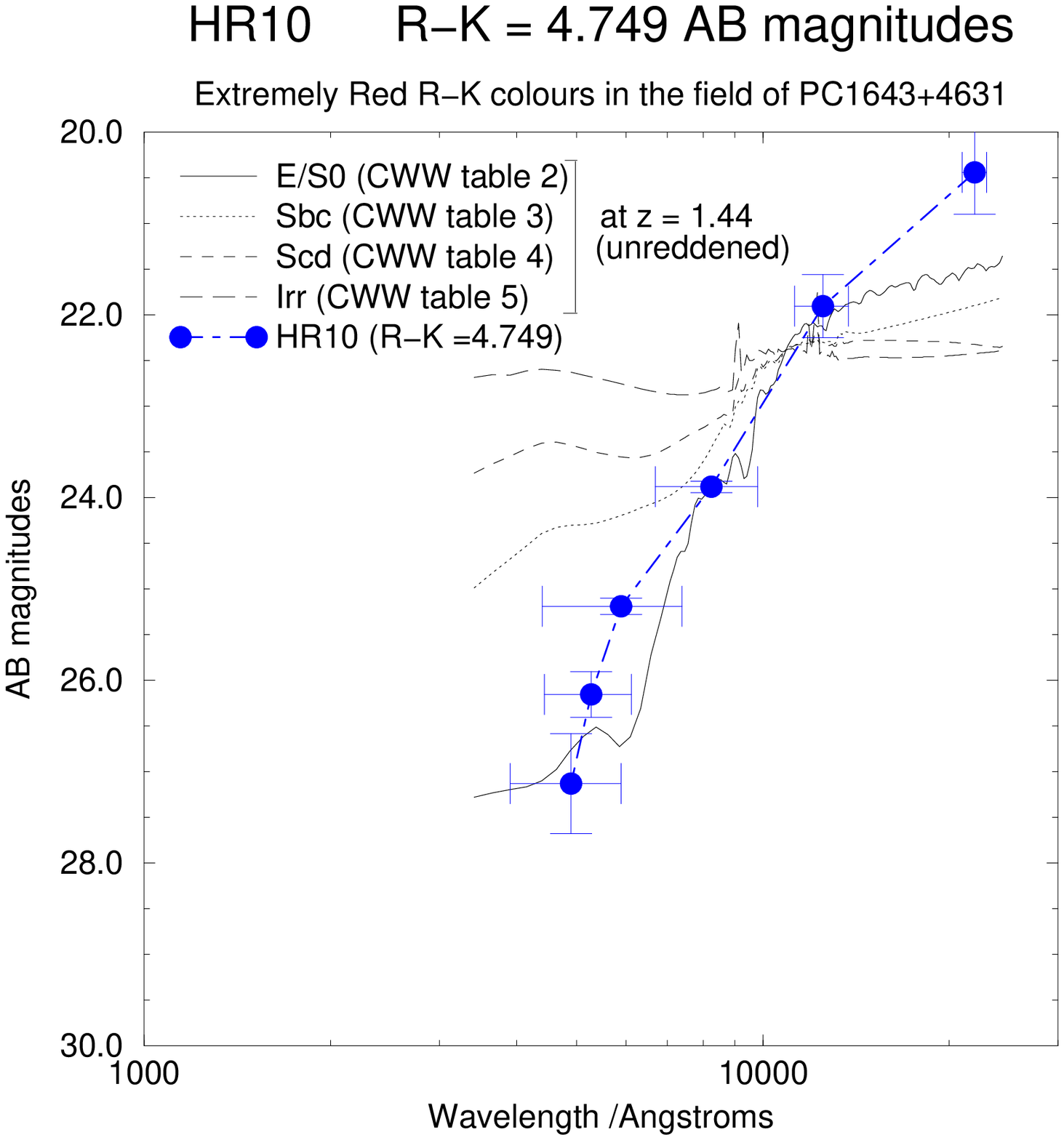,angle=270,width=0.8\linewidth}
\caption{The photometry for HR10 compared against the redshifted CWW SEDs
\label{fig:HR10sed}
}
\end{center}
\end{figure}

The $z=1.44$ redshift of Graham \& Dey \nocite{GD96} for HR 10 is
consistent with these colours. In Figure~\ref{fig:HR10sed} we plot the
SED from HR10 against the redshifted SEDs from CWW. The extremely good
agreement of the HR10 colours with that of the E/S0 galaxy is
intriguing. Compared with the B\&C models of galaxy colours for a
galaxy forming at $z=5$, one would expect the colours of an unreddened
$z=1.5$ galaxy are $U-G\approx0$ and $G-R\approx0$ -- ie extremely
blue regardless of morphology. Dust scattering models where
$E_\lambda\sim\lambda^{-1}$ result in reddening with
\((\Delta(U-G))/(\Delta(G-R)) \approx 1.8\). For
$E_\lambda\sim\lambda^{-4}$, equivalent to Rayleigh scattering,
reddening is more severe in $U-G$: \((\Delta(U-G))/(\Delta(G-R))
\approx 4.4\). The non-detection of HR10 in the $U$ image is
sufficient to give $U-G>2.0$ at the $1\sigma$ level.

The PC1643 data reveal two new objects with $R-K$ colours even more
extreme than HR10. Comparing the SEDs of these against the CWW models
suggest that they are also well fitted by an E/S0 spectrum at
redshifts of $z\approx1.7$. However, it is not possible to determine
whether these objects belong to a cluster of galaxies at this
redshift: the redshift determination, even with 7 broadband filters
stretching from 3000\AA\ to 24000\AA, cannot be done to better than
$\Delta z\approx 0.1$ and therefore determining whether these objects
could be physically associated in some structure cannot be done with
any accuracy from these data.

Little can be said about the morphology of the EROs observed here, as
the seeing is too great to resolve a significant fraction. All have
angular sizes of $1.5''$ or less, consistent with the FWHM$=0.7''$ for
HR10 and HR14 (\cite{HuR94}).

\begin{figure}
\begin{center}
\epsfig{figure=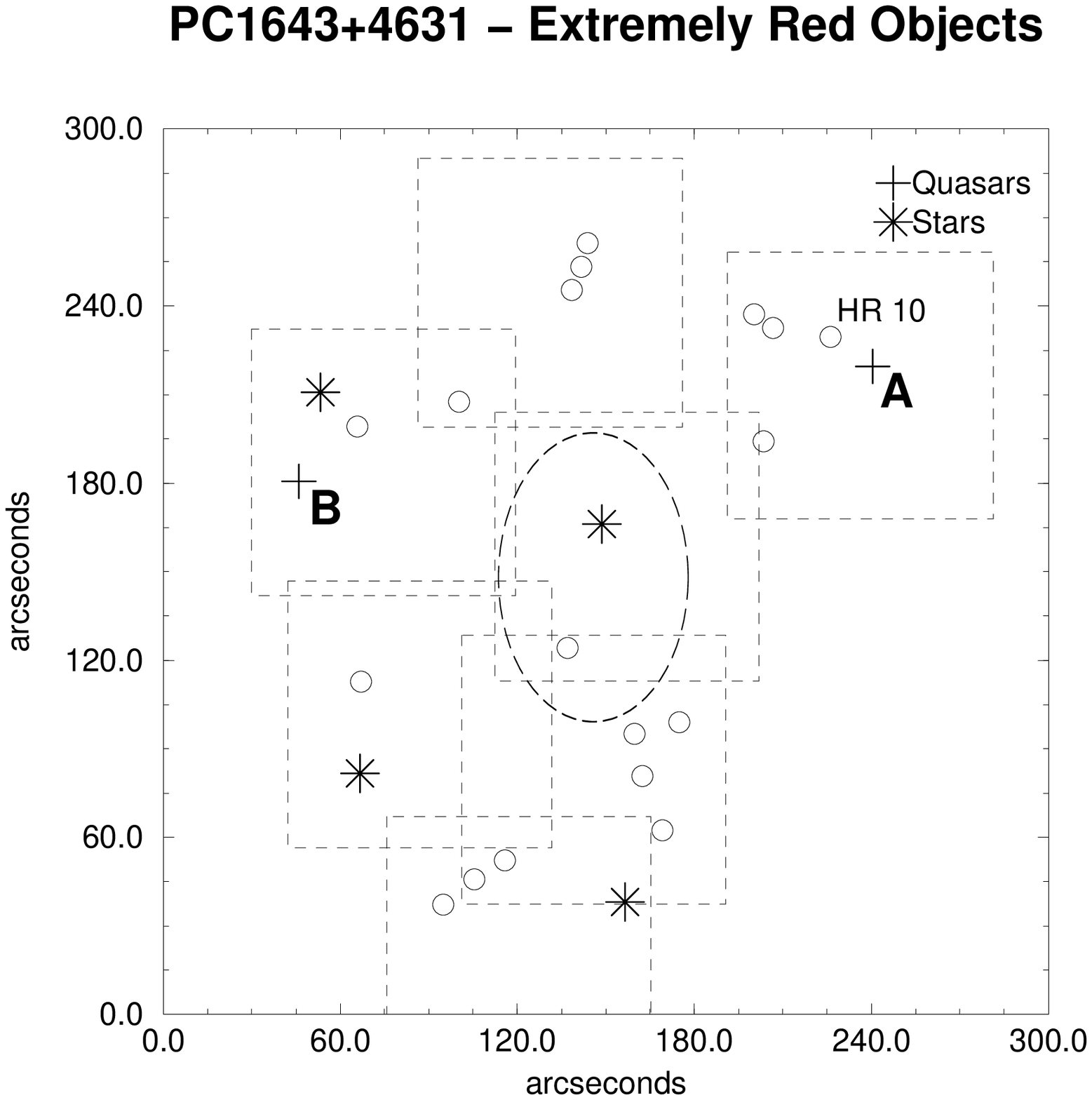,width=0.8\linewidth,angle=270}

\end{center}

\caption[Sky positions of the $R-K>3$ candidates]{Sky positions of the
$R-K>3$ candidates. The dashed ellipse shows the $1\sigma$ limits of
the position of the centre of the S-Z decrement. Note that the sky
coverage in $K$, indicated by the dashed boxes, covers only part of
the field. \label{fig:ero.space} }
\end{figure}

Figure~\ref{fig:ero.space} indicates that the distribution of the EROs
on the sky is not uniform but rather appears to be clustered into two
main groups. The first group is in the upper half of the field and
includes HR10, while the second group is in the lower half. Since the
$K$ and $J$ observations do not cover the whole area of these optical
observations, it is difficult to draw strong conclusions from this
distribution. However, treating each of the $K$ images as a
randomly--chosen independent area of sky (which is almost true -- only
the small areas of overlap are a problem here), one can estimate
whether this distribution is consistent with a random
distribution. Table~\ref{tab:erostat} shows the number of candidates
in each field. The field numbers used here are the same as in
\cite{Saun97}.

There are 24 objects apparent in 7 fields -- therefore the mean number
of objects per square is $24/7 \approx 3.4$. Using this as the
expected number of objects per field, we obtain a value of $\chi^2 =
15.15$. There are 6 degrees of freedom, and hence the probability that
this field is sampled from a uniform random distribution is 0.019 --
ie this distribution is not consistent with a uniform random
distribution at the 98\% confidence level.

\afterpage{\clearpage}

\section{Additional discussion}

From the full colour image alone, the lack of any visually obvious
clustering of similarly coloured objects is enough to suggest that
whatever system we are dealing with here is either extremely faint,
with the majority of the member galaxies having $R\gtrsim24$, or has
only a few luminous members which are confused with the other galaxies
in the field of view. Given the mass estimate derived from the S--Z
detection, a normal Abell-like cluster of the same mass as Coma should
be a distinct feature in these images if it were at $z\approx1$ (cf
\cite{LK97}).

In section \ref{sec:steidel}, we have found a 3-$\sigma$ excess in the
number of $z\approx3$ galaxies in the field, as well as an indication
of a diminution between the quasars. There would seem to be some bias
operating in this field. Gravitational lensing by a system of
$10^{15}M_\odot$ at $z\approx2$, that also produces the CMB decrement,
would produce these features. The seeing ($\approx 1''$) of the
present observations would mask the weak shear of the background
galaxies that would be present.

We also consider the possibilty that a proportion of these
high--redshift objects is also at $z\approx3.8$. Given the magnitude
limits of these observations, it is unlikely that many galaxies at
such a redshift would be visible in our images unless there is
significantly greater star--formation at these redshifts that that
found in previous high--redshift Lyman--break samples.

If the cluster of galaxies responsible for the S-Z decrement is dark,
other techniques for its detection must be examined. A rich cluster of
galaxies will gravitationally lens any background objects in the line
of sight and affect the differential galaxy counts as a result
(\cite{Kamp97},\cite{BTP95}). Given the variations of real galaxy counts
from field to field, and along the line--of--sight, it is in this case
impractical to apply this sort of analysis to this data. Since this
system lies at high redshift, the difficulty lies in identifying what
the real background population of galaxies at suitable redshifts (ie
beyond $z\sim2$) must be -- the photometric uncertainties at these
faint levels consistent with this population ($R\gtrsim25$) are enough
to statistically invalidate any strong hypothesis based on galaxy
counts or colour--colour information.

Following the gravitational lensing theory further, the same
background galaxies should show some shear distortion about the
cluster position. Out of the images presented here, only the $R$ image
is sufficiently deep to provide any hope of detecting shear in a high
redshift population. However, these galaxies are all small -- typical
aperture size at $R=25$ is of the $\lesssim30$ pixels at a 3-$\sigma$
threshold isophotal level -- and the seeing is too poor to extract any
useful shear information out of the field.

\section{Conclusions}

We have obtained deep multicolour images of an area of $5'\times5'$ in
the direction of the double quasar pair PC1643+4631 A\& B and the CMB
decrement. We have produced differential galaxy counts in five
broadband filters, which are consistent with other published results,
to $I<25.5$, $R<26.0$, $V<26.0$, $G<27.0$ and $U<28.0$. In doing so:

\begin{enumerate}[(1)]

\item{We find no cluster evident in contrast with the background. The
    distribution of galaxies in the image appears to be uniform within
    a 95\% confidence limit. Given that the CMB decrement is most
    probably caused by a cluster of galaxies, such a cluster must
    either: (a) be indistinguishable, either in colour or
    distribution, from the other galaxies in the line of sight
    requiring that the over--density of galaxies in the cluster is a
    small signal; or (b) be at lower redshift and consist of too few
    luminous members to provide any contrast.}

\item{Colour selection indicates there are some 500 intermediate
    redshift ($z\approx1.5$) galaxies candidates in the field. The
    distribution of these galaxies on the sky appears uniform.}

\item{We have detected 27 Lyman--break galaxies at $z\approx3$ with
$R<25.5$, of which 16 have $25.0<R<25.5$. This represents a 3-$\sigma$
excess over that expected for a $5'\times5'$ field {\em assuming}
Poisson-distributed galaxies with Steidel's apparent average surface
density of $0.7~\textrm{gals}/\textrm{arcmin}^2$ for $R<25.5$ and
$0.4~\textrm{gals}/\textrm{arcmin}^2$ for $R<25.0$, even though the
criteria we use to select the Lyman--break galaxies is at $least$ as
conservative as Steidel et al.}

\item{However, the distribution of the Ly--break galaxies is
inconsistent with a uniform one at the 2-$\sigma$ level. Rather, there
appears to be a hole in their distribution positioned approximately
midway between the two quasars; certainly there is no concentration of
Ly-break candidates towards the either the quasar midpoint or the CMB
decrement. The two--point correlation function for the Lyman--break
candidates in this field is consistent with other published results on
scales $<60$ arcsecs.}

\item{Points (3) \& (4) are consistent with a model in which the S--Z
    effect is caused by a massive system of $10^{15}M_\odot$ at
    $z\approx2$. This would also result in gravitational lensing of
    the background objects, including the quasars and the Ly--break
    galaxies, and is consistent with the distribution of the Ly--break
    galaxies in this field. Points (3) \& (4) are also consistent with
a genuine clustered system of Ly-break galaxies.}

\item{In a search for a third image of the quasars, several faint
    candidates were identified with consistent colours in the images;
    none of the objects is within the expected area of the sky for a
    gravitationally lensed third image. However, this does not affect
    the gravitational lensing hypothesis: the third image may be
    confused with or reddened by some object in the line--of--sight,
    or be too faint to detect.}

\item{A search for galaxies at $z=3.14$ and $z=3.81$ using
    custom--built narrow--band filters identified 3 and 6 faint
    candidates respectively.}

\item{We identified 18 EROs with $R-K>3$~AB~magnitudes, and the
    distribution of these on the sky does not appear to be uniform.
    We find evidence for a population of red galaxies consistent with
    those found in sub--mm observations but reddened out of surveys in
    optical wavelengths.}

\item{The galaxy counts for this field, which are compatible with
    other published results, including the Hubble Deep Field. The
    $U_{300}$ counts from the HDF match the PC1643 raw $U$ counts more
    closely than the corrected $U$ counts of Hogg et al, and
    others. We suggest that this may be due to assumptions made by the
    algorithms used to correct for completeness, as in \cite{Hogg97}
    and \cite{SCL90}.}

\end{enumerate}

\noindent We have carried out simulations to measure the performance of commonly
used object--finding algorithms and to investigate the possible biases
in galaxy counts and catalogues. These show that:

\begin{enumerate}[(1)]
\setcounter{enumi}{9}

\item{simulations testing the completeness of the catalogues show that
    the magnitude at which the characteristic turn--over in the raw
    differential galaxy counts occurs due to incompleteness can be
    estimated from such simulations.}

\item{To obtain sensible estimates of the corrections needed to the
    differential counts, such simulations should go at least a
    magnitude deeper than the faint limit of the plate. Estimations of
    the ``true'' differential galaxy counts are biased towards the
    assumptions used to create the simulations. Therefore, important
    features in the differential galaxy counts will not be seen unless
    the raw differential counts are effectively complete at those
    magnitudes.}

\item{FOCAS appears to be more efficient at detecting faint objects
    than SExtractor. FOCAS also appears to be superior at subdividing
    composite objects where faint components adjoin a much brighter
    companion. However, SExtractor's morphological classification
    appears to be more reliable than FOCAS, particularly for faint
    objects near the resolution limit.}

\item{Choice of isophotal aperture appears not to have a strong effect
    on the detection of Lyman--break galaxies, with both $R$ and $G$
    isophotal apertures giving similar results. Similar results are
    also observed with different colour selection criteria, suggesting
    that the object shape and size is not a strong function of
    colour.}

\end{enumerate}

\noindent Finally, we have analysed the amount of flux lost from objects
measured using isophotal apertures. We used two sets of simulations
involving recovery of artificial galaxies from both the real image of
the field and a noise--only image. These show that:

\begin{enumerate}[(1)]
\setcounter{enumi}{13}

\item{recovery of artificial galaxies from the noise--only image
significantly over--estimates the flux lost from the object, and we
find that corrections made using such a technique suffer a significant
systematic error ($\approx0.4$ magnitudes) as a result.}
 
\end{enumerate}

\section{Acknowledgements}

We would like to acknowledge the support from the staff at the
WHT. The WHT receives funding from PPARC. TH acknowledges a PPARC
studentship. GC acknowledges a PPARC Postdoctoral Research
Fellowship. We would like to thank Richard McMahon for the loan of
the $G$ filter used in these observations.

\clearpage

\bibliographystyle{apalike}
\bibliography{general}

\begin{thebibliography}{}

\bibitem[{Bertin} and {Arnouts}, 1996]{Sext}
{Bertin}, E. and {Arnouts}, S. (1996).
\newblock {SE}xtractor: Software for source extraction.
\newblock {\em Astronomy and Astrophysics Supplement Series}, 117:393--404.

\bibitem[{Brainerd} and {Smail}, 1998]{BS98}
{Brainerd}, T.~G. and {Smail}, I. (1998).
\newblock {A} {C}onstant {C}lustering {A}mplitude for {F}aint {G}alaxies?
\newblock {\em \apjl}, 494:L137--+.

\bibitem[{Broadhurst} et~al., 1995]{BTP95}
{Broadhurst}, T.~J., {Taylor}, A.~N., and {Peacock}, J.~A. (1995).
\newblock Mapping cluster mass distributions via gravitational lensing of
  background galaxies.
\newblock {\em \apj}, 438:49--61.

\bibitem[{Bruzual} and {Charlot}, 1993]{BC93}
{Bruzual}, G.~A. and {Charlot}, S. (1993).
\newblock Spectral evolution of stellar populations using isochrone synthesis.
\newblock {\em \apj}, 405:538--553.

\bibitem[{Casertano} et~al., 1995]{Cas95}
{Casertano}, S., {Ratnatunga}, K.~U., {Griffiths}, R.~E., {Im}, M.,
  {Neuschaefer}, L.~W., {Ostrander}, E.~J., and {Windhorst}, R.~A. (1995).
\newblock Structural parameters of faint galaxies from prerefurbishment
  {H}ubble {S}pace {T}elescope {M}edium {D}eep {S}urvey {O}bservations.
\newblock {\em \apj}, 453:599+.

\bibitem[{Coleman} et~al., 1980]{CWW}
{Coleman}, G.~D., {Wu}, C.~C., and {Weedman}, D.~W. (1980).
\newblock Colors and magnitudes predicted for high redshift galaxies.
\newblock {\em \apjs}, 43:393--416.

\bibitem[{Cowie} and {Hu}, 1998]{CHu98I}
{Cowie}, L.~L. and {Hu}, E.~M. (1998).
\newblock {H}igh-z {L}yman-alpha {E}mitters. {I}. {A} {B}lank-{F}ield {S}earch
  for {O}bjects {N}ear {R}edshift z=3.4 in and around the {H}ubble {D}eep
  {F}ield and the {H}awaii {D}eep {F}ield {S}{S}{A}22.
\newblock {\em \aj}.
\newblock In press.

\bibitem[{Driver} et~al., 1994]{Dri94}
{Driver}, S.~P., {Phillipps}, S., {Davies}, J.~I., {Morgan}, I., and {Disney},
  M.~J. (1994).
\newblock Multicolour faint galaxy number counts with the hitchhiker parallel
  {CCD} camera.
\newblock {\em \mnras}, 266:155+.

\bibitem[{Driver} et~al., 1995]{Dri95}
{Driver}, S.~P., {Windhorst}, R.~A., {Ostrander}, E.~J., {Keel}, W.~C.,
  {Griffiths}, R.~E., and {Ratnatunga}, K.~U. (1995).
\newblock The morphological mix of field galaxies to m {I} = 24.25 mag (b$_{J}$
  approximately 26 mag) from a {D}eep {H}ubble {S}pace {T}elescope {WFPC}2
  image.
\newblock {\em \apjl}, 449:L23--+.

\bibitem[{Frayer} et~al., 1994]{FBB94}
{Frayer}, D.~T., {Brown}, R.~L., and {Vanden Bout}, P.~A. (1994).
\newblock Co emission from the z = 3.137 damped {L}y-alpha system toward {P}{C}
  1643+4631{A}.
\newblock {\em \apjl}, 433:L5--L8.

\bibitem[Giavalisco et~al., 1998]{GSA98}
Giavalisco, M., Steidel, C.~C., Adelberger, K.~L., Dickinson, M.~E., Pettini,
  M., and Kellogg, M. (1998).
\newblock {T}he {A}ngular {C}lustering of {L}yman-{B}reak {G}alaxies at
  {R}edshift z=3.
\newblock {\em \aj}.
\newblock In press.

\bibitem[{Glazebrook} et~al., 1995]{Gla95}
{Glazebrook}, K., {Ellis}, R., {Santiago}, B., and {Griffiths}, R. (1995).
\newblock The morphological identification of the rapidly evolving population
  of faint galaxies.
\newblock {\em \mnras}, 275:L19--L22.

\bibitem[{Graham} and {Dey}, 1996]{GD96}
{Graham}, J.~R. and {Dey}, A. (1996).
\newblock The redshift of an {E}xtremely {R}ed {O}bject and the nature of the
  very red galaxy population.
\newblock {\em \apj}, 471:720+.

\bibitem[{Gunn} and {Stryker}, 1983]{GS83}
{Gunn}, J.~E. and {Stryker}, L.~L. (1983).
\newblock Stellar spectrophotometric atlas, wavelengths from 3130 to 10800
  {\aa}.
\newblock {\em \apjs}, 52:121--153.

\bibitem[{Hall} and {Mackay}, 1984]{HM84}
{Hall}, P. and {Mackay}, C.~D. (1984).
\newblock Faint galaxy number-magnitude counts at high galactic latitude. {I}.
\newblock {\em \mnras}, 210:979--992.

\bibitem[{Hawkins} et~al., 1997]{HCF97}
{Hawkins}, M. R.~S., {Clements}, D., {Fried}, J.~W., {Heavens}, A.~F., {Veron},
  P., {Minty}, E.~M., and {Van Der Werf}, P. (1997).
\newblock The double quasar {Q}2138-431: lensing by a dark galaxy?
\newblock {\em \mnras}, 291:811--818.

\bibitem[{Hogg} et~al., 1997]{Hogg97}
{Hogg}, D.~W., {Pahre}, M.~A., {McCarthy}, J.~K., {Cohen}, J.~G., {Blandford},
  R., {Smail}, I., and {Soifer}, B.~T. (1997).
\newblock Counts and colours of faint galaxies in the {U} and {R} bands.
\newblock {\em \mnras}, 288:404--410.

\bibitem[{Hu}, 1998]{Hu98}
{Hu}, E.~M. (1998).
\newblock Pushing back studies of {G}alaxies toward the dark ages:
  {H}igh-{R}edshift {L}yman alpha {E}mission-{L}ine {G}alaxies in the field.
\newblock {\em astroph-9801170}.
\newblock preprint.

\bibitem[{Hu} and {Ridgway}, 1994]{HuR94}
{Hu}, E.~M. and {Ridgway}, S.~E. (1994).
\newblock Two extremely red galaxies.
\newblock {\em \aj}, 107:1303--1306.

\bibitem[{Jarvis} and {Tyson}, 1981]{FOCAS81}
{Jarvis}, J.~F. and {Tyson}, J.~A. (1981).
\newblock {FOCAS} - {F}aint {O}bject {C}lassification and {A}nalysis {S}ystem.
\newblock {\em \aj}, 86:476--495.

\bibitem[{Jones} et~al., 1997]{Jones97}
{Jones}, M.~E., {Saunders}, R., {Baker}, J.~C., {Cotter}, G., {Edge}, A.,
  {Grainge}, K., {Haynes}, T., {Lasenby}, A., {Pooley}, G., and {Rottgering},
  H. (1997).
\newblock Detection of a {C}osmic {M}icrowave {B}ackground {D}ecrement toward
  the z = 3.8 quasar pair {PC} 1643+4631{A},{B}.
\newblock {\em \apjl}, 479:L1--+.

\bibitem[{Keeton} and {Kochanek}, 1998]{KK98}
{Keeton}, C.~R. and {Kochanek}, C.~S. (1998).
\newblock Gravitational lensing by spiral galaxies.
\newblock {\em \apj}, 495:157+.

\bibitem[{Koo} et~al., 1986]{Koo86}
{Koo}, D.~C., {Kron}, R.~G., {Nanni}, D., {Trevese}, D., and {Vignato}, A.
  (1986).
\newblock A multicolor photometric catalog of galaxies and stars in the field
  of the rich cluster {II} {ZW} 1305.4 + 2941 at z = 0.24.
\newblock {\em \aj}, 91:478--493.

\bibitem[{Le Fevre} et~al., 1995]{LeF95}
{Le Fevre}, O., {Crampton}, D., {Lilly}, S.~J., {Hammer}, F., and {Tresse}, L.
  (1995).
\newblock The {C}anada-{F}rance {R}edshift {S}urvey. {II}. {S}pectroscopic
  program: Data for the 0000-00 and 1000+25 fields.
\newblock {\em \apj}, 455:60+.

\bibitem[{Luppino} and {Kaiser}, 1997]{LK97}
{Luppino}, G.~A. and {Kaiser}, N. (1997).
\newblock Detection of weak lensing by a cluster of galaxies at z = 0.83.
\newblock {\em \apj}, 475:20+.

\bibitem[{Metcalfe} et~al., 1996]{MSC96}
{Metcalfe}, N., {Shanks}, T., {Campos}, A., {Fong}, R., and {Gardner}, J.~P.
  (1996).
\newblock Galaxy formation at high redshifts.
\newblock {\em \nat}, 383:236+.

\bibitem[{Oke} and {Gunn}, 1983]{OG83}
{Oke}, J.~B. and {Gunn}, J.~E. (1983).
\newblock Secondary standard stars for absolute spectrophotometry.
\newblock {\em \apj}, 266:713--717.

\bibitem[Pettini et~al., 1997]{PSA97}
Pettini, M., Steidel, C.~C., Adelberger, K., Kellogg, M., Dickinson, M., and
  Giavalisco, M. (1997).
\newblock {T}he {D}iscovery of {P}rimeval {G}alaxies and the {E}poch of
  {G}alaxy {F}ormation.
\newblock In Shull, J., Woodward, C., and Thronson, H., editors, {\em Origins},
  {ASP} {C}onference {S}eries.

\bibitem[{Saunders} et~al., 1997]{Saun97}
{Saunders}, R., {Baker}, J.~C., {Bremer}, M.~N., {Bunker}, A.~J., {Cotter}, G.,
  {Eales}, S., {Grainge}, K., {Haynes}, T., {Jones}, M.~E., {Lacy}, M.,
  {Pooley}, G., and {Rawlings}, S. (1997).
\newblock Optical and infrared investigation toward the z = 3.8 quasar pair
  {PC} 1643+4631{A}, {B}.
\newblock {\em \apjl}, 479:L5--+.

\bibitem[{Schechter} et~al., 1998]{SGB98}
{Schechter}, P.~L., {Gregg}, M.~D., {Becker}, R.~H., {Helfand}, D.~J., and
  {White}, R.~L. (1998).
\newblock The first {FIRST} gravitationally lensed quasar: {FBQ} 0951+2635.
\newblock {\em Submitted to \aj}.
\newblock Preprint astro-ph/9710120.

\bibitem[{Schneider} et~al., 1994]{ptgs94}
{Schneider}, D.~P., {Schmidt}, M., and {Gunn}, J.~E. (1994).
\newblock Spectroscopic {CCD} surveys for quasars at large redshift. 3: The
  {P}alomar {T}ransit {G}rism {S}urvey catalog.
\newblock {\em \aj}, 107:1245--1269.

\bibitem[{Smail} et~al., 1995]{SHY95}
{Smail}, I., {Hogg}, D.~W., {Yan}, L., and {Cohen}, J.~G. (1995).
\newblock Deep optical galaxy counts with the {K}eck telescope.
\newblock {\em \apjl}, 449:L105--+.

\bibitem[{Songaila} et~al., 1990]{SCL90}
{Songaila}, A., {Cowie}, L.~L., and {Lilly}, S.~J. (1990).
\newblock Galaxy formation and the origin of the ionizing flux at large
  redshift.
\newblock {\em \apj}, 348:371--377.

\bibitem[{Steidel} et~al., 1998]{SAD98}
{Steidel}, C.~C., {Adelberger}, K.~L., {Dickinson}, M., {Giavalisco}, M.,
  {Pettini}, M., and {Kellogg}, M. (1998).
\newblock A large structure of galaxies at redshift z approximately 3 and its
  cosmological implications.
\newblock {\em \apj}, 492:428+.

\bibitem[{Steidel} et~al., 1996]{SGP96}
{Steidel}, C.~C., {Giavalisco}, M., {Pettini}, M., {Dickinson}, M., and
  {Adelberger}, K.~L. (1996).
\newblock Spectroscopic confirmation of a population of normal star-forming
  galaxies at redshifts z $>$ 3.
\newblock {\em \apjl}, 462:L17--+.

\bibitem[{Steidel} and {Hamilton}, 1993]{StII}
{Steidel}, C.~C. and {Hamilton}, D. (1993).
\newblock Deep imaging of high redshift {QSO} fields below the {L}yman limit.
  {II} - {N}umber counts and colors of field galaxies.
\newblock {\em \aj}, 105:2017--2030.

\bibitem[{Surdej} et~al., 1997]{SCR97}
{Surdej}, J., {Claeskens}, J.~F., {Remy}, M., {Refsdal}, S., {Pirenne}, B.,
  {Prieto}, A., and {Vanderriest}, C. (1997).
\newblock {HST} confirmation of the lensed quasar {J}03.13.
\newblock {\em \aap}, 327:L1--L4.

\bibitem[{Thompson} et~al., 1995]{TDT95}
{Thompson}, D., {Djorgovski}, S., and {Trauger}, J. (1995).
\newblock A narrow band imaging survey for primeval galaxies.
\newblock {\em \aj}, 110:963+.

\bibitem[{Tody}, 1993]{IRAF93}
{Tody}, D. (1993).
\newblock {IRAF} in the nineties.
\newblock {\em Astronomical Data Analysis Software and Systems {II},
  {A}.{S}.{P}. {C}onference {S}eries, {V}ol. 52, 1993, {R}. J. {H}anisch, {R}.
  J. {V}. Brissenden, {a}nd {J}eannette {B}arnes, eds., p. 173.}, 2:173+.

\bibitem[{Tyson}, 1988]{Tyson88}
{Tyson}, J.~A. (1988).
\newblock Deep {CCD} survey - {G}alaxy luminosity and color evolution.
\newblock {\em \aj}, 96:1--23.

\bibitem[{van Kampen}, 1997]{Kamp97}
{van Kampen}, E. (1997).
\newblock Cluster mass estimation from lens magnification.
\newblock In {\em {L}arge {S}cale {S}tructure: tracks and traces}. {P}otsdam,
  {W}orld {S}cientific.
\newblock Preprint astro-ph/9711178.

\bibitem[{Williams} et~al., 1996]{HDF96}
{Williams}, R.~E., {Blacker}, B., {Dickinson}, M., {Dixon}, W. V.~D.,
  {Ferguson}, H.~C., {Fruchter}, A. w.~S., {Giavalisco}, M., {Gilliland},
  R.~L., {Heyer}, I., {Katsanis}, R., {Levay}, Z., {Lucas}, R.~A., {McElroy},
  D.~B., {Petro}, L., and {Postman}, M. (1996).
\newblock The {H}ubble {D}eep {F}ield: {O}bservations, data reduction, and
  galaxy photometry.
\newblock {\em \aj}, 112:1335+.

\bibitem[{Witt} et~al., 1992]{WTC92}
{Witt}, A.~N., {Thronson}, Harley~A., J., and {Capuano}, John~M., J. (1992).
\newblock Dust and the transfer of stellar radiation within galaxies.
\newblock {\em \apj}, 393:611--630.

\end{thebibliography}

\clearpage

\begin{table*}[hbt]
\begin{center}
\begin{tabular}{crrrr} 
 
Filter 	& \multicolumn{4}{c}{Exposure time /s} \\ 
	& 15 April & 16 April & 17 April & 18 April \\
 
$U$      & 6,300 &  ---  & 3,600 & 2,700 \\     
$G$      &  ---  & 3,000 &  ---  &  ---  \\     
$V$      &  ---  & 1,200 &  ---  &   993 \\     
$R$      & 2,700 &  ---  & 4,500 &  ---  \\     
$I$      &  ---  &  ---  &  ---  & 2,700 \\     
$L5840$  & 1,800 & 2,400 &  ---  &  ---  \\     
$S5040$  &  ---  &  ---  &  ---  & 1,800 \\   
\end{tabular}                                   
\smallskip
\caption{Observing log for 15 -- 18 April 1996 \label{tab:obslog} }
\end{center}
\smallskip
\end{table*}

\clearpage
\clearpage
\begin{table*}
\begin{center}

\begin{tabular}{ccccc} 
 Filter & Exposure   & Zero--point & RMS noise      & 3$\sigma$ faint magnitude  \\
       & time (total) & magnitude & of image       & limit for $2''$ radius \\
       & /second     &            & /ADU$^a$       & circular aperture \\  
$U$      & 12,600      & 25.067     & 5.99 $\pm$ 0.2 & 27.01 \\
$G$      &  3,000      & 25.938     & 14.3 $\pm$ 1.0 & 26.49 \\
$V$      &  2,194      & 26.117     & 17.5 $\pm$ 1.2 & 26.01 \\
$R$      &  6,600      & 26.324     & 20.1 $\pm$ 1.2 & 26.51 \\
$I$      &  2,700      & 25.719     & 22.2 $\pm$ 0.9 & 25.04 \\
$L5840$  &  4,200      & 23.861     & 13.5 $\pm$ 0.8 & 24.92 \\
$S5040$  &  1,800      & 23.769     & 5.16 $\pm$ 0.3 & 25.43 \\ 
\end{tabular}                                   
\smallskip
\caption{Calibration information for the PC1643+4631 images presented
here. Note that seeing is $\approx$ 1.1 arcseconds FWHM for all images
except $U$ where it is $\approx$ 1.5 arcseconds. 
\label{tab:calibrat} 
}
\end{center}
\small{$^a$ Analogue to Digital Unit -- the number of electrons per ADU
is described by the gain, which is 1.6~$e^-$/ADU for all observations,
except those in $U$ where the gain is 1.2~$e^-$/ADU}
\smallskip
\end{table*}

\clearpage
\begin{table*}
\centering
\begin{tabular}{cc} 
Filter & 50\% complete   \\
       & magnitude limit \\  
$U$      &  26.0           \\
$G$      &  25.5           \\
$V$      &  25.3           \\
$R$      &  25.6           \\
$I$      &  24.6           \\ 
\end{tabular}
\smallskip
\caption{Completeness limits for the broadband filters
\label{tab:complete}
}
\smallskip
\end{table*}

\clearpage
\begin{table*}
\centering
\begin{tabular}{cc} 
Isophotal & Number of intermediate  \\
aperture  & redshift candidates$^a$ \\  
$U$         & 501  \\
$G$         & 592  \\
$V$         & 547  \\
$R$         & 494  \\
$I$         & 468  \\ 
Self        & 460  \\ 
\end{tabular}
\smallskip
\caption{Completeness limits for the broadband filters
\label{tab:intersel}
}
\smallskip
$^a$ \small{Using $R<26.5$, $0<G-R<0.5$ \& $0<U-G<2$ in all cases}
\end{table*}

\clearpage
\begin{table*}
\centering
\begin{tabular}{ccc} 
Isophotal & Number of candidates matched & Number fulfilling \\
aperture  & to isophotal apertures       & colour criteria   \\ 
          & within 1.6''                 & \\  
$U$       &          11                  &      5            \\
$G$       &          26                  &     20            \\
$V$       &          26                  &     14            \\
$I$       &          22                  &     10            \\ 
Self      &          27                  &     18            \\ 
\end{tabular}
\smallskip
\caption{Comparison of the positions and colour selection criteria of
high--redshift galaxies, based on the criteria outlined in
section~\ref{sec:steidel}
\label{tab:steidel.diff}
}
\smallskip
\end{table*}

\clearpage
\begin{table}
\begin{center}
\begin{tabular}{cr@{.}l@{ $\pm$ }r@{.}l}
Filter	&\multicolumn{4}{c}{Magnitude /AB} \\  
$U$	& $>$27&7 & 0&7 \\
$G$	& 26&5  & 0&5 \\
$V$	& $>$26&0 & 0&5 \\
$R$	& 25&6  & 0&4 \\
$I$	& 24&5  & 0&4 \\ 
\end{tabular}
\end{center}
\caption{AB magnitudes for HR10
\label{tab:hr10}
}
\end{table}

\clearpage
\begin{table}
\begin{center}
\begin{tabular}{cc} 
Field   & Number of candidates  \\

0       &        1  \\
1       &        7  \\
2       &        3  \\
3       &        1  \\
4       &        4  \\
5       &        2  \\
6       &        4  \\

Apparent total & 24  \\

\end{tabular}
\caption{\label{tab:erostat}
There are 7 slightly overlapping observations in K. These have the
following number of EROs (classified as having $R-K > 3$). There are
18 individual candidates, some of which appear in more than one frame,
hence the larger apparent total.}
\end{center}
\end{table}

\clearpage
\clearpage

\clearpage

\clearpage

\clearpage

\clearpage

\clearpage

\clearpage

\clearpage
\begin{figure*}
\begin{center}

This figure is avaliable at
ftp.mrao.cam.ac.uk:/pub/PC1643/paper1.figure18.ps

\caption{Full colour picture of PC1643+4631, comprising $U$,$G$,$V$,$R$ and $I$
images
\label{fig:fullcolour}
}
\end{center}
The field of view is $5'\times5'$. The two quasars in the field are
marked A \& B.  In order to represent the full spectrum, the five
filters have been combined as follows:- \\
\smallskip
\begin{center}
\begin{tabular}{r@{$=$}c@{+}c}
Red   &      $I$      &$\frac{2}{3} R$  \\
Green & $\frac{1}{3}R$&$V+\frac{1}{3} G$ \\
Blue  & $\frac{2}{3}G$&$U$         \\
\end{tabular}
\end{center}
\end{figure*}

\clearpage
\clearpage
\clearpage
\clearpage
\clearpage

\clearpage

\clearpage

\clearpage

\clearpage

\clearpage

\clearpage

\clearpage

\clearpage

\clearpage
\clearpage

\clearpage

\clearpage

\clearpage

\clearpage

\clearpage

\clearpage

\clearpage

\end{document}